\documentclass[usenatbib]{mn2e}
\usepackage{epsfig}
\usepackage{subfigure}
\usepackage{pstricks}

\def\apj{ApJ}
\def\aap{A\&A}
\def\aaps{A\&AS}
\def\apjl{ApJL}

\def\mnras{MNRAS}
\def\aj{AJ}

\def\apss{Ap\&SS}
\def\nat{Nature}
\def\pasj{PASJ}

\title{The dynamics of tidal tails from massive satellites}

\author[J.-H. Choi et al.]{
Jun-Hwan Choi, Martin D. Weinberg \& Neal Katz
\\ 
Department of Astronomy, University of Massachusetts, Amherst,  MA 01003
\\
jhchoi@nova.astro.umass.edu, weinberg@astro.umass.edu, nsk@kaka.astro.umass.edu
}

\begin{document}

\maketitle

\begin{abstract}
  We investigate the dynamical mechanisms responsible for producing
  tidal tails from dwarf satellites using N-body simulations.  We
  describe the essential dynamical mechanisms and morphological
  consequences of tail production in satellites with masses greater
  than 0.0001 of the host halo virial mass.  We identify two important
  dynamical co-conspirators: 1) the points where the attractive force
  of the host halo and satellite are balanced (X-points) do not occur
  at equal distances from the satellite centre or at the same
  equipotential value for massive satellites, breaking the
  morphological symmetry of the leading and trailing tails; and 2) the
  escaped ejecta in the leading (trailing) tail continues to be
  decelerated (accelerated) by the satellite's gravity leading to
  large offsets of the ejecta orbits from the satellite orbit.  The
  effect of the satellite's self gravity decreases only weakly with a
  decreasing ratio of satellite mass to host halo mass, proportional to
  $(M_s/M_h)^{1/3}$, demonstrating the importance of these effects over
  a wide range of subhalo masses.  Not only will the morphology of the
  leading and trailing tails for massive satellites be different, but
  the observed radial velocities of the tails will be displaced from
  that of the satellite orbit; both the displacement and the maximum
  radial velocity is proportional to satellite mass.  If the tails are
  \emph{assumed} to follow the progenitor satellite orbits, the tails
  from satellites with masses greater than 0.0001 of the host halo
  virial mass in a spherical halo will \emph{appear} to indicate a
  flattened halo.  Therefore, a constraint on the Milky Way halo shape
  using tidal streams requires mass-dependent modelling.  Similarly,
  we compute the the distribution of tail orbits both in
  $E_{r}-r^{-2}$ space (Lynden-Bell \& Lynden-Bell 1995) and in $E-L_{z}$ space
  (Helmi \& de Zeeuw 2000), advocated for identifying satellite stream relics.
  The acceleration of ejecta by a massive satellite during escape
  spreads the velocity distribution and obscures the signature of a
  well-defined ``moving group'' in phase space.  Although these
  findings complicate the interpretation of stellar streams and moving
  groups, the intrinsic mass dependence provides additional leverage
  on both halo and progenitor satellite properties.
\end{abstract}


\section{Introduction}
\label{sec:intro}

According to the currently favoured galaxy formation scenario, the cold
dark matter (CDM) cosmogony, galaxies are built up from the assembly
of small structures.  In this paradigm the assembly mechanism plays a
key role in understanding the formation history of galaxies.  Recent
cold dark matter cosmological numerical simulations predict the
existence of a large population of \emph{subhalos}.  Comparisons with
the observed population of dwarf galaxies and detailed predictions of the
present-day subhalo population, dark or luminous, have become
important tests of the CDM galaxy formation paradigm
\citep{Ghigna98, Ghigna00, DLucia04, Diemand04, Gao04, OL04}.
Most studies to date use large cosmological simulations and
classify their properties statistically. However, to properly investigate these
processes, one needs to perform high resolution idealised simulations
of subhalo evolution within the CDM paradigm \citep{Hayashi03}.
Alternatively, in this study, we investigate one important consequence 
of subhalo disruption: the formation and evolution of tidal tails.  By
adopting initial conditions motivated by the CDM simulations, we can
focus our computational resources on understanding the dynamical mechanism.

Satellite galaxy tidal tails are an important observable
fossil signature to help understand the formation history of the Milky
Way and to test CDM theory as a consequence.  Tails and streams
provide information about the Galactic halo mass model as well as the
evolutionary history of the observed satellite galaxy.  In the CDM
model, galaxies are embedded in massive dark matter halos.  Estimating
dark matter halo structure is essential to understand galaxy formation
and tidal tail morphology probes halo structure
\citep{Johnston99,HdZ00,Ibata01a,Ibata01b}.  Several space missions,
for example the ESA astrometric satellite GAIA \citep{Lindegren96,
  Perryman01}, are planned to measure the position and motion of stars
in the Milky Way with very high accuracy, in the near future.
Together with ground-based radial velocity experiments, e.g. RAVE \footnote
{See http://www.rave-survey.org}\citep{Steinmetz06}, these surveys
will provide full phase space information.  Accurate six dimensional
phase space information of Milky Way stars will provide observational
information of the tidal tail and hence the formation history of the
Milky Way.  The time is ripe to carry out a detailed theoretical study
of satellite galaxy disruption and the induced tidal tail morphology.
  
In this study we perform numerical simulations of satellite galaxy
disruption and its induced tidal tail morphology within the CDM
cosmogony.  The objective of this study is to understand the physical
processes responsible for satellite galaxy disruption rather than
reproducing the evolutionary history of any individual Milky Way
satellite galaxy.  In particular, satellite disruption in N-body
simulations is produced by escaping satellite particles.  In addition,
the gravitational shock, which is caused by the slowly varying host halo
potential as the satellite goes through its orbit, changes the
satellite's internal structure.  An initially stable satellite galaxy
and accurate numerical integration of a satellite particle's orbit are
necessary to represent these physical processes correctly.  We
investigate satellite galaxy evolution by performing high resolution
and low noise N-body simulations with such stable initial conditions.

In addition, we can estimate any trends of satellite tidal tail
morphology with satellite properties from our simulations, even though
we do not reproduce the evolutionary history of any specific Milky Way
satellite.  Our simulation results show that the gravity of the satellite
alters the location of the tidal tails relative to the satellite
orbit.  The satellite decelerates (accelerates) the leading (trailing)
tail beyond the tidal radius, which is proportional to the satellite
mass.  For more massive satellites, this results in the leading tail
being located well inside and the trailing tail being located well
outside of the satellites orbit, rather than tracking the original
satellite orbit \citep{Johnston96, Johnston01, MD94}.  Since the
satellite torques the tidal tail, the distribution of the tidal tail
in the observational plane is rather different from predictions that
exclude such satellite torquing.  In addition, the simulations provide
six-dimensional phase-space information of the tidal tail to compare
with upcoming astrometric measurements.

In Section, \ref{sec:simulation} we describe how we make stable satellite
initial conditions and provide a brief overview of the simulation
algorithm.  In Section, \ref{sec:satellite} we investigate satellite
disruption and the formation of the tidal tail, including the effects
of the satellite potential on the tidal tail, and in
Section \ref{sec:observation}, we investigate the observational
consequences.  In Section \ref{sec:summary}, we summarise our results and
discuss their importance.

\section{Initial conditions and N-body methodology}
\label{sec:simulation}

\begin{table}
\caption{Initial properties of the three satellite models}
\label{tbl:Satellites}
\begin{tabular}{cccc}
\hline
Satellite  & $M/M_{vir,host}$ & $R/R_{vir,host}$ & $V_{max}/V_{max,host}$  \\
\hline
Massive    & $1.9\times 10^{-2}$  & $9.02 \times 10^{-2}$  & 0.45 \\
Low-Mass   & $9.0\times 10^{-4}$  & $3.38 \times 10^{-2}$  & 0.16 \\
Tiny-Mass  & $9.9\times 10^{-5}$  & $1.66 \times 10^{-2}$  & 0.08 \\
\hline
\end{tabular}

\end{table}

The initial conditions of our simulations are motivated by the CDM
cosmology.  CDM cosmological simulations suggest that dark matter
halos have a universal density profile \citep[hereafter NFW]{NFW97},
$\rho(r) \propto r^{-1}(r+r_{s})^{-2}$, where $r_{s}$ is a scale
length characterised by the concentration parameter $c=R_{vir}/r_{s}$
and $R_{vir}$ is the virial radius of the halo.  Although there are
some disagreements regarding the accuracy of this simple formula in
describing halos in numerical simulations and in comparisons with
observed galaxies, it remains the accepted CDM halo density profile
and we adopt it for our study.

We represent the host halo by a static NFW halo potential.  Hence our
simulations ignore the effects of dynamical friction and the
subsequent reaction of the host halo. Obviously this is not realistic
but since the satellite masses of interest are often much smaller than
the host halo mass, the consequences are minor.  Moreover, the prime
motivation of this study is to understand the physical processes
responsible for satellite disruption and tidal tail formation and not
the evolutionary history of the satellite.  Dynamical friction is not
vital to understand these processes.

\begin{figure}
\centerline{\includegraphics[width=1.0\columnwidth,angle=0] {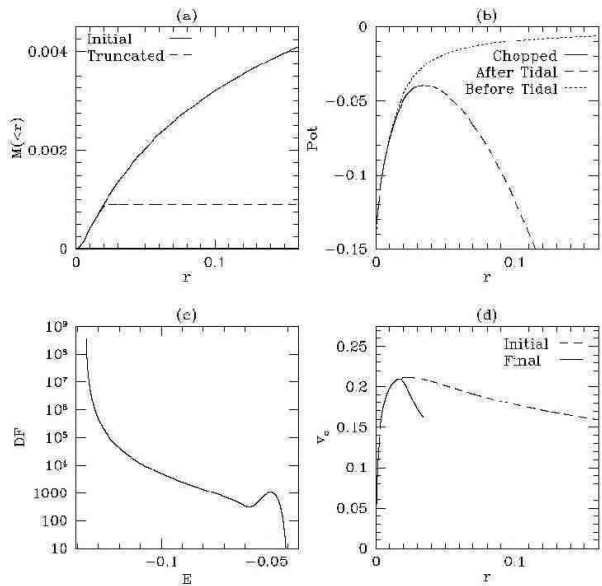}}
\caption{The effect of our truncation procedure (see text) on a satellite's
  initial NFW profile. (a) The enclosed mass profile.
  (b) The effective potential profile.
  (c) The Distribution function versus energy.
  (d) The circular velocity profile.
  We use \emph{system} units unless otherwise specified: $G=1$, $M_{vir, host}=1$,
  and $R_{vir, host}=1$.
}
\label{fig:procedure}
\end{figure}

Before tidal truncation, the satellites have an NFW density profile.  However, the
NFW profile extends to infinity and real astronomical systems have a
finite size.  The conventional solution limits the size of an isolated
dark matter halo to its virial radius \citep{GG72,BN98}.  The host
halo's tidal field then determines the size of satellite halo.  The
host halo's tidal field affects a satellite halo even before the
satellite halo passes within the host halo's virial radius.
As a result, it is computationally expensive to simulate the 
entire evolution of a satellite.
Remember, the objective of our study is to understand the physical
processes responsible for tidal tail formation not the reproduction of
a particular tail feature. Therefore, we place the satellite in the
host halo on the desired orbit to start and 
include the host halo's tidal field when we generate a 
satellite's initial phase-space distribution.   It is natural to
characterise the tidal length scale in the satellite by the radius of the 
X-point that this satellite would have at some fiducial galactocentric radius
in its orbit.
We call this fiducial radius the \emph{tidal distance}.  In other
words, the satellite on a circular orbit at the tidal distance would
have the X-point $r_\times$.  At this point, the gravitational force from
the satellite exactly balances the gravitational force from the host
halo and non-inertial centrifugal force.  

The details of the
iterative procedure that we use to generate the satellite's initial
condition is as follows.  First, we
truncate the virial radius limited NFW satellite halo at the X-point
radius, $r_\times$, using the error function, $\{1-\hbox{erf}[(r-r_\times)/s]\}$.
We then compute the distribution
function using Eddington inversion and calculate a new satellite
density profile by integrating the distribution function over
velocity.  
The parameter $s$ in the error function truncation formula is
increased 
from zero until a
smooth phase-distribution function results.  We iterate these steps
until the density--distribution-function pair is converged.  Figure
\ref{fig:procedure} shows an example of how this procedure modifies an
initial satellite halo. At the conclusion of the procedure, the
effective tidal radius is approximately 75\% of the initial X-point
radius radius.  We characterise a satellite halo by its \emph{initial} maximum
circular velocity; Figure \ref{fig:procedure} demonstrates that this
velocity is only weakly affected by the truncation procedure.  
We denote the outer radius of the satellite 
after the truncation procedure as the \emph{effective tidal radius}.
It is smaller than $r_\times$ owing to the truncation with the error
function and the Eddington inversion process.
Finally, we use an acceptance-rejection algorithm to generate
each particle halo realization. 
The satellite is made up $10^6$ particles.
Since the initial satellites are
already truncated, satellite particles are ejected only through
interactions with the host halo during the simulation.

We simulate a set of satellite realizations with the same tidal
distance but different initial maximum rotation velocities.  We
investigate the effects of a satellite's size and orbit on its tidal
tail morphology by varying them separately but keeping the other
parameters fixed.  In detail, we choose a $c=15$ NFW model for both
the host halo potential and for the satellites.  We generate three
different size satellites, which we refer to as the massive satellite,
the low-mass satellite, and the tiny-mass satellite.  We use the
maximum rotation velocity, $V_{max, sat}$, as a measure of satellite
size since the continuous mass loss makes mass an inexact measure.  We
use a $V_{max, sat}$ of 0.45, 0.16, and 0.08 times $V_{max, host}$ for
the massive, low-mass, and tiny satellite, respectively.  The
tidal distance for all three satellites is $0.4R_{vir}$.  We also set
the galactocentric orbital radius to $0.4R_{vir}$ for our circular orbit 
simulations. After
our truncation procedure is complete, the initial mass of the massive
satellite is 0.018 $M_{host}$, the low-mass satellite 0.001
$M_{host}$, and the tiny-mass satellite 0.0001 $M_{host}$.  Converting
our simulation units to a Milky Way size galaxy system and evolving
for a few satellite orbits, the low-mass satellite roughly corresponds
in mass to the Sagittarius dwarf galaxy halo \citep{Majewski04,LJM05}.
The massive and tiny-mass satellites are an order of magnitude more
and less massive, respectively.  The properties of these satellite
halos is summarised in Table \ref{tbl:Satellites} in units of the
virial quantities of the host halo. All satellite initial conditions
in this study have $10^{6}$ particles.

We evolve each of the three satellites on three different orbits with
the same energy but with different eccentricities.  We define the
eccentricity of the orbits as $e\equiv(r_{a}-r_{p})/(r_{a}+r_{p})$
where $r_{a}$ and $r_{p}$ are the apocentre and the pericentre of a
satellite.  The first orbit is circular ($e=0$) at $0.4R_{vir}$.  The
second orbit has an $e=0.5$ with a pericentre of $0.2R_{vir}$ and an
apocentre of $0.6R_{vir}$, and the third orbit has $e=0.74$ with a
pericentre of $0.1R_{vir}$ and an apocentre of $0.67R_{vir}$.  The
third orbit is particularly relevant cosmologically since its
\emph{circularity} ($\kappa$) \footnote{$\kappa \equiv J/J_{c}$,where $J$ is the
  angular momentum and $J_{c}$ is the angular momentum of a circular
  orbit with the same energy.} is 0.5, which is the median
$\kappa$ of subhalos in a sample taken from recent cosmological
simulations \citep{Ghigna98, Zentner05}.  We quote results using the
following \emph{system} units unless otherwise specified: $G=1$,
$M_{vir, host}=1$, and $R_{vir, host}=1$.  The timestep for our N-body
simulations is $2.5\times 10^{-4}$ system time units.  Therefore, one
circular orbit is made of about $8000$ timesteps because one circular
orbit period is $T_{period} \sim 2.0$

\begin{figure}
\centerline{\includegraphics[width=1.0\columnwidth,angle=0]
{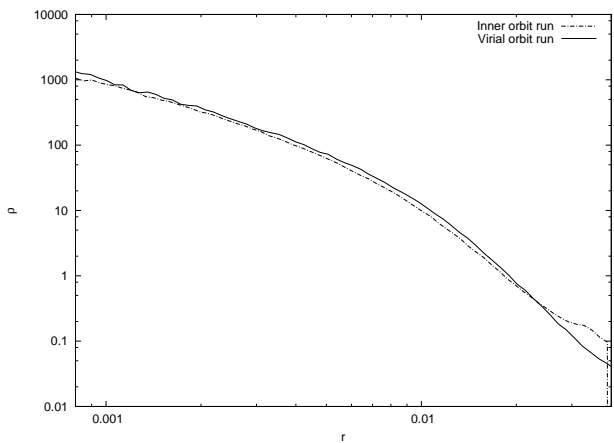}}
\caption{Comparison of the evolved density profiles of two identical,
  low mass satellite halos evolved with different starting radii.  The
  inner orbit begins with $r=0.67R_{vir}$ and a tidal distance
  corresponding to $0.4R_{vir}$ with an eccentricity $e=0.73$.  The
  virial radius orbit begins with $r=1.0R_{vir}$ and a tidal distance
  corresponding to $1.0R_{vir}$ with $e=0.5$.  Owing to gravitational
  heating, the evolution of the two satellites is different but the
  profiles approximately agree when the total mass lose is the same.
}
\label{fig:profile_comp}
\end{figure}

Our satellite subhalo is initially truncated without considering its
evolutionary history and without including any gravitational heating.
This is crudely consistent with our initial condition generation procedure
that assumes an equilibrium configuration at some radius inside the
host halo to start.  Fortunately, this idealised setup does not
produce an unrealistic satellite mass loss history.
\citet{Stoehr02} and \citet{Hayashi03} performed a quantitative study of NFW subhalo
evolution in a host halo using idealised N-body simulations and
claim that satellites on two different
orbits have similar mass and velocity profiles after losing the same
amount of mass.  To check this, we
compared the evolved density profiles of two low-mass satellites on
orbits with $e=0.74$ but at two
different tidal distances: 1) the radius of a circular orbit with
the same energy; and 2) the host halo virial radius, $R_{vir}$.
The first test describes the satellite evolution scenario adopted for
this study. The second test describes the cosmologically-motivated
scenario of a satellite entering the host halo for the first time.
Certainly, these two satellites have quite different evolutionary
histories. However, when the bound mass of the two satellites is scaled to the
same value, their evolved density profiles are similar, as shown in 
Figure \ref{fig:profile_comp}. This test, together with the results of
\citet{Stoehr02} and \citet{Hayashi03}, suggests that our tidally truncated
satellite models are a fair representation of CDM subhalos \citep[see 
Figure 10 in][]{Hayashi03}.  In addition, although tidal heating, which is
sensitive to a satellite's structure, plays an important role in the
satellite disruption process, we will show that the tidal tail
morphology does not depend on a satellite's inner structure but only on
a satellite's mass and orbit.

For the gravitational potential solver, we use a three-dimensional
self-consistent field algorithm \citep[SCF, also known as an \emph
  {expansion} algorithm, e.g.,][]{cbrock72, cbrock73, ho92,
  Weinberg99}.  This algorithm produces a bi-orthogonal basis set of
density-potential pairs from which it computes the gravitational
potential of a N-body system, given the mass and positions of the
particles.  For an arbitrary basis, e.g. spherical Bessel functions,
the expansion generally requires a large number of terms to achieve
convergence, which introduces small-scale noise as well as requiring
greater computational expense.  The situation was dramatically
improved by \citet{Weinberg99} using a numerical solution of the
Sturm-Liouville equation to match the lowest-order pair to the
equilibrium profile, and therefore, the expansion series converges
rapidly.  Here, we use the current density profile as the zero-order
basis function.

For our purposes, this expansion algorithm is attractive for two
reasons.  First, the expansions can be chosen to follow structure over
an interesting range of scales and simultaneously suppresses
small-scale noise.  In contrast, noise from two-body scattering can
arise at all scales in direct-summation, tree algorithm, and mesh
based codes.  Small-scale scattering can give rise to a diffusion in
conserved quantities, which can lead to unphysical outcomes
particularly for studies of long-term galaxy evolution
\citep[see][]{WK07a,WK07b}.  Second, the expansion
algorithm is computationally efficient; the computational time only
increases linearly with particle number.  Hence, the expansion
algorithm permits the use of a much larger number of particles than
most other algorithms for the same computational cost.

An accurate potential solver for a cuspy halo demands a precise
determination of the expansion centre, ${\bf C}$.  This is the major
disadvantage of the expansion algorithm relative to a Lagrangian
potential solver such as a tree code.  We developed and
tested the following algorithm for evolving cuspy dark matter halos 
with an expansion code:
\begin{enumerate}
\item At time step $n$, we compute ${\bf C}_n$ from the centre
  of mass of the $N_{min}$ most bound particles;
\item To evaluate the expansion centre at time step $n+1$, a predicted
  centre ${\bf C}_{pred,n+1}$ is estimated from a linear least squares solution
  using the previous $N_{keep}$ centres: $\{{\bf
    C}_j|n-N_{keep}< j\le n\}$;
\item For $n<2$, we set ${\bf C}_{pred,n+1}={\bf C}_n$.
\item To reduce truncation error, we separately track the motion
  relative to the satellite's centre and the motion of the centre itself.
\end{enumerate}
The linear least 
squares estimator for the expansion centre ${\bf C}_{pred}$ 
reduces the Poisson noise from the $N_{min}$ particles used to determine
each of the ${\bf C}_n$.
For our simulations we have adopted $N_{min}=512$ and $N_{keep}=10$ and 
have verified that this centring scheme maintains the cusp while the
satellite orbits in a host halo for situations where the tidal field
is insignificant.

\section{The morphology of satellite tidal tails}
\label{sec:satellite}

Time-dependent forcing by the host halo's tidal field adds energy to
the satellite, driving mass loss and, ultimately, disruption.  These
forces are a combination of the differential force from the host halo
and the non-inertial forces from the satellite orbit.  The work done
against the satellite's gravitational potential results in mass loss.
In addition, these forces \emph{deform} the outer density contours of
the satellite.  To understand the evolution of the ejecta, one must
also consider the gravitational field of the satellite.  The
gravitational force from the satellite decelerates (accelerates) the
leading (trailing) tail, modifying the energy and angular momentum of
the ejecta well past the point of escape.  The conserved quantities of
the ejecta, then, may be dramatically different than that of the
satellite centre of mass.  The strength of the satellite gravity
increases with satellite mass, of course.  These effects combine to
make the morphology of tidal tails more complicated than previously
suggested \citep{MD94,IL98, Johnston01,HW99,Mayer02}, especially for a
massive satellite.  We investigate the causes of these effects and
their consequences in detail below.

\subsection{Satellite disruption}
\label{sec:disruption}

\begin{figure}
\centerline{
\includegraphics[width=0.475\columnwidth,angle=0]{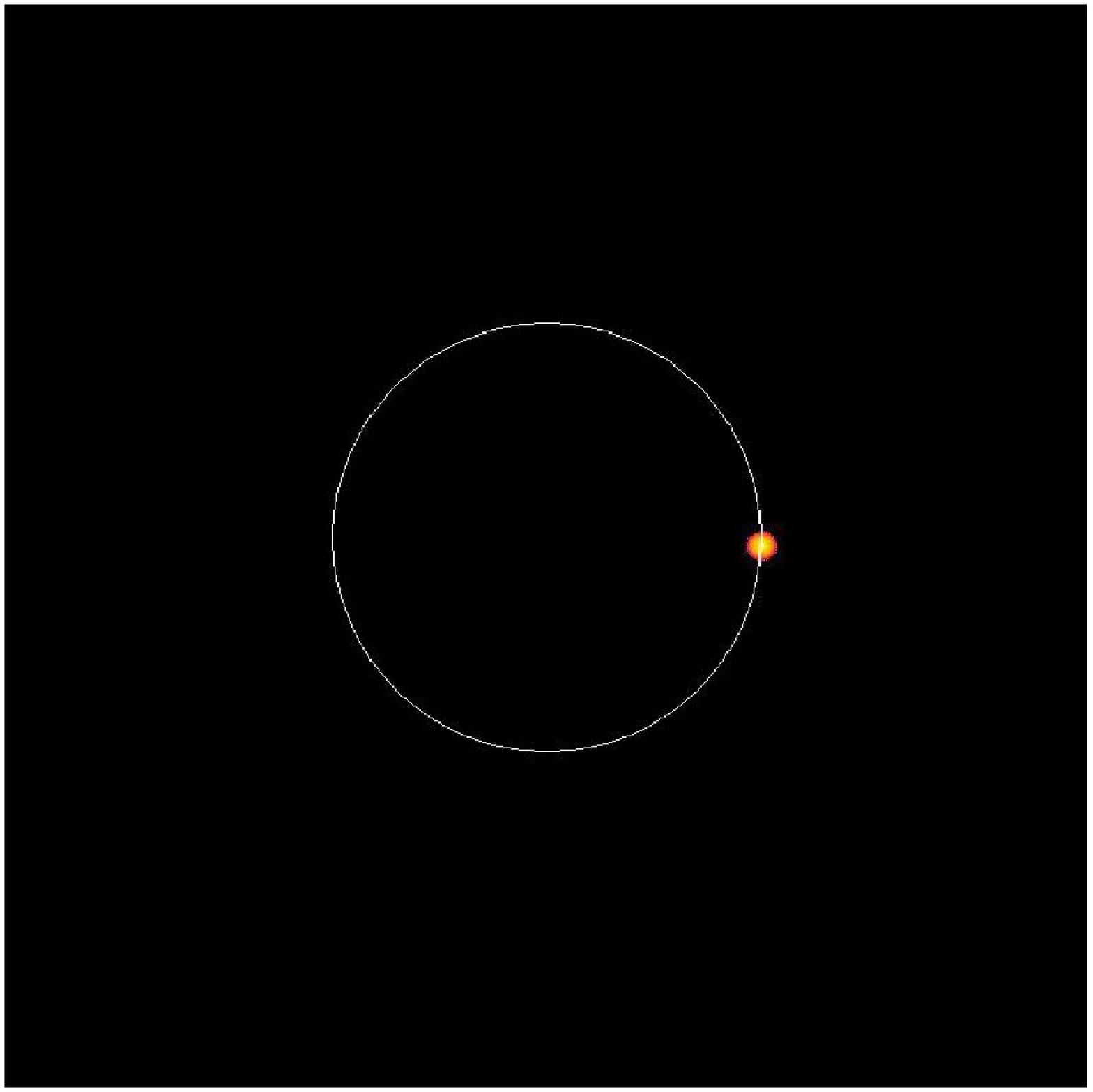}
\includegraphics[width=0.475\columnwidth,angle=0]{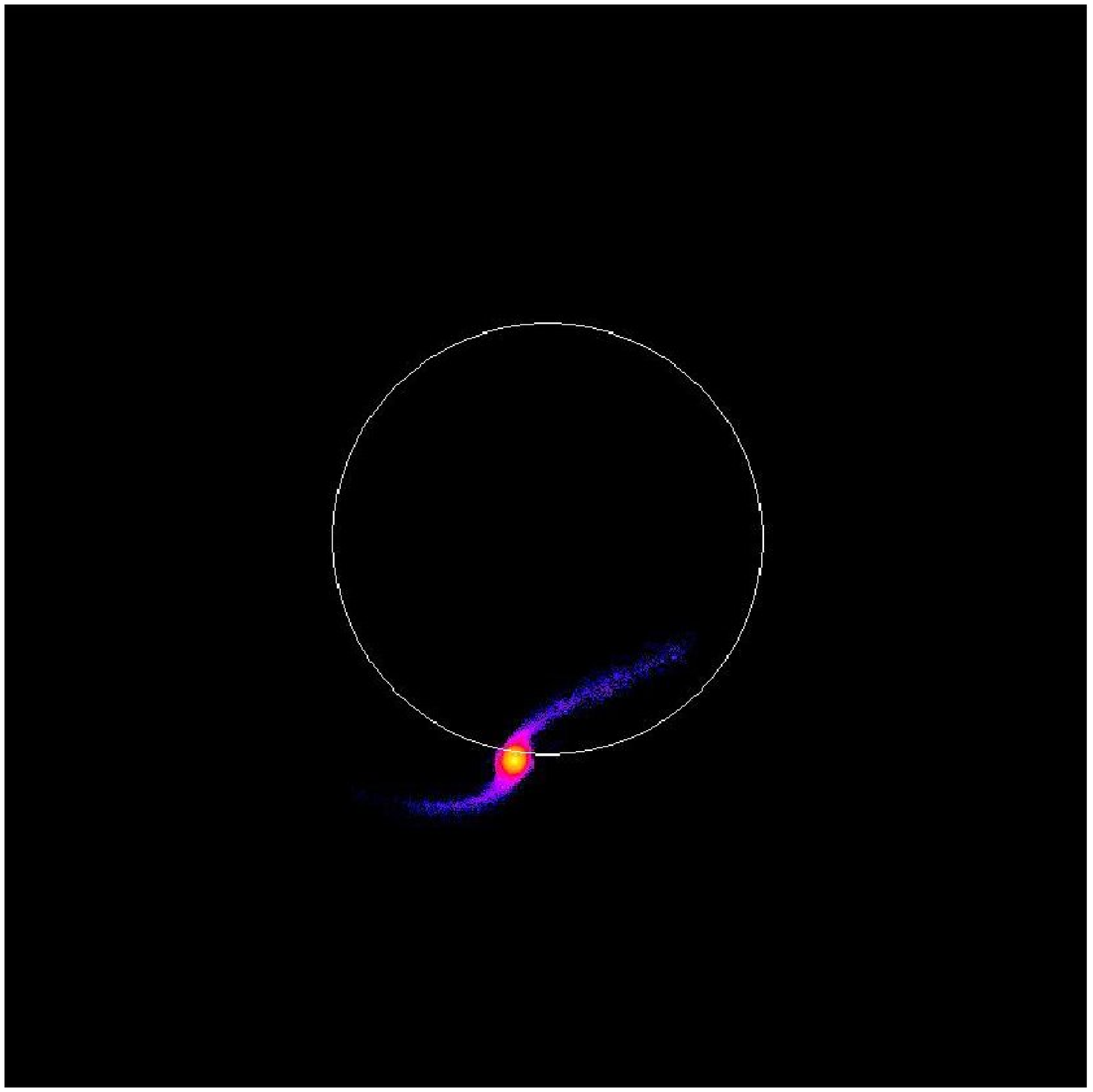} }
\centerline{
\includegraphics[width=0.475\columnwidth,angle=0]{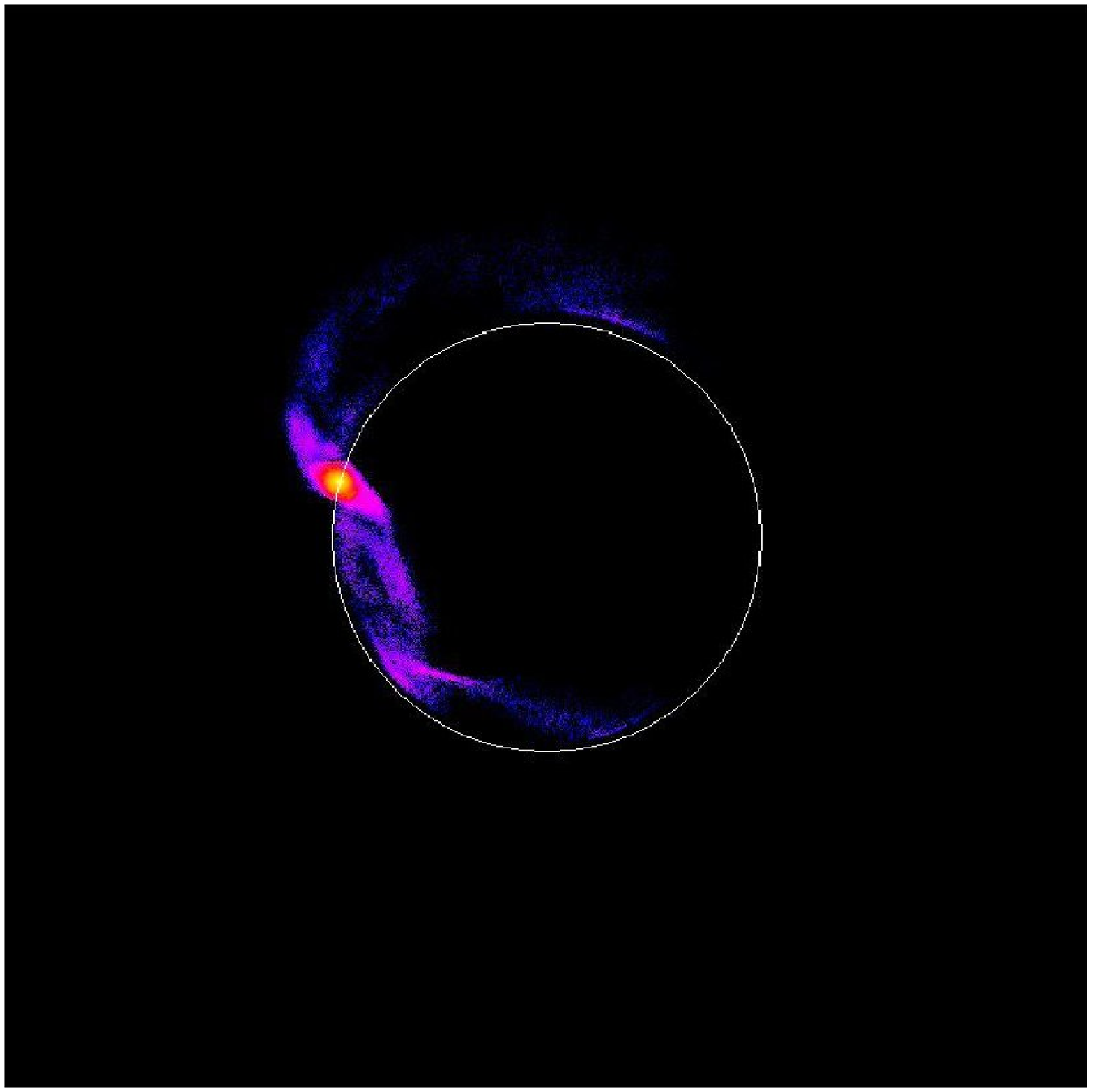}
\includegraphics[width=0.475\columnwidth,angle=0]{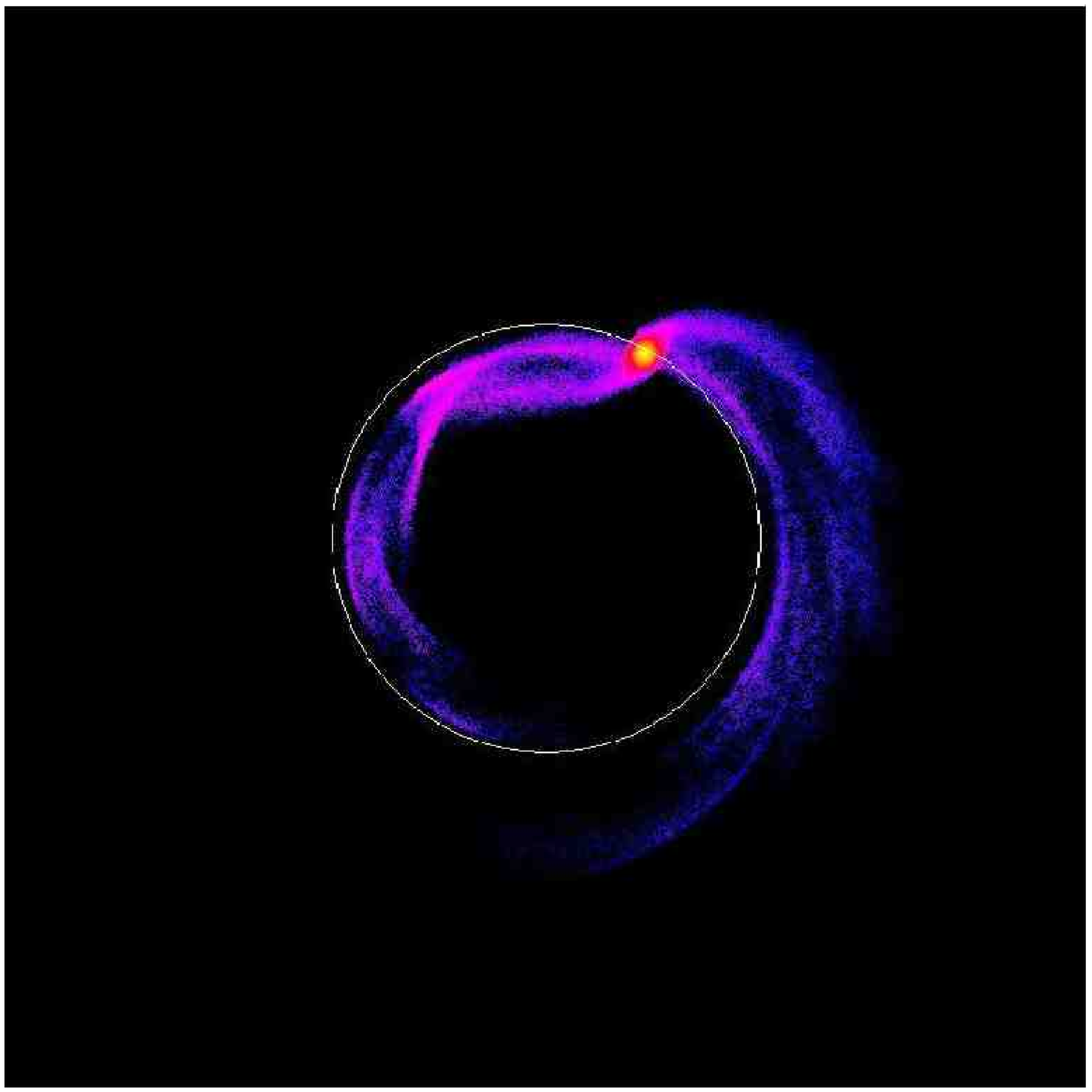} }
\caption{The mass density of the low-mass satellite on a circular
  orbit with $r=0.4R_{vir}$ at $T=0.0$ (top-left), $1.5$ (top-right),
  $3.0$ (bottom-left), and $4.5$ (bottom-right).  
  Recall that the orbital period for the circular orbit in this simulation 
  is $T_{period} \approx 2.0$.
  The colour scale is
  logarithmic in the dark matter mass density, increasing from blue to
  red, and is fixed for all times $T$.  Each panel has a linear size
  of 2 host-halo virial radii. For this simulation, the tail particles
  do not feel the gravitational force of the satellite after escape.
  The circles show the satellite orbit.  The multiple streams in the
  tail owe to phase crowding near apocentre for initially prograde and
  retrograde orbits.}
\label{fig:FramesLTC3}
\end{figure}

We begin by describing the dynamics and morphology of the tidal tails
in a simulation that \emph{ignores} the gravitational field of the
satellite past the tidal radius.  Figure \ref{fig:FramesLTC3} shows a
sequence of snapshots of the low-mass satellite ($0.001 M_{host}$) on
a circular orbit at $0.4R_{vir}$.  We use units where $M_{vir}=1$,
$R_{vir}=1$, and Newton's gravitational constant $G=1$, together which
defines a natural time unit. Scaled to the Milky Way, 0.5 natural time
units is approximately 1 Gyr. Appealing to the standard zero-velocity
Roche potential, which balances the effective gravitational potential
in the rotating frame with the halo potential, we expect the mass to
become unbound in the vicinity of the Lagrange or X-points.  Indeed,
we observe the double cometary appearance of tails leading and
trailing the satellite, enforced by the conservation of angular
momentum.  For this halo model, the leading ejecta orbits faster than
the satellite and has a position angle of $300^\circ$ measured
from the positive vertical axis, the direction of satellite's 
instantaneous motion.
The trailing tail moves slower than the satellite and has a position
angle of $120^\circ$.
Since the simulation in Figure
\ref{fig:FramesLTC3} ignores the satellite's gravity beyond the tidal
radius, the orbit of the tidal tail merely represents the kinematic
condition of the tail material just when it escapes from the
satellite.

\begin{figure}
\centerline{\includegraphics[width=0.5\columnwidth,angle=-90] {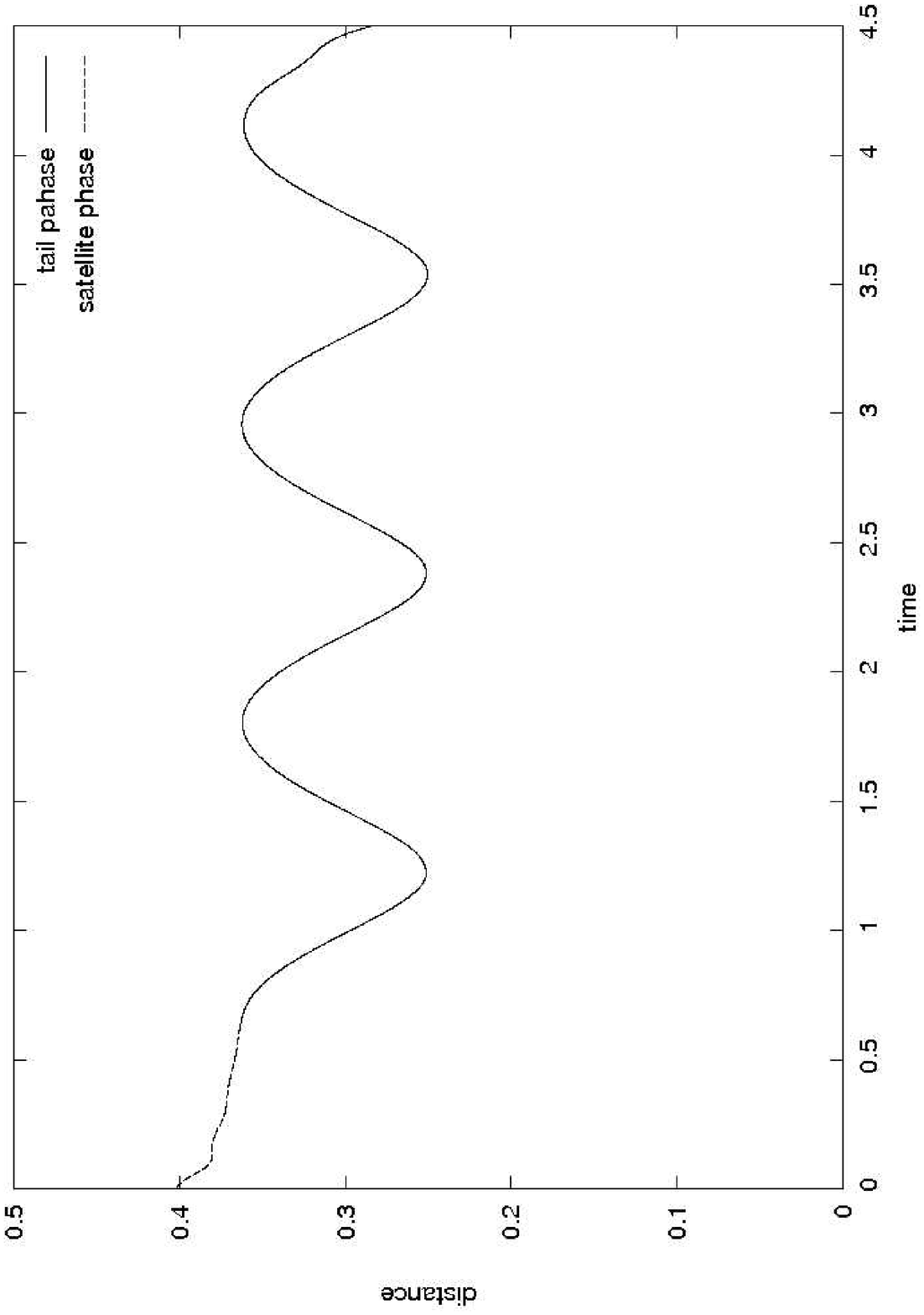}}
\centerline{\includegraphics[width=0.5\columnwidth,angle=-90] {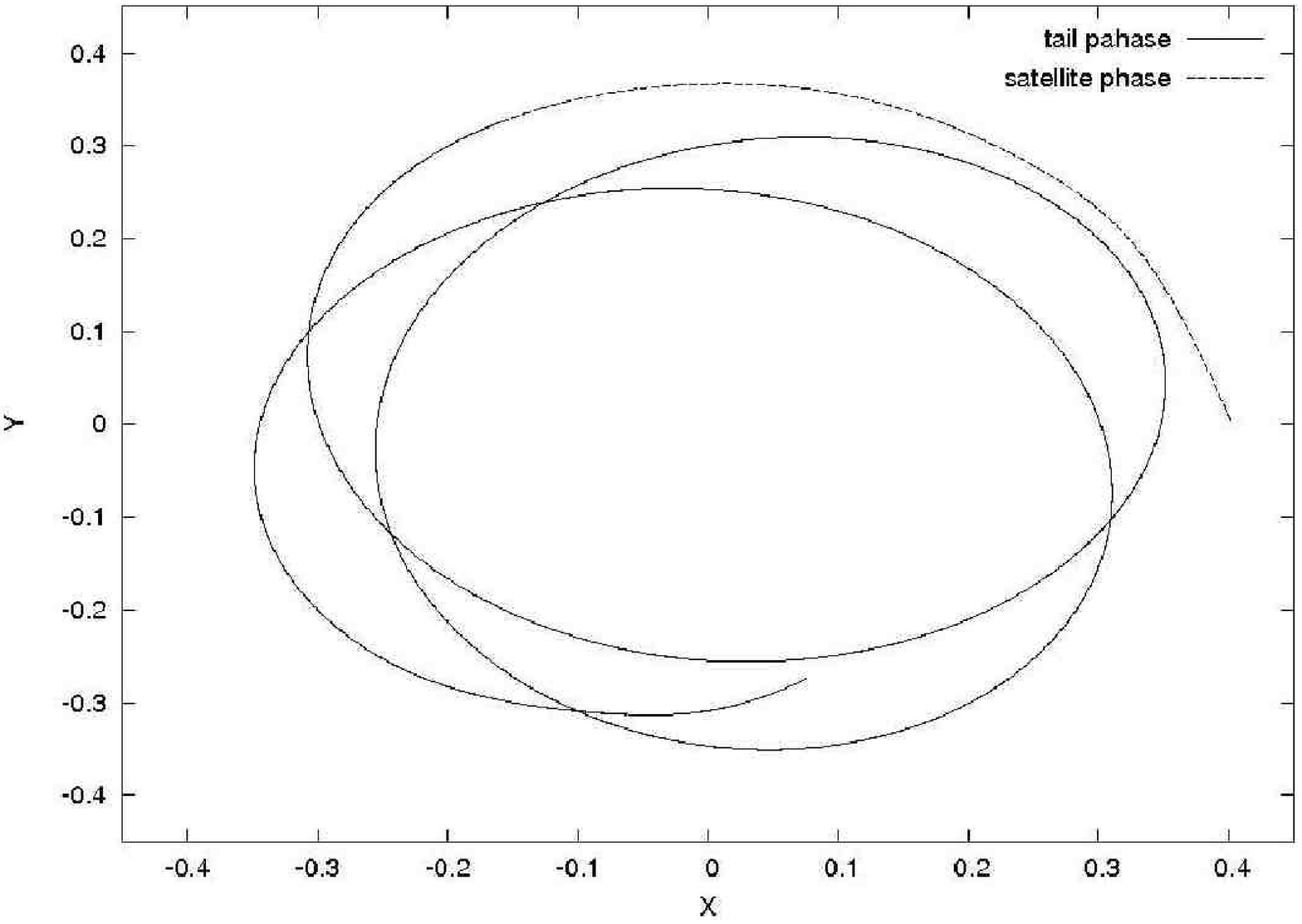}}
\caption {The orbit of a random particle in the leading tail for the
  simulation presented in Figure \ref{fig:FramesLTC3}.  The top panel plots
  galactocentric radius versus time and the bottom panel shows
  the trajectory in the orbital plane.  In both panels, the trajectory
  is plotted as a dashed line when it is still bound to the satellite and as
  a solid line after escape.  The particle describes a rosette with
  its apocentre near the satellite orbital radius after escape.}
\label{fig:OrbTC3}
\end{figure}

An example of a randomly chosen orbit in the leading tail is shown in
Figure \ref{fig:OrbTC3}.  As expected, the orbit describes a rosette
with its apocentre at the radius of the satellite orbit.  The energy and angular
momentum lost during the escape changes the conserved quantities of
the ejecta orbits from that of the satellite orbit; the leading (trailing)
ejecta lose (gain) energy during deformation.  Moreover, the
distribution of the tails fills a wide region about the satellite
orbit.  This reflects the broad distribution of phases for orbits at
escape.  Hence, the width of the tail is nearly the same as the distance
between the apocentre and the pericentre of a typical rosette orbit.  Each
tail has several distinct \emph{streamers} filling a common envelope.
The two primary streams in each tail demarcate the escape of the most
extreme prograde and retrograde orbits.  The originally prograde
orbits have lower specific angular momentum and, therefore, smaller
pericentres and larger epicyclic amplitudes.  In contrast, originally
retrograde orbits have larger pericentres and smaller epicyclic
amplitudes.  Distinct streamers result from the phase caustics near
apocentre, similar to shells in elliptical galaxies caused by merger
ejecta with a velocity dispersion much smaller than its new orbital
velocity.  This mechanism, illustrated in Figure \ref{fig:FramesLTC3}, is
the \emph{massless} description of tail formation. This massless description 
assumes that the orbital energy and angular momentum of a tail is the same as
those of a satellite; this yields a simple easy-to-compute
prescription for the tails' location.

\begin{figure}
\centerline{
\includegraphics[width=0.475\columnwidth,angle=0]{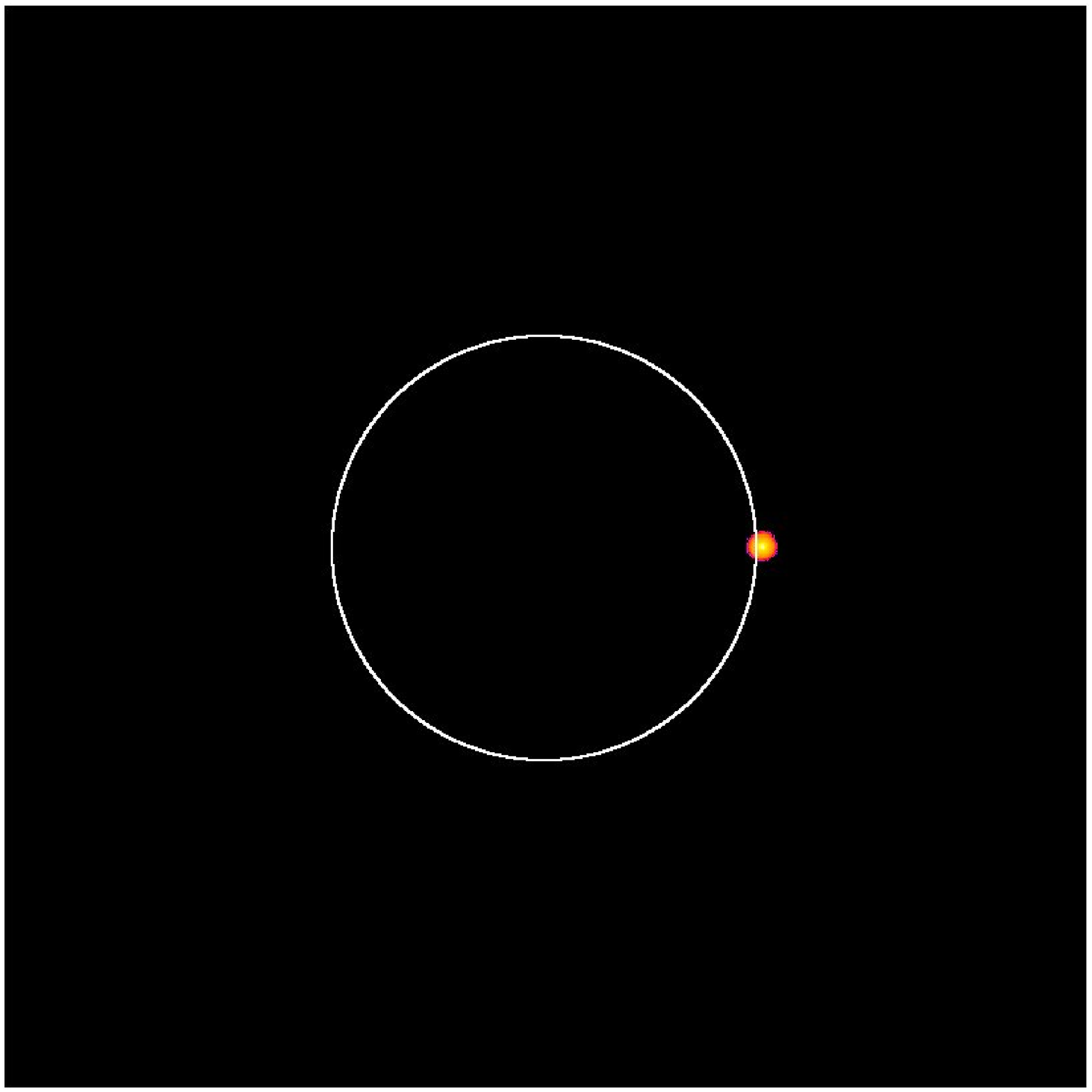}
\includegraphics[width=0.475\columnwidth,angle=0]{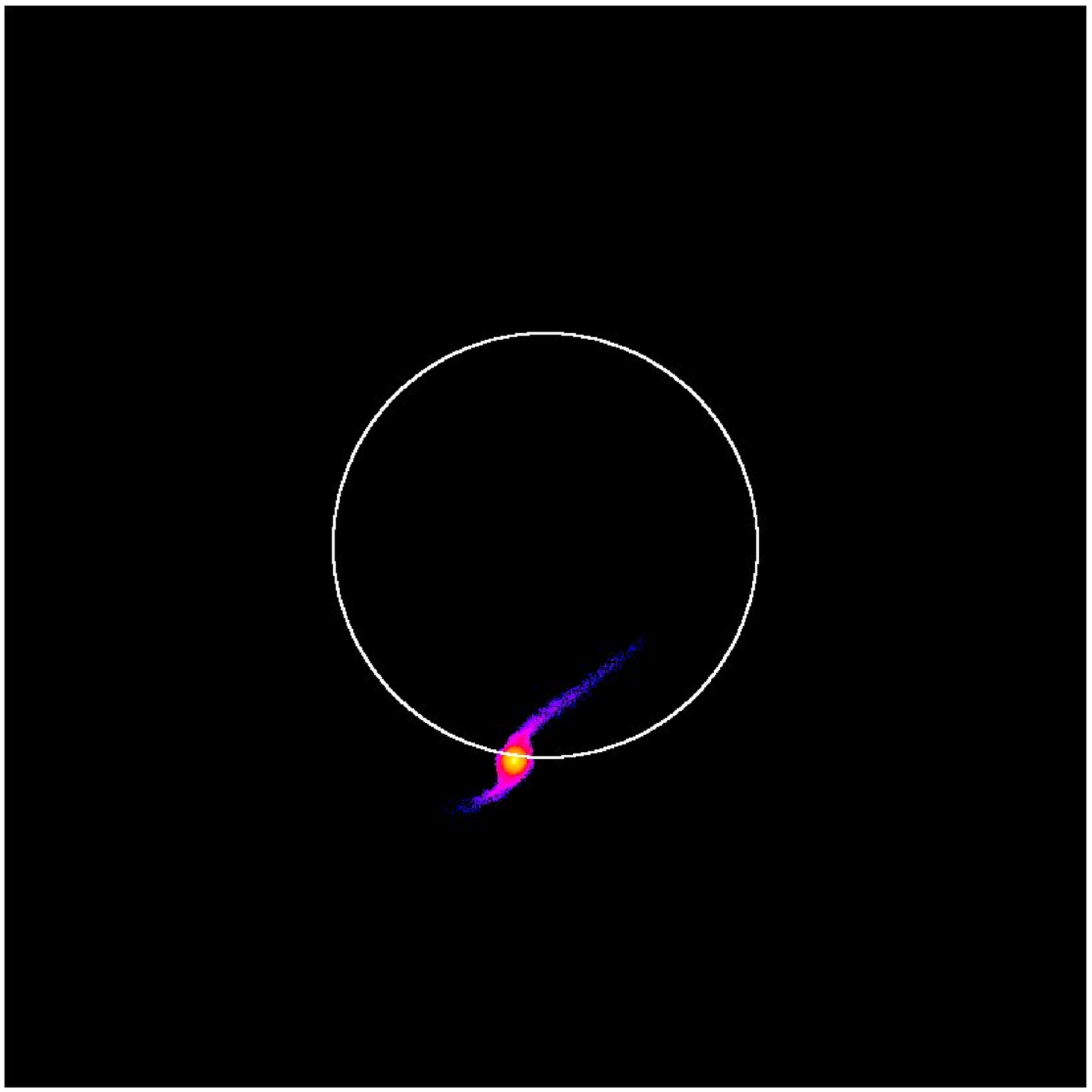}
}
\centerline{
\includegraphics[width=0.475\columnwidth,angle=0]{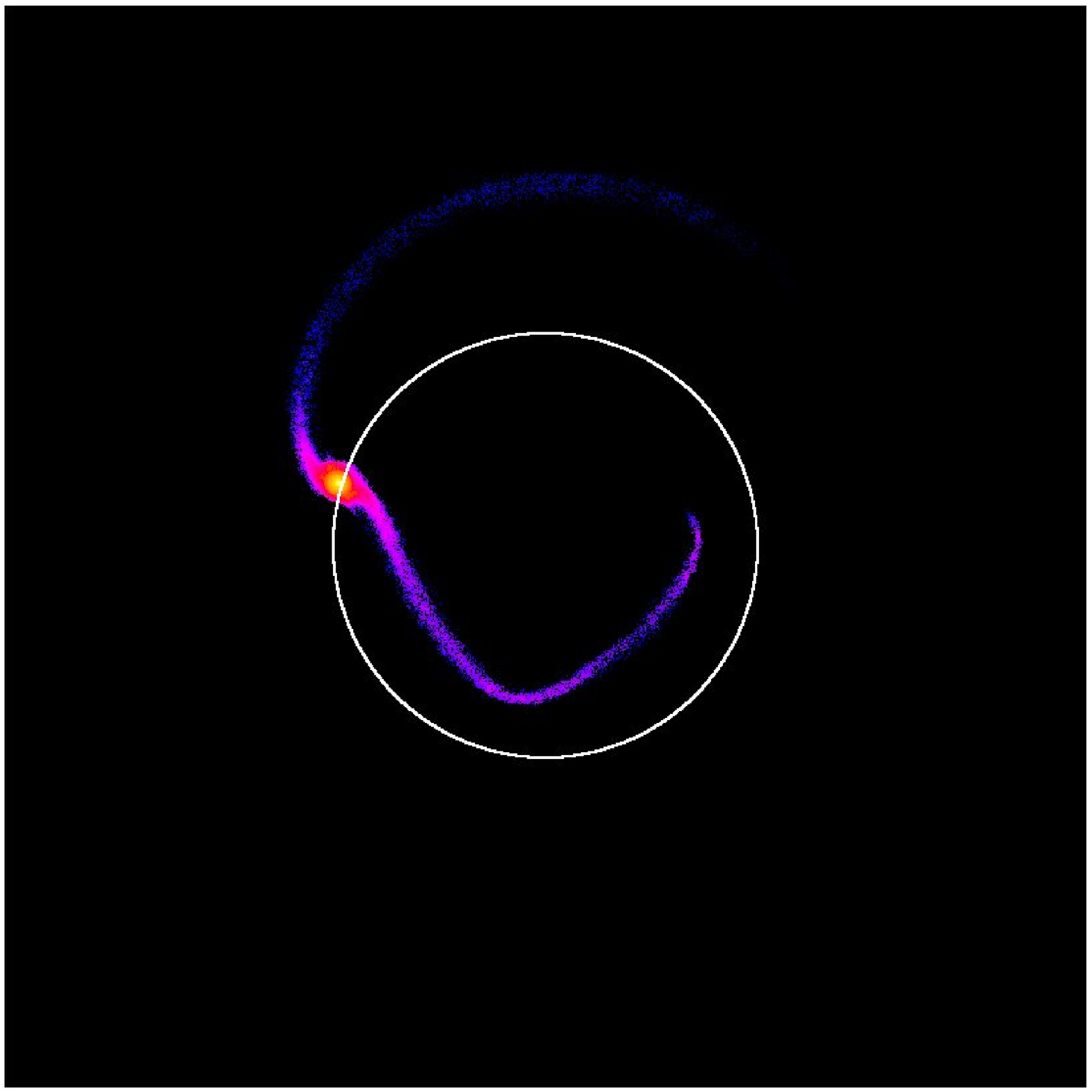}
\includegraphics[width=0.475\columnwidth,angle=0]{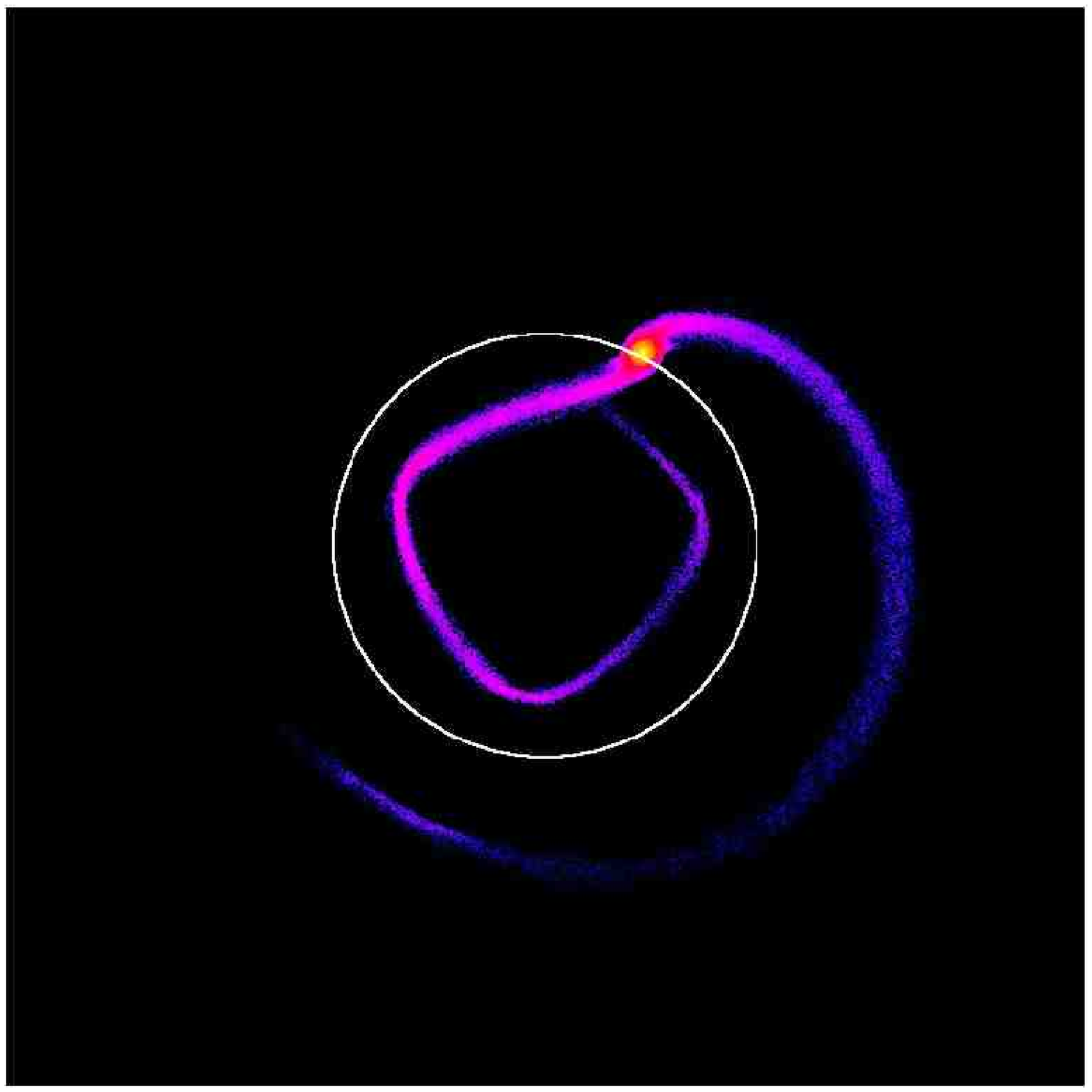}
}
\caption{As in Figure \ref{fig:FramesLTC3} but including the
  gravitational attraction of the satellite on the ejecta at all times.}
\label{fig:FramesLTC2}
\end{figure}

\begin{figure}
\centerline{
\includegraphics[width=0.475\columnwidth,angle=0]{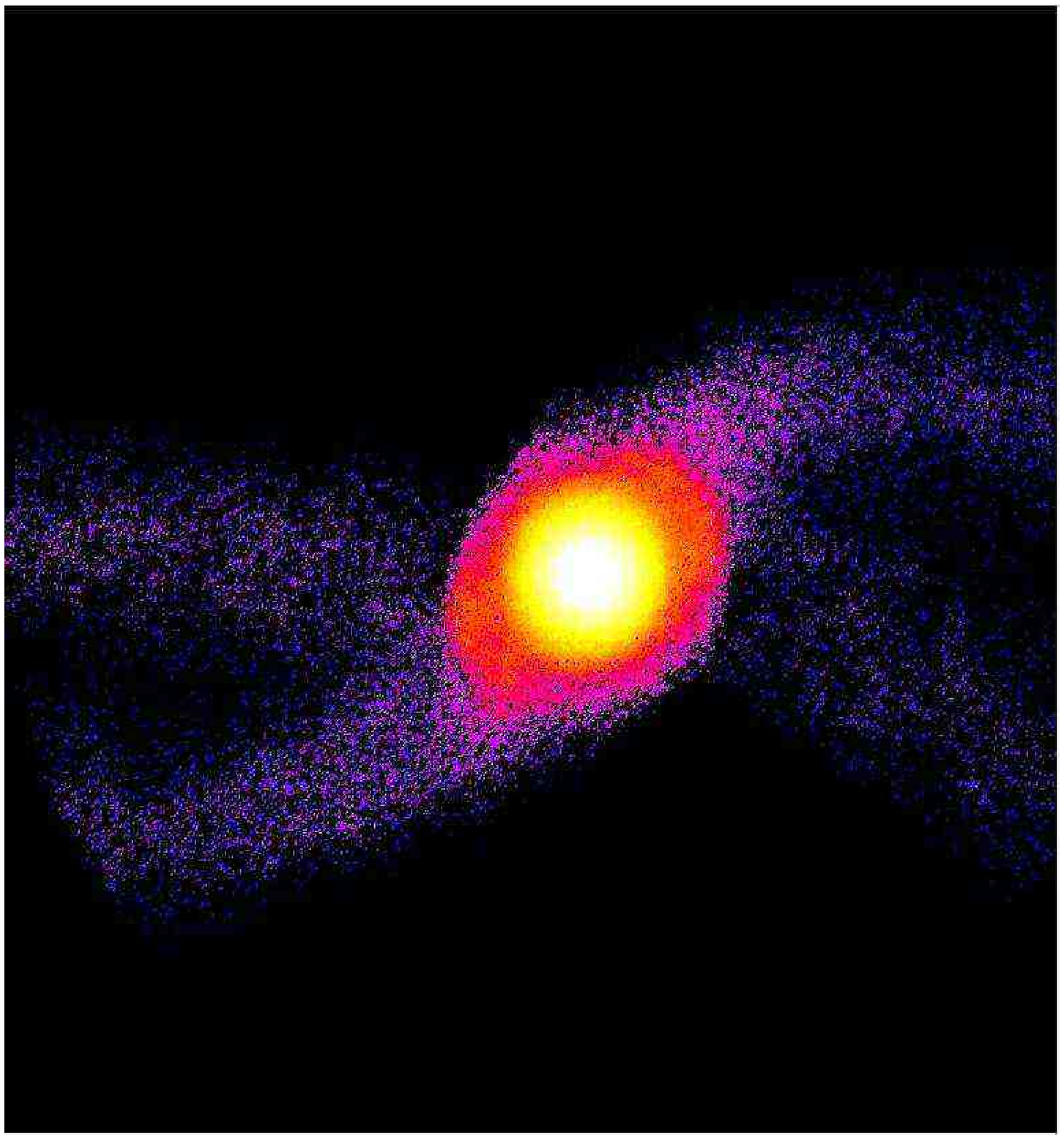}
\includegraphics[width=0.475\columnwidth,angle=0]{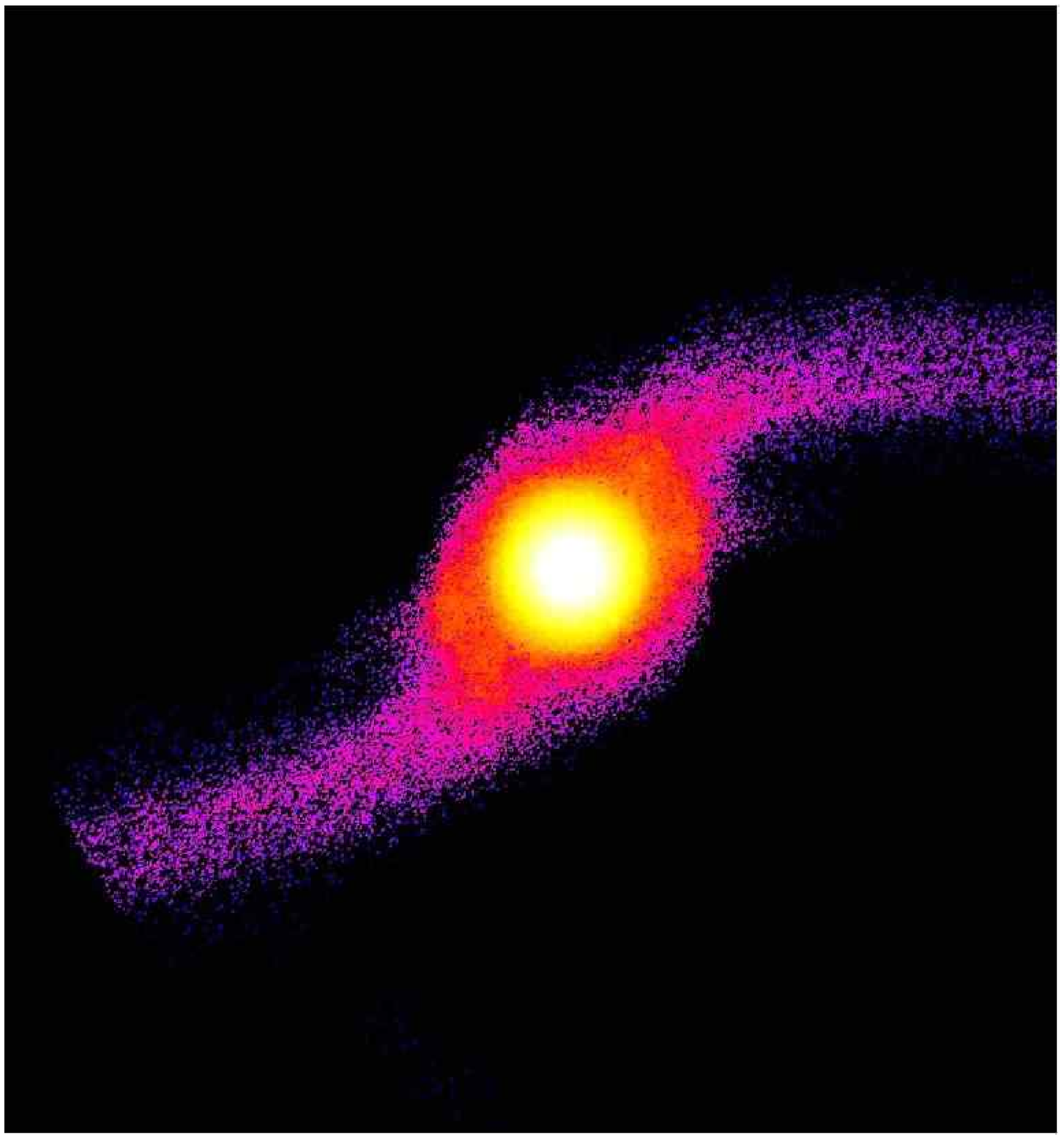} }
\caption {A high-resolution view of the tail streamers in the low-mass
  satellite simulation without (left) and with (right) the
  gravitational acceleration by the satellite at $T=4.5$ (compare with
  Figure \ref{fig:FramesLTC3} left panel, and Figure
  \ref{fig:FramesLTC2} middle panel, respectively).  The two steamers
  are clear in left panel but very weak in the right panel.  Although
  very weak, the second streamers can be identified near the
  tidal radius.  }
\label{fig:FramesLSate}
\end{figure}

In contrast, Figure \ref{fig:FramesLTC2} repeats the simulation
including the gravity of both the halo and the satellite at all
times.  At early times (upper-right panel), the evolution is similar.
However, at later times (lower panels), the effects of the satellite
gravity are marked.  The continued acceleration of the tail by the
satellite after escape decreases the internal velocity dispersion and
narrows or focuses the tail as a consequence.  The streamers in Figure
\ref{fig:FramesLTC3} become less distinct when accelerated by the
gravity of the satellite and the host halo together for the same
reason (see Figure \ref{fig:FramesLSate}).  
Similarly, the acceleration of the ejecta by the satellite
also decreases the angular separation between the streamers.  
Although the multi-streamer feature is diminished as the satellite 
gravitational field accelerates the ejecta, the feature can still be seen
very close to the tidal radius.

\subsection{Tail evolution}
\label{sec:tail}

\subsubsection{Circular orbits}
\label{sec:circtail}

\begin{figure*}
\centerline{
\includegraphics[width=0.4\textwidth,angle=0]{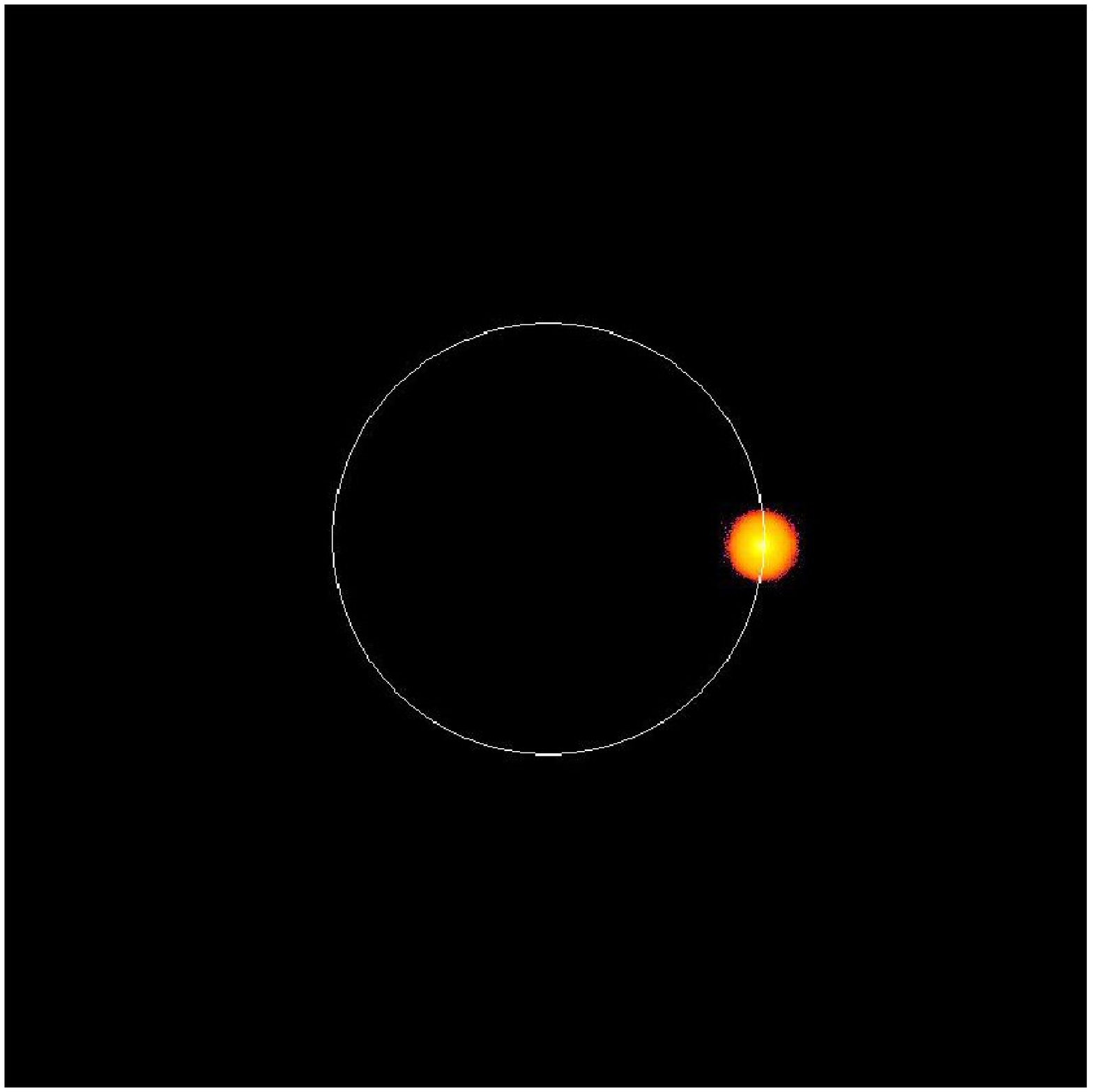}
\includegraphics[width=0.4\textwidth,angle=0]{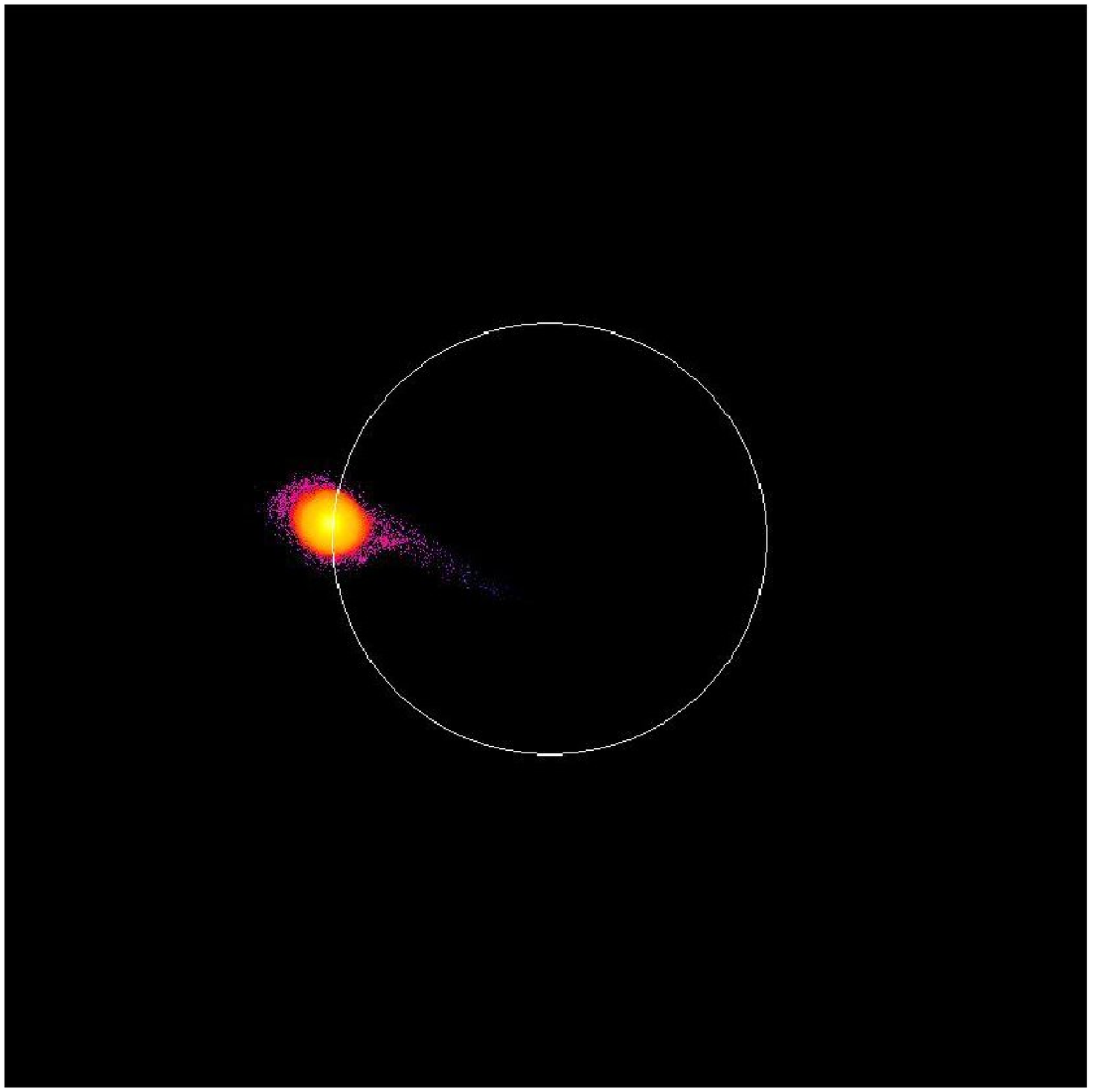}
}
\centerline{
\includegraphics[width=0.4\textwidth,angle=0]{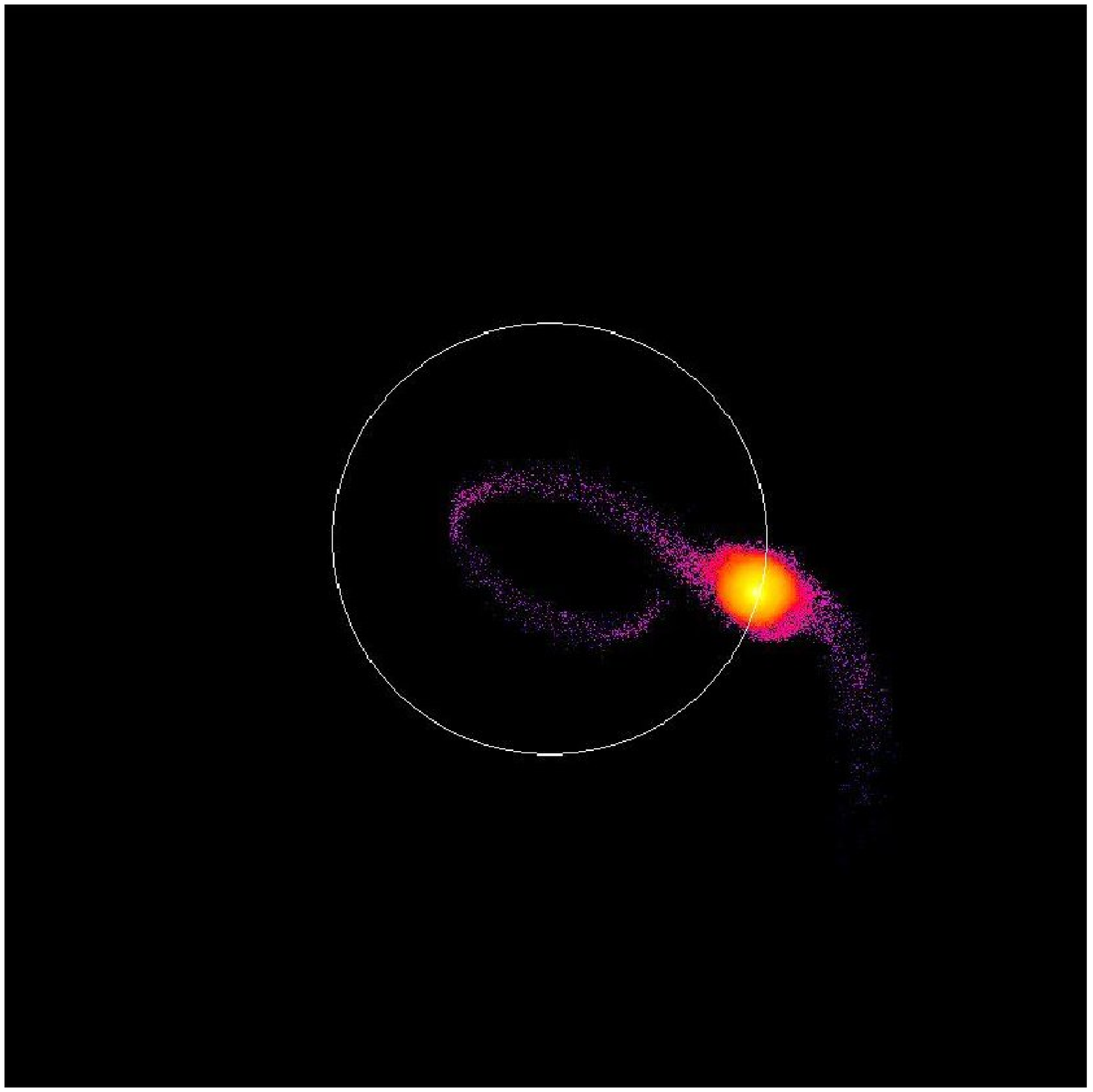}
\includegraphics[width=0.4\textwidth,angle=0]{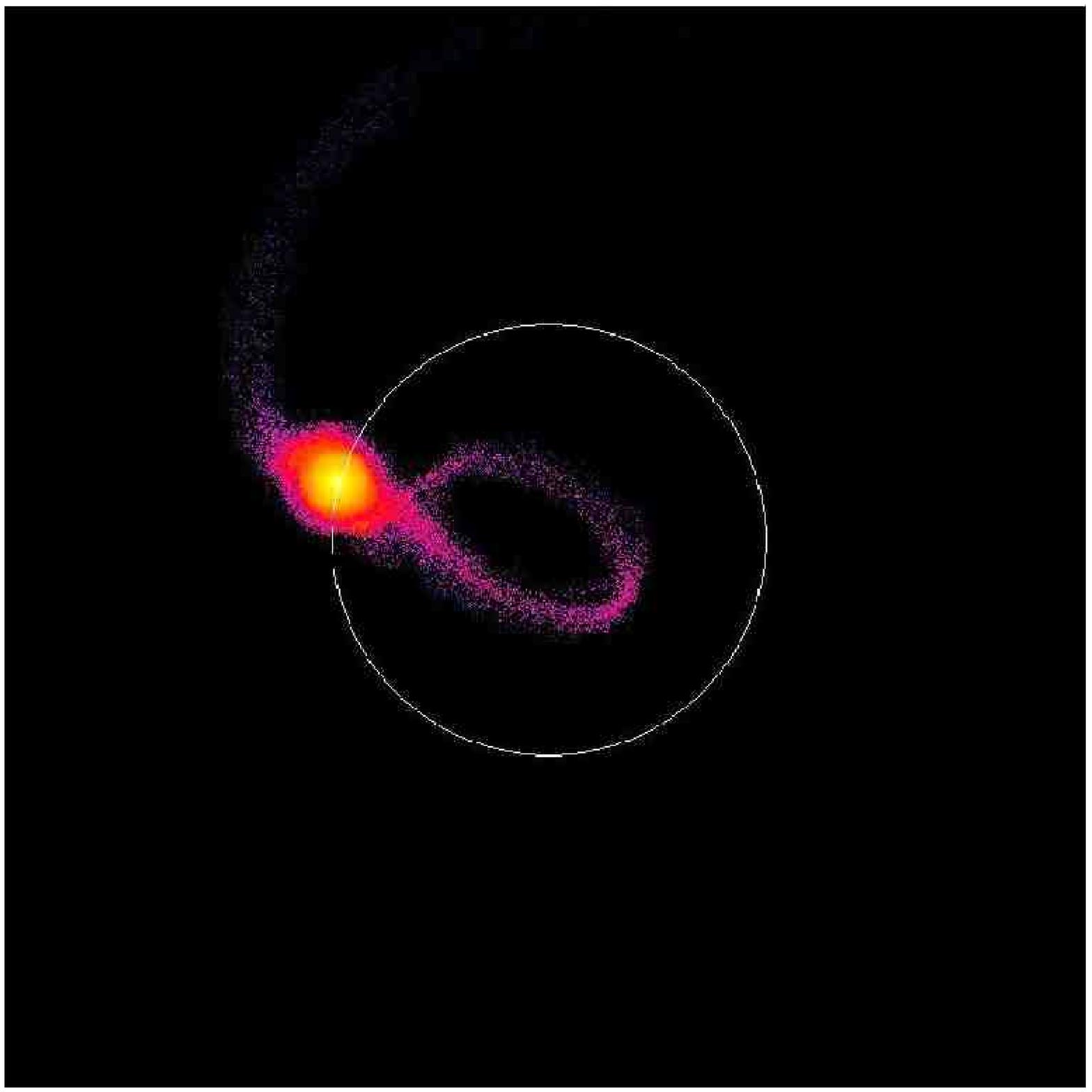} 
}
\centerline{
\includegraphics[width=0.4\textwidth,angle=0]{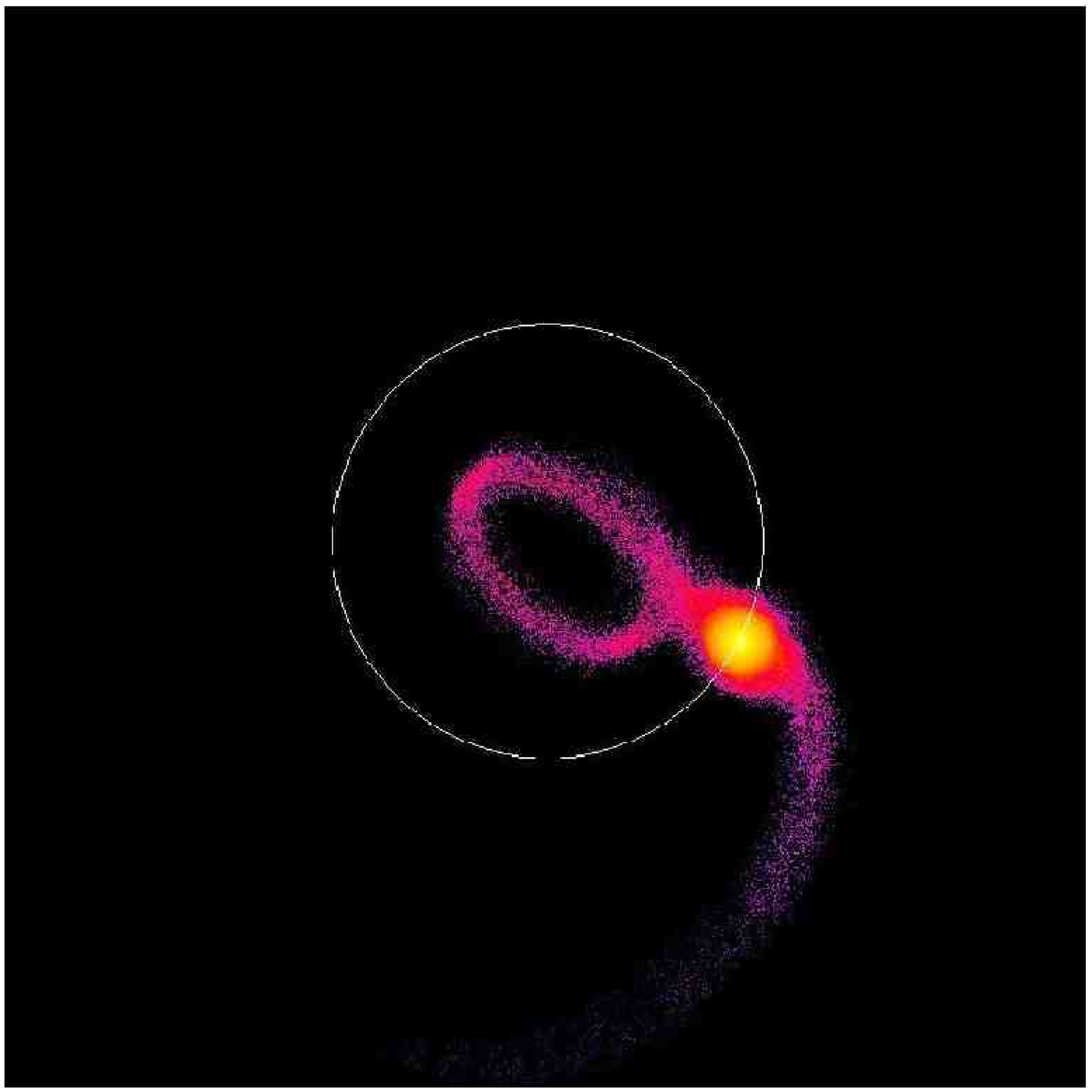}
\includegraphics[width=0.4\textwidth,angle=0]{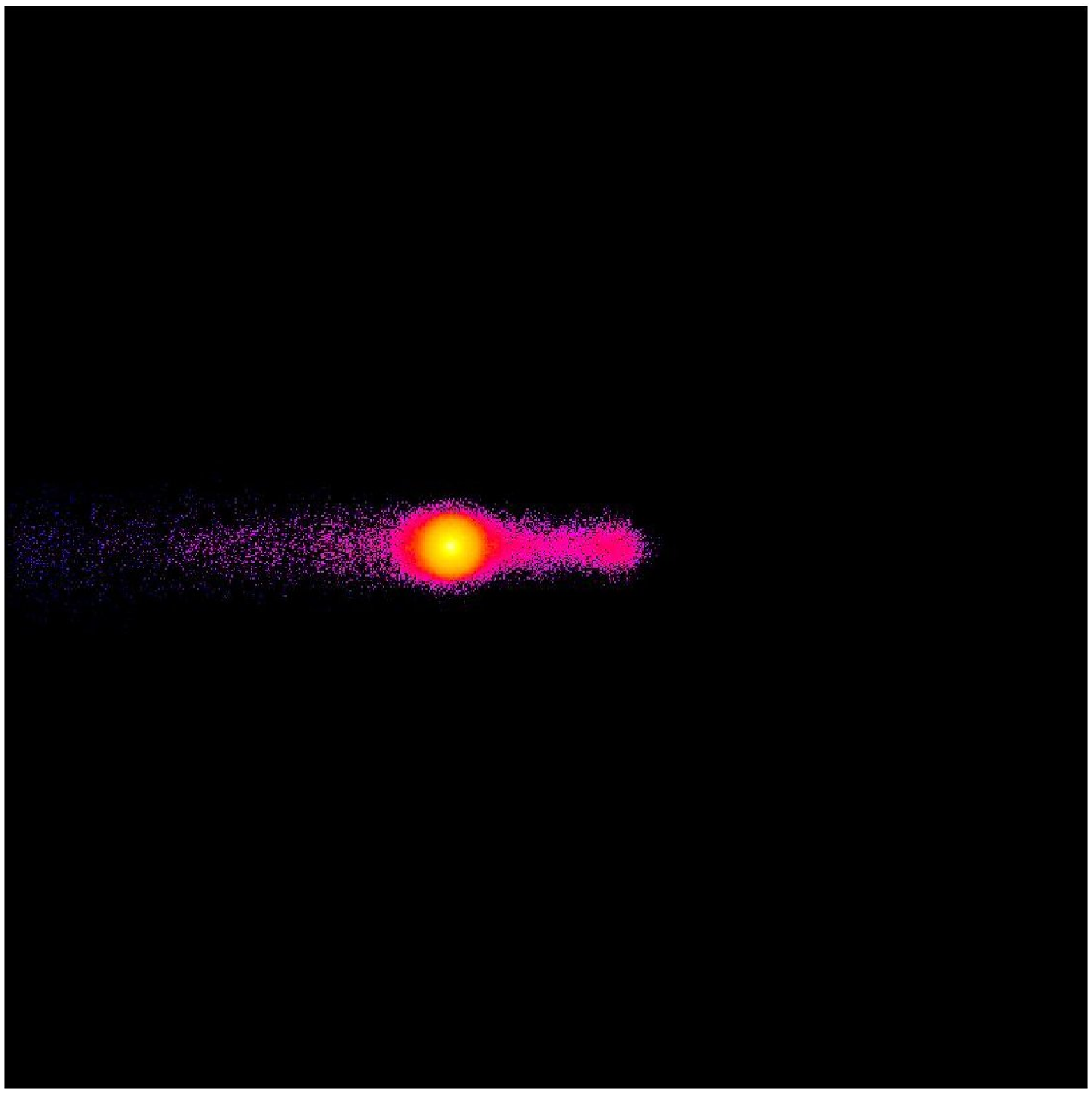}
}
\caption {The mass density in the orbital plane for a massive
  satellite halo on a circular orbit with $r=0.4 R_{vir}$ at $T=0.0$,
  $1.0$, $2.0$, $3.0$, and $4.0$ in the top-left, top-right,
  middle-left, middle-right, and bottom-left panels, respectively.
  Recall that the orbital period of the circular orbit
  is $T_{period} \sim 2.0$.
  The bottom-right panel shows the edge on view at $T=4.0$. The colour
  scale is logarithmic in the dark matter mass density from blue to
  red.  The colour scale is fixed for all snapshots (as described in
  Figure  \ref{fig:FramesLTC3}).  The circles show the satellite orbit.
  The tail remains confined to the orbital plane as expected
  (lower-right panel).}
\label{fig:FramesMTC1}
\end{figure*}

The importance of a satellite's gravity increases with mass and,
therefore, we begin with a study of the tail produced by a massive
satellite.  Figure \ref{fig:FramesMTC1} shows snapshots
of a massive satellite ($0.018 M_{host}$) on an circular orbit at
$0.4R_{vir}$, where once again the tail particles always feel the
gravitational force from the satellite.  The overall
evolution of the satellite and its disruption time is similar to the
less massive satellite shown in Figure \ref{fig:FramesLTC2}. However, the
long-term acceleration of the ejected material by the remaining
satellite significantly alters these orbits.  As the tail continues to
lose mass, the leading and trailing tails evolve to positions that are
well inside and well outside the satellite's orbit and hence does not trace
the satellite orbit at all \citep{Johnston01,MD94}.  The leading tail
significantly tilts toward the centre of the halo and almost points directly
there at late times.  The trailing tail is 
distributed throughout a wide annulus in the outer halo.  This
difference results from the torque applied by the satellite \emph{well}
after escape.  Orbits in the leading tail that lose energy and angular
momentum fall toward the centre of halo, while orbits in the trailing tail
gain energy and angular momentum and spread over a wide range of radii
in the outer halo.

\begin{figure*}
\centerline{
\includegraphics[width=0.3\textwidth,angle=0]{frame.MTC1.Or.00032.eps}
\includegraphics[width=0.3\textwidth,angle=0]{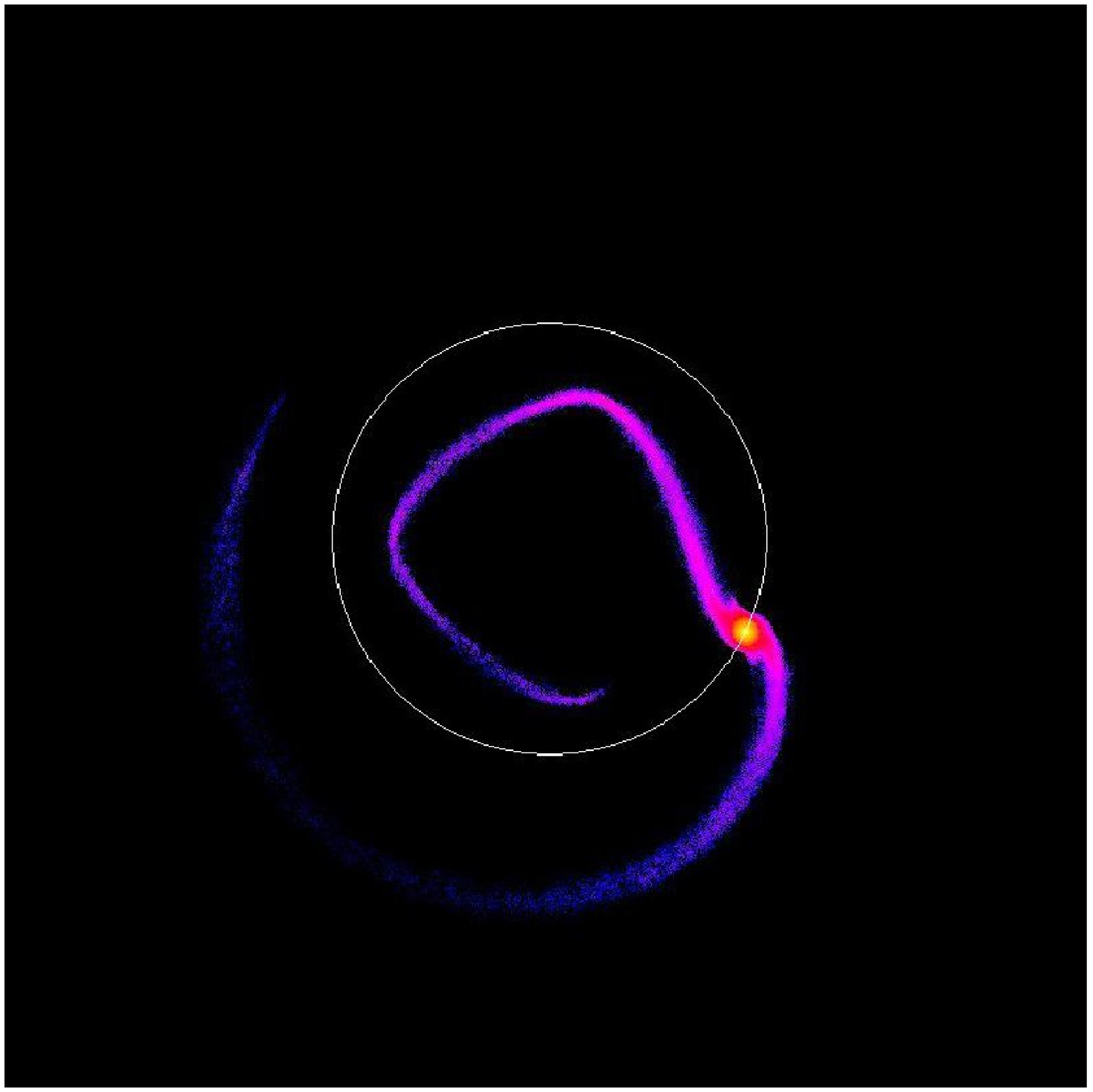}
\includegraphics[width=0.3\textwidth,angle=0]{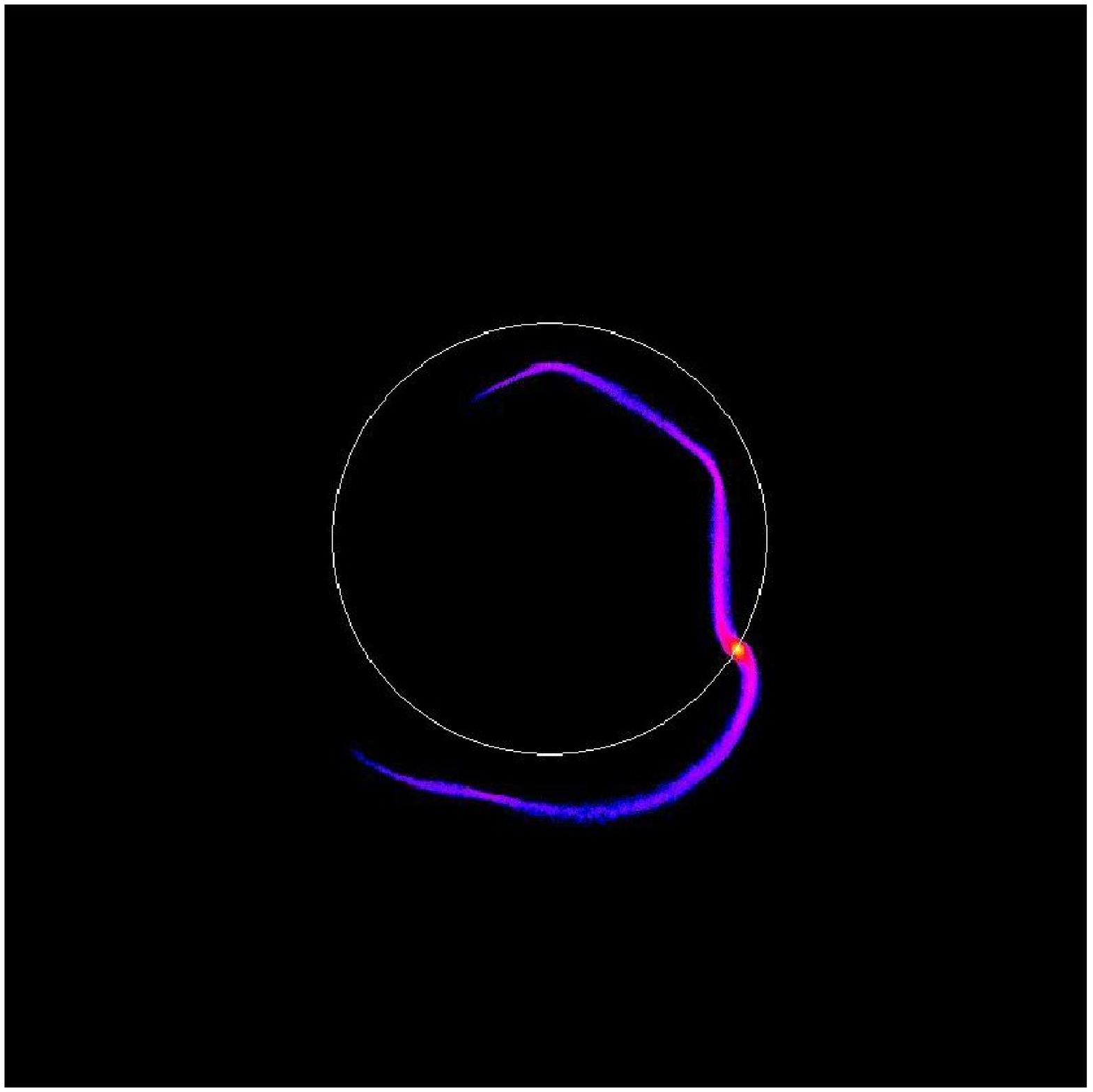} 
}
\caption {As in Figure \ref{fig:FramesMTC1} but comparing the ejecta at
  $T=4.0$ for the massive, low-mass, and tiny-mass satellites
  from left to right, respectively.}
\label{fig:FramesCirc}
\end{figure*}

We show the tail morphology for our satellites with three different
masses (see Table \ref{tbl:Satellites}) on circular orbits at $T=4$ in
Figure \ref{fig:FramesCirc}.  The tidal tails in the low-mass and
tiny-mass satellites ($0.001$ and $0.0001 M_{host}$, respectively)
very roughly follow the satellite orbit, with the leading and trailing
tail located inside and outside of the satellite orbit. Compared to
Figure \ref{fig:FramesLTC3}, it is clear that the differences decrease
with the satellite mass.  As we described in Section \ref{sec:simulation},
the low-mass satellite corresponds to the Sagittarius dwarf spheroidal
galaxy halo and the tiny-mass satellite corresponds to the Draco dwarf
spheroidal galaxy halo.

\begin{figure}
\centerline{\includegraphics[width=0.5\columnwidth,angle=-90] {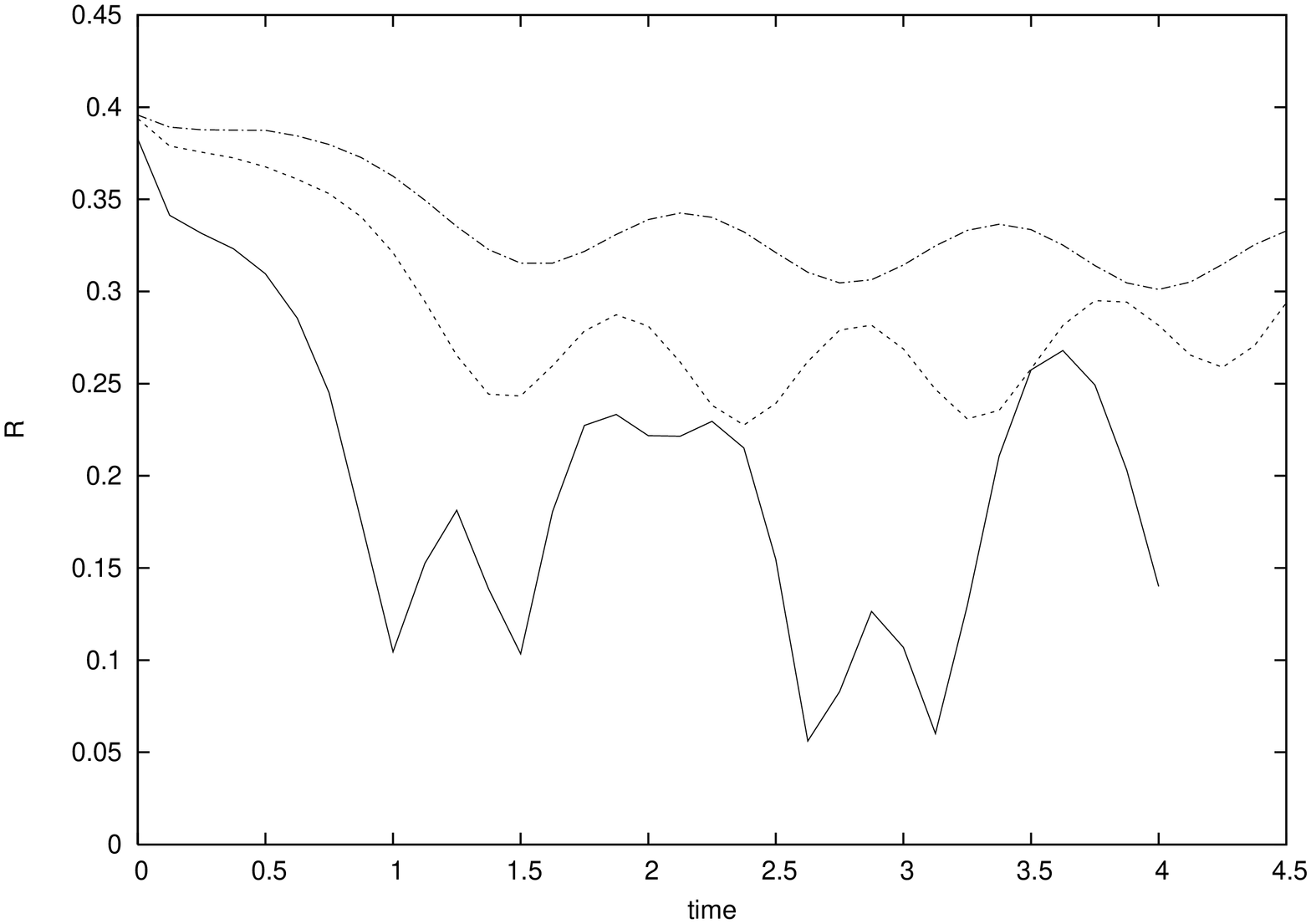}}
\centerline{\includegraphics[width=0.5\columnwidth,angle=-90] {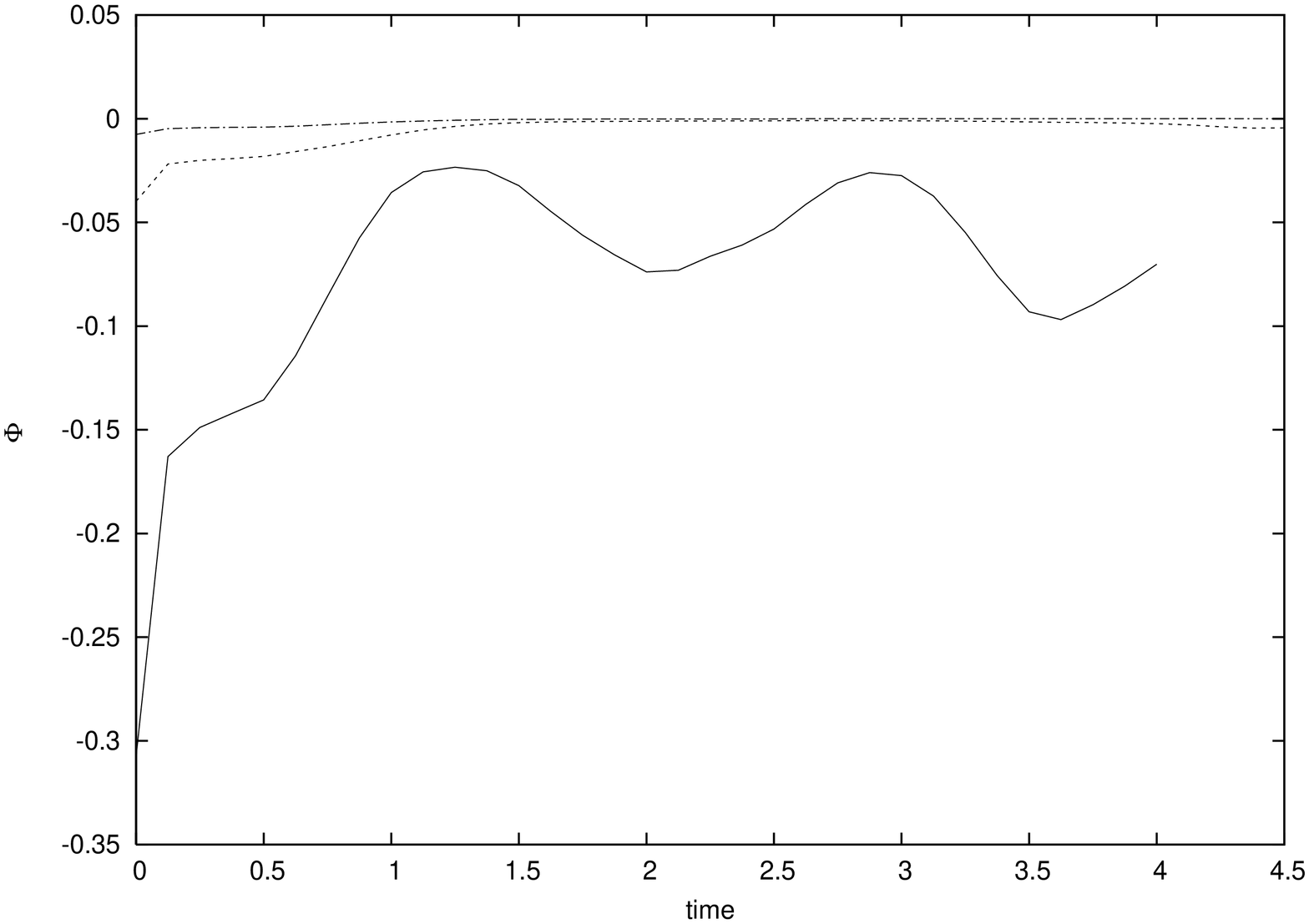}}
\caption {The evolution of the mean radius (top panel) and satellite
  potential (lower panel) for an ensemble of particles randomly selected
  from the leading tails in Figure \ref{fig:FramesCirc} for the massive (solid),
  low-mass (dotted), and tiny-mass (dash-dot) satellites.
}
\label{fig:CompCirc}
\end{figure}

Figure \ref{fig:CompCirc} shows the evolution of the
distance from the host halo centre
and the satellite's gravitational potential for an
ensemble average of 10 randomly sampled particles near the tip of the
leading tail in the three satellites.
The tail from a massive satellite receives a larger torque
and a larger shift to smaller energies and angular momentum than the
tail from a lower-mass satellite.  The bottom panel in Figure
\ref{fig:CompCirc} shows that the decay results from interactions with the
satellite potential.  Figure \ref{fig:CompCirc} also shows that the
satellite potential remains important in the low-mass and tiny-mass
satellites when the tail is close to the satellite but it is unimportant
when the tail is far from satellite.  The satellite potential always remains
significant for the massive satellite tail.  The long-term influence of the
satellite on the tail morphology makes any inference of the satellite
orbit from the tidal tail impractical, especially for satellites on
non-circular orbits (see Section \ref{sec:ecctail}).

\begin{figure}
\centerline{\includegraphics[width=0.5\columnwidth,angle=-90] {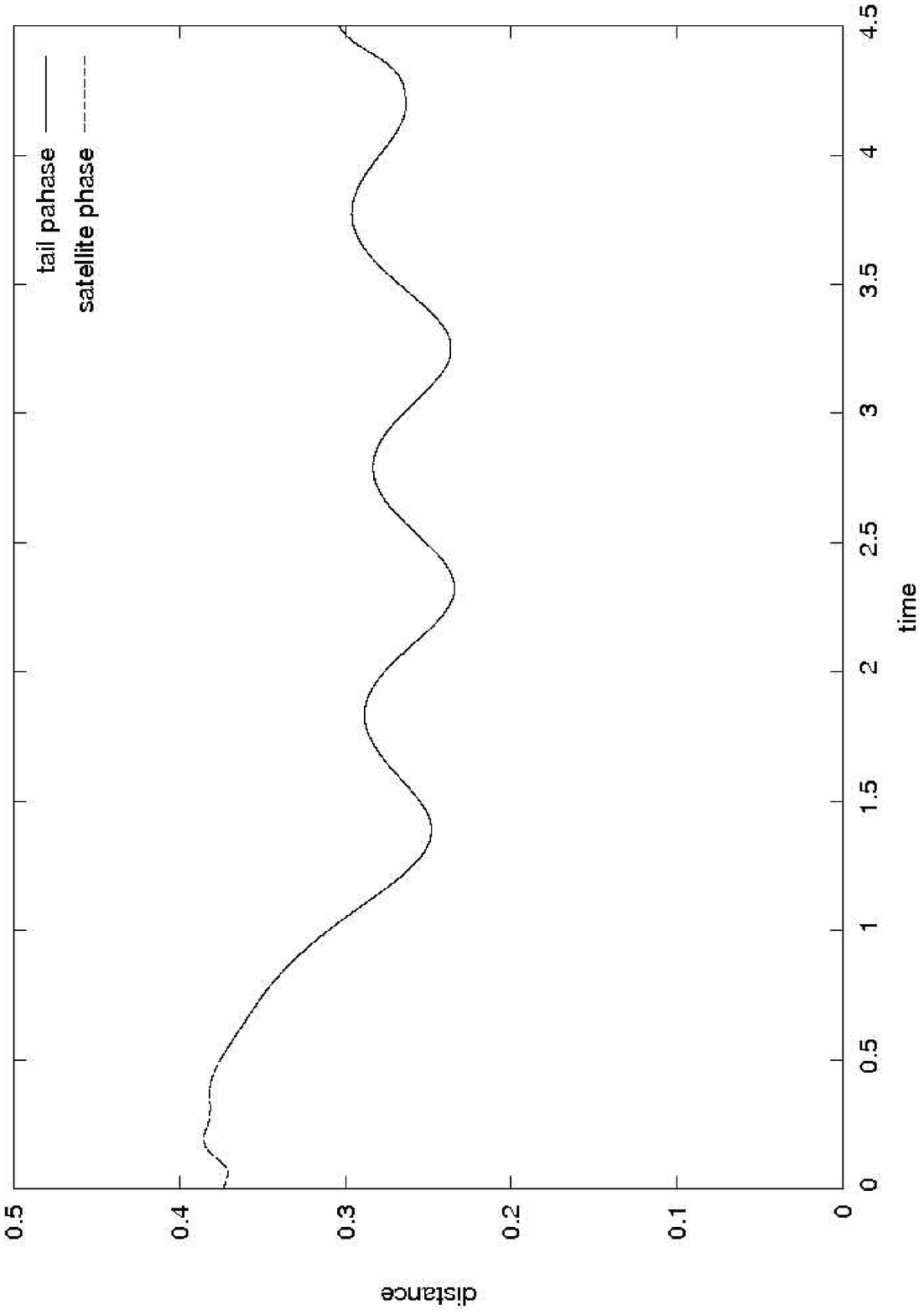}}
\centerline{\includegraphics[width=0.5\columnwidth,angle=-90] {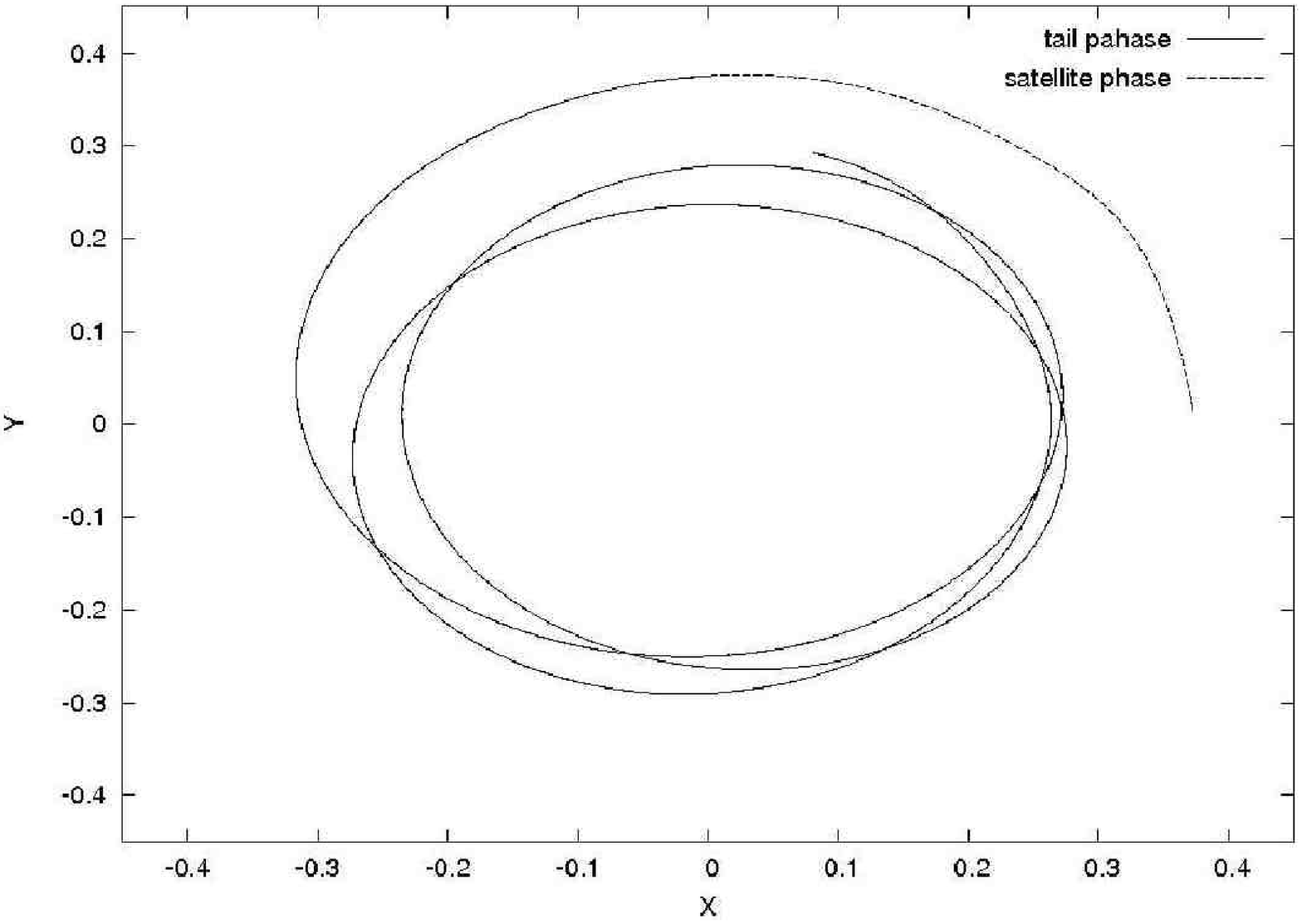}}
\caption {As in Figure \ref{fig:OrbTC3} but now \emph{including} the
  gravitational force of the satellite at all times.  This force
  significantly lowers the energy and angular momentum of the leading
  orbit, decreasing its mean and apocentric radius.  This initial
  period of deceleration ($T<1$) is responsible for the observed
  `kink' in the tail (see Figure \ref{fig:FramesCirc}).}
\label{fig:OrbTC2}
\end{figure}

The leading tail from the low-mass and the tiny-mass satellites in
Figure \ref{fig:FramesCirc} exhibits kinks.  The kinks are a
consequence of the epicyclic motion of the tail orbits and of
acceleration by the satellite at subsequent apocentres.  Figure
\ref{fig:OrbTC2} shows the ensemble averaged distance and positions
for a sample of leading tail particles orbits taken from the
low-mass satellite simulation shown in Figure \ref{fig:FramesLTC2}.
The kink occurs at the first apocentre of the ejecta, after it is
decelerated by the satellite during and subsequent to its escape.  The
deceleration during escape tends to correlate the phases of the
ejected orbits and results in a narrowing of the tidal tail's width.
In contrast, the satellite potential accelerates the trailing tail
particles, which increases the peri- and apocentres of the trailing
tail.  The analogous kink in the trailing tail is not so obvious
because of its lower orbital frequencies.  However, a plot analogous
to Figure \ref{fig:OrbTC2} does show a similar oscillation with lower
angular frequency.

\begin{figure}
  \subfigure[$5\times10^{-4}$]{
    \includegraphics[width=0.75\columnwidth]{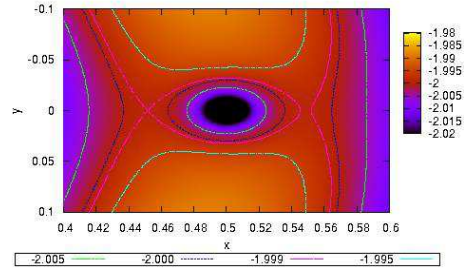}}
    \subfigure[$5\times10^{-2}$]{
    \includegraphics[width=0.75\columnwidth]{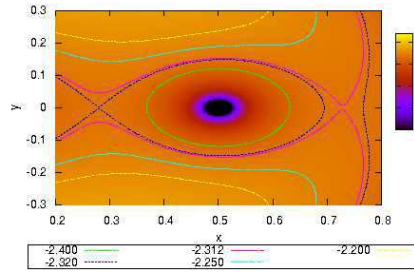}}
  \caption{Contours of the Jacobi constant for two satellites with
    different masses as labelled.  They follow a circular orbit with a
    radius of $q_s=0.4$ in a $c=15$ NFW halo of $M_h=1$ and a virial
    radius $R_h=1$.  The $x$-axis describes the distance between the
    host halo centre and the satellite centre and the $y$-axis
    describes the location in the direction of orbital motion.  Note
    the strong asymmetry in $x$ about the centre for the higher mass
    satellite in Panel (b).}
  \label{fig:jacobi}
\end{figure}

The large changes in the orbits of escaping particles orbits are easily
understood
using a restricted three-body approach.  Consider a satellite of
mass $M_s$ in circular orbit at galactocentric radius $r_s$ in a halo
of mass $M_h$.  In the frame of reference moving with a satellite of
vanishingly small mass, the effective potential is symmetric about the
satellite centre.  Although orbital energy and angular momentum are
not conserved, this system admits a conserved quantity, the Jacobi
constant:
\begin{equation}
E_{J} = E -\vec{\Omega_s} \cdot \vec{L}
\label{eq:jacobi}
\end{equation}
where $E$ and $\vec{L}$ are the orbital energy and angular momentum
and $\vec{\Omega_s}$ is the satellite's angular frequency about the host
halo.  This expression is easily derived by identifying a perfect time
derivative in the inner product of the velocity vector and Newton's
equations of motion in the rotating frame of reference \citep[ Section
3.3.2]{BT87}.
An isocontour of the Jacobi constant passes
through the X-points, $r_\times$, and demarcates the bound and unbound
trajectories as shown in Figure \ref{fig:jacobi}a.  As the satellite mass
increases, the inversion symmetry about the satellite centre is broken
and the unstable points separate as shown in Figure \ref{fig:jacobi}b.  For
small-mass satellites, therefore, the tidal force is symmetric 
about the satellite centre leading to symmetric tidal tails as seen in
globular clusters.  However, for large-mass satellites, the asymmetry
in the tidal force leads to asymmetric mass loss.

Now consider the mass lost through the inner (outer) critical point,
$r_\times$.  Such orbits will have an inward (outward) velocity and
unbound values of the Jacobi constant.  The force from the satellite
continues to affect the orbit beyond the tidal radius in this
restricted problem as in the N-body simulations.  Moreover, the
smaller the mass of the satellite, the closer the radius is to that of
the satellite, and the ejected orbit \emph{lingers} near the original
satellite orbit, partly offsetting the smaller gravitational force.
For this reason, the orbit does not take on the orbital actions of the
satellite but continues to be torqued by the satellite.  One may
estimate the scaling of this energy change by computing the work done
in the satellite frame on the escaping tail particle; this naturally
takes into account the lingering.  Begin with the standard restricted
three-body problem with generalised forces.  Assuming that the
satellite orbits in the $x$-$y$ plane and using Hamilton's Equations,
one may compute the $z$-component torque on an escaping particle and
the change in angular momentum of the escaping particle after an interval $T$ becomes
\begin{equation}
\Delta L_z = \int_0^T dt\,\left(-{\partial H\over\partial\phi}\right) = 
  \int_0^T dt\,\left(-{\partial V_s\over\partial\phi}\right)
\label{eq:ejtorque}
\end{equation}
where $H$ is the Hamiltonian, $\phi$ is the azimuthal coordinate
conjugate to $L_z$ and
\[
V_s = -{GM_s\over|{\bf r} - {\bf r}_s(t)|}
\]
is the gravitational potential of the satellite.  The second equality
in Equation (\ref{eq:ejtorque}) owes to the $\phi$ independence of
all the other terms in $H$.  We may consider an escaping orbit in the
limit that the mass of the satellite $M_s$ is much smaller than the
mass of halo $M_h$ and use perturbation theory to evaluate Equation
(\ref{eq:ejtorque}).  To do this, let the unperturbed orbit be the
circular orbit that passes through the X-point, $r_\times$ at $t=0$.  
Expanding to lowest contributing order in $M_s/M_h$, after some 
straightforward algebra and taking the limit $T \rightarrow \infty$, 
one may show that
\[
\Delta L_z = {GM_s\over r_\times^2}
  \left(\left.{\partial\Omega_\phi\over\partial r}\right|_{r_s}\right)^{-1}
\]
where $\Omega_\phi$ is the azimuthal orbital frequency.  Finally, it
follows that $\Delta E = \Omega_s\Delta L_z$ from the conservation of
the Jacobi constant (Equation \ref{eq:jacobi}) which yields:
\begin{equation}
\Delta E = - 2 \Omega_s^2 r_s^2
\left(-\left.{\partial\ln\Omega_\phi^2\over\partial\ln r}\right|_{r_s}\right)^{-1/3}
\left({M_s\over M_h}\right)^{1/3}.
\label{equ:scale}
\end{equation}
Since $G$, $r_s$, and $\Omega_s$ are constant, Equation
(\ref{equ:scale}) implies that the work done is proportional to
$(M_s/M_h)^{1/3}$. In other words, the change in the orbital energy of
the escaping particle decreases as the satellite mass decreases but
only weakly!

\begin{figure}
  \subfigure[inner]{
    \includegraphics[width=0.43\columnwidth]{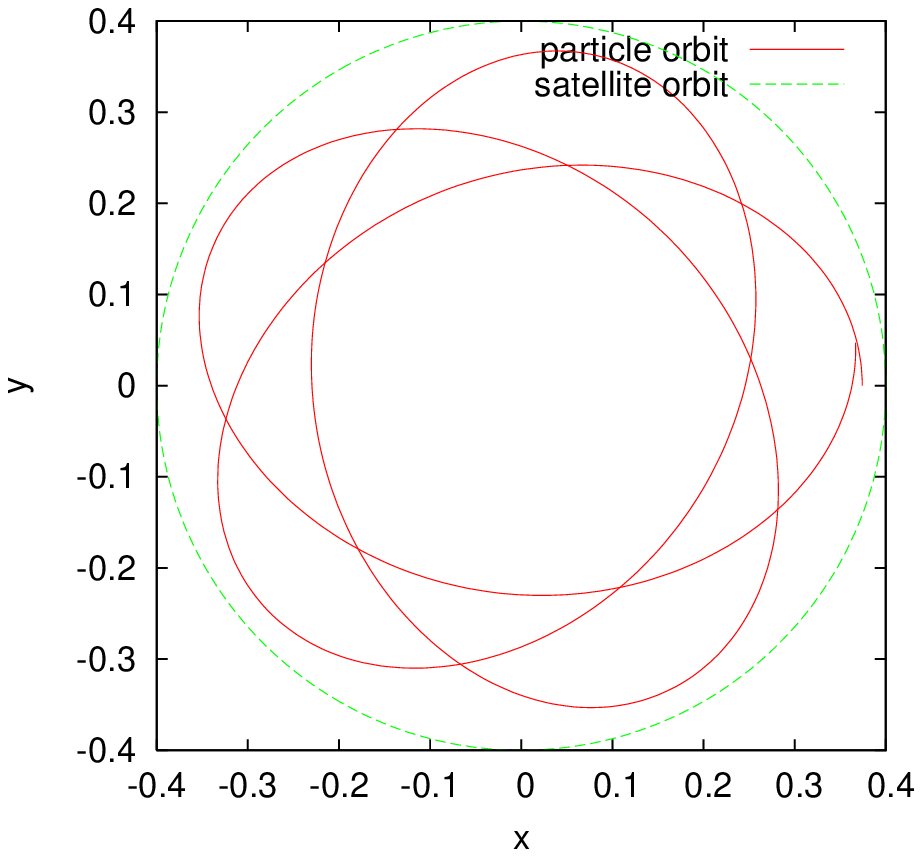}
    \includegraphics[width=0.57\columnwidth]{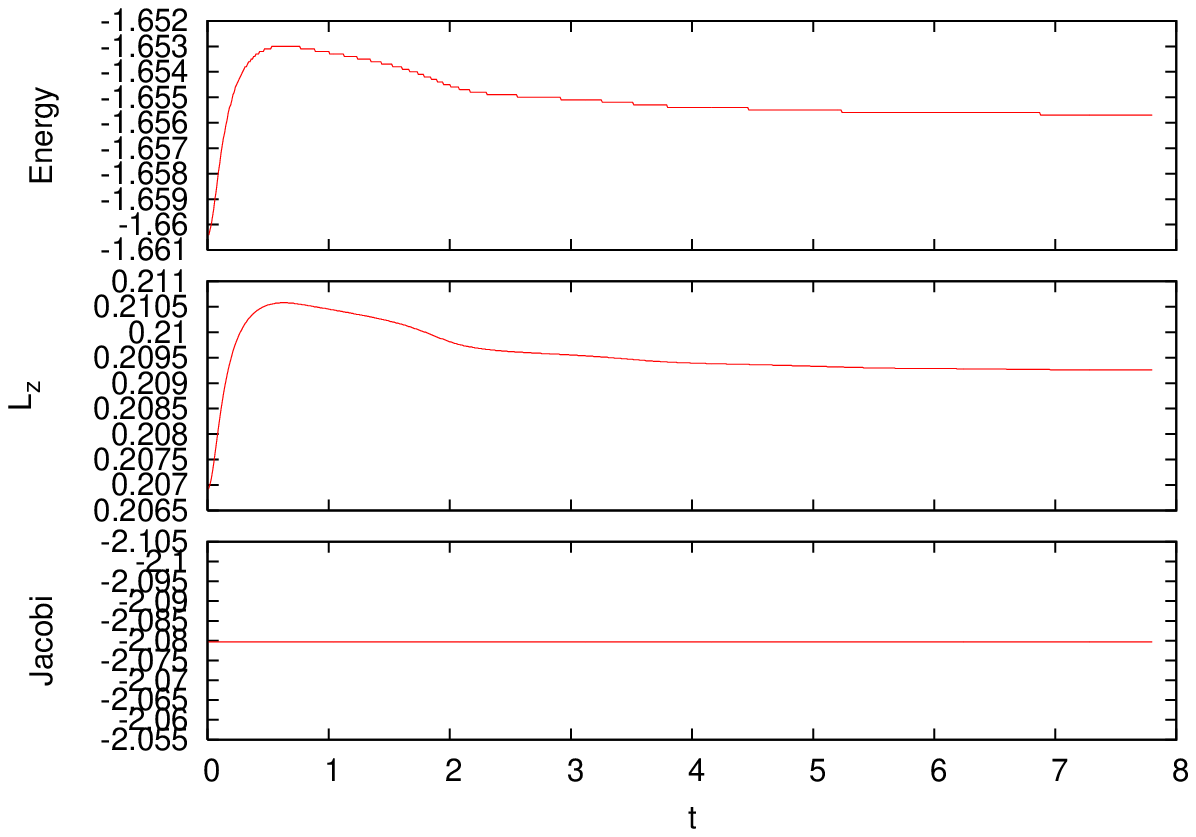} }
    \subfigure[outer]{
    \includegraphics[width=0.43\columnwidth]{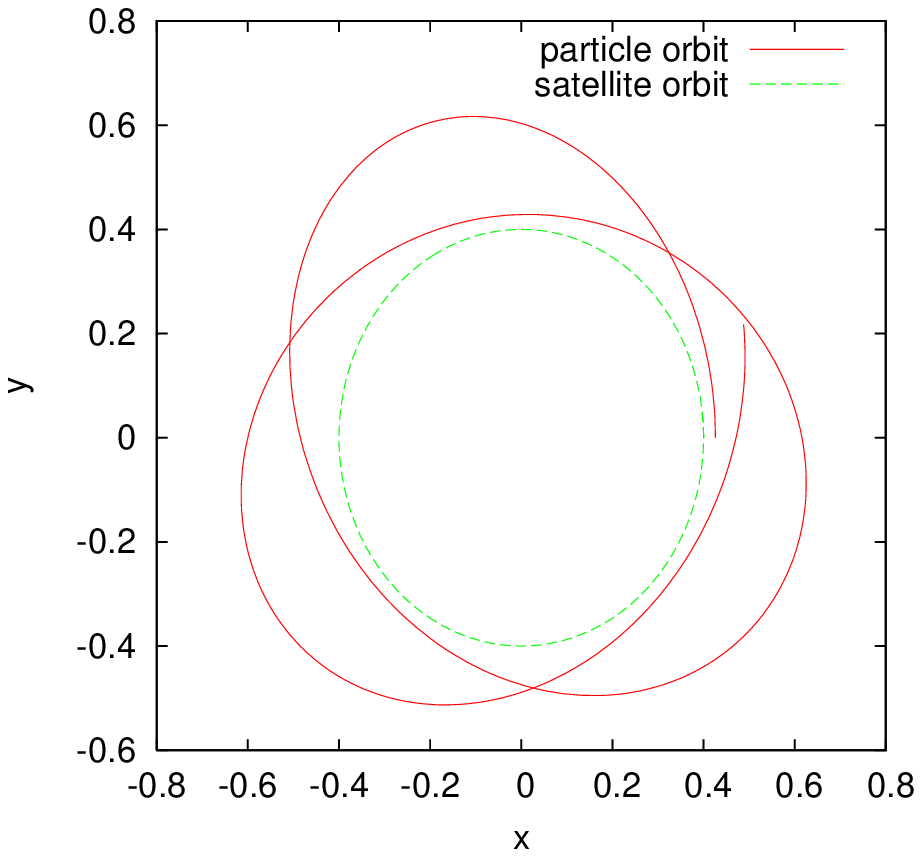}
    \includegraphics[width=0.57\columnwidth]{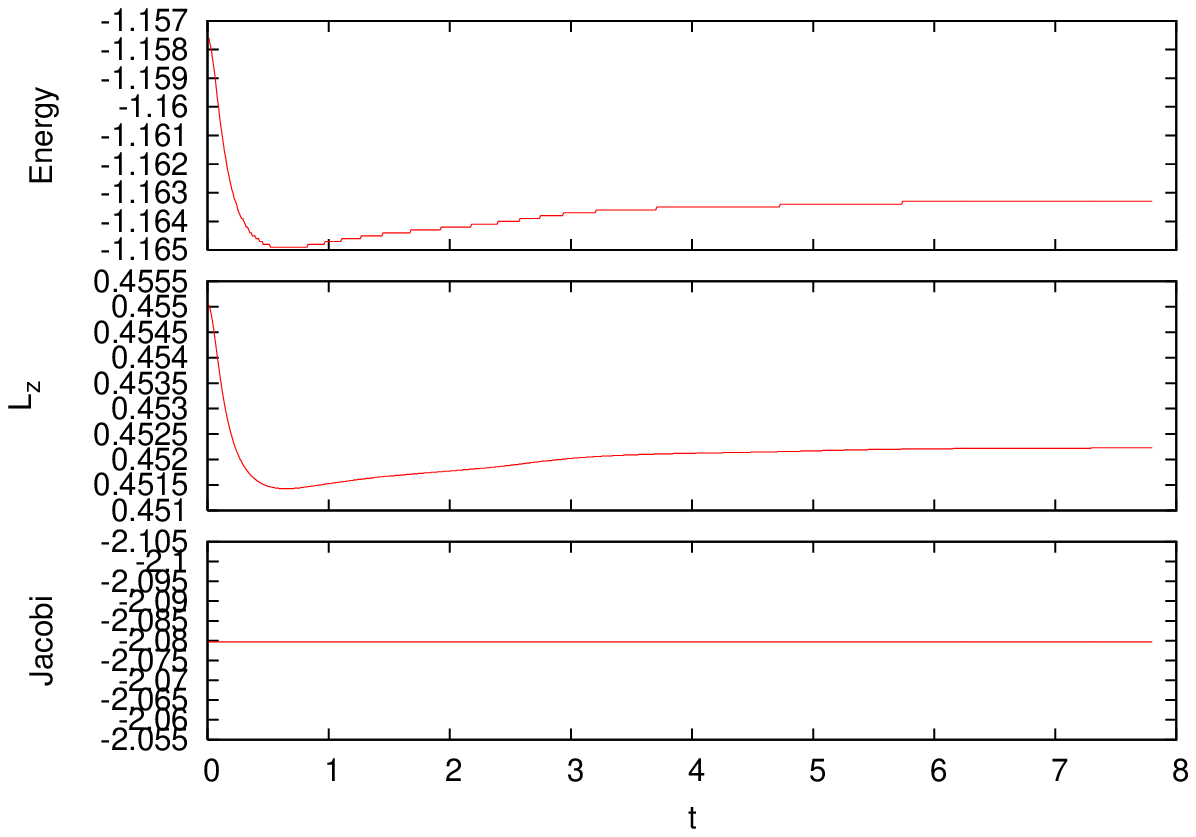} }
  \caption{The evolution of two orbits escaping from the inner (upper panels)
  and the outer (lower panels) tidal radii for $M_s/M_h=10^{-4}$ where $M_s$ is
  the satellite mass and $M_h$ is the host halo mass.  The left panels show
    the orbital plane and the right panels show the evolution of energy,
    angular momentum, and the
    Jacobi constant.  The value of the Jacobi constant is conserved as
    expected.}
\label{fig:m10em4}
\end{figure}

\begin{figure}
  \subfigure[inner]{
    \includegraphics[width=0.43\columnwidth]{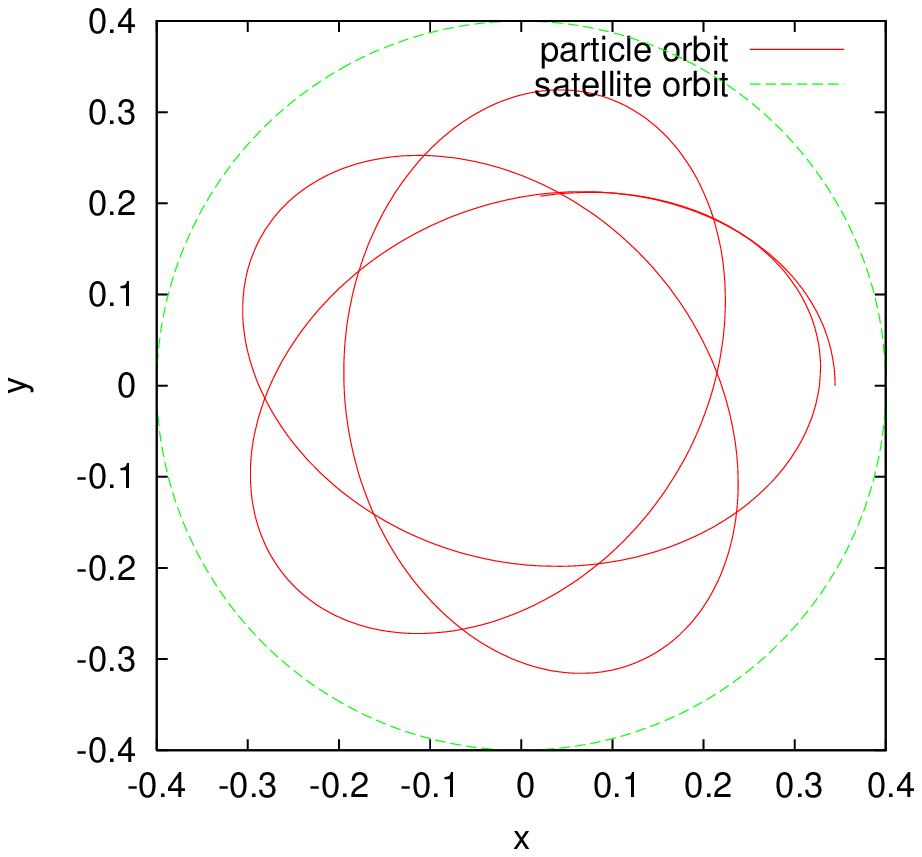}
    \includegraphics[width=0.57\columnwidth]{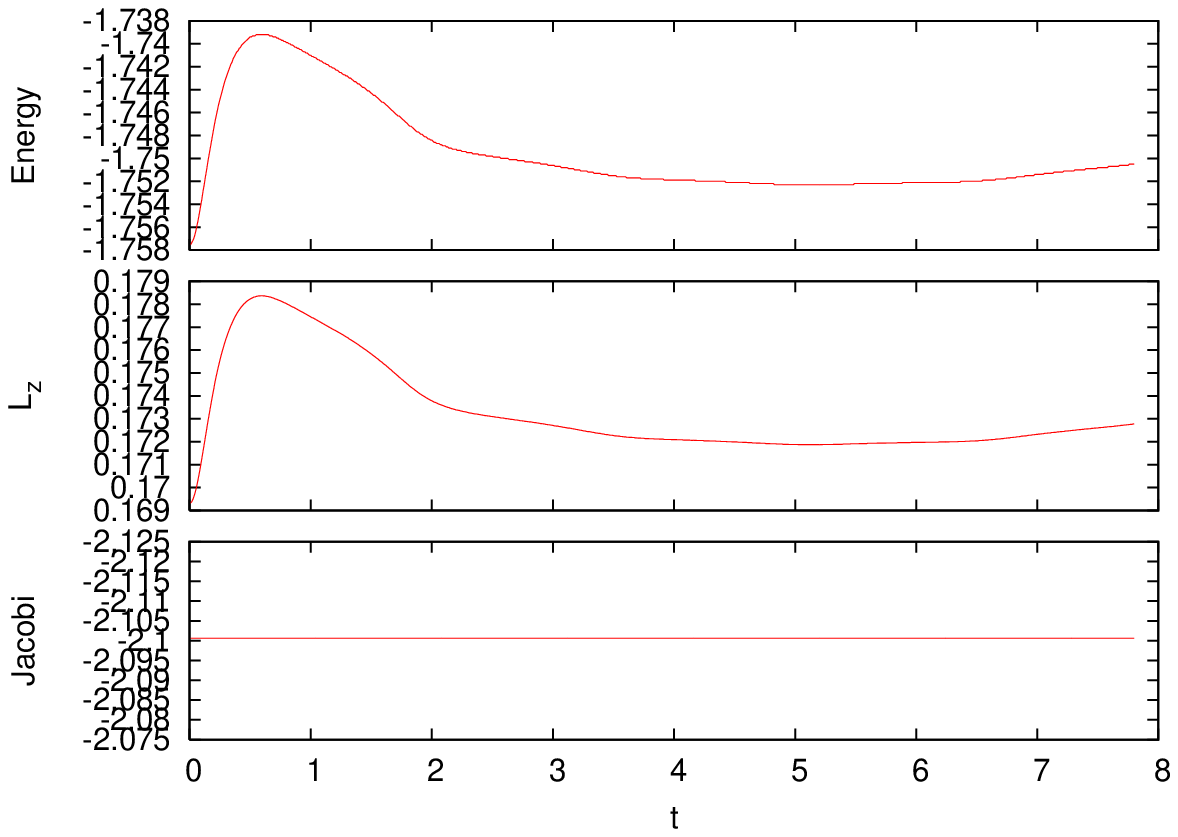} }
    \subfigure[outer]{
    \includegraphics[width=0.43\columnwidth]{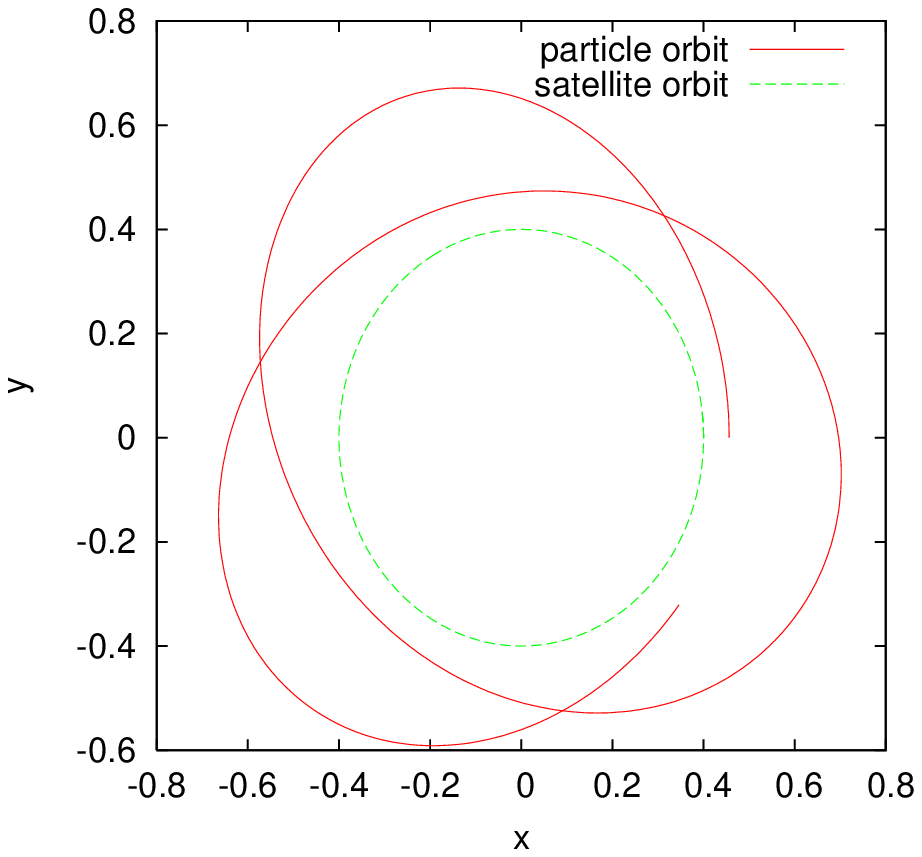}
    \includegraphics[width=0.57\columnwidth]{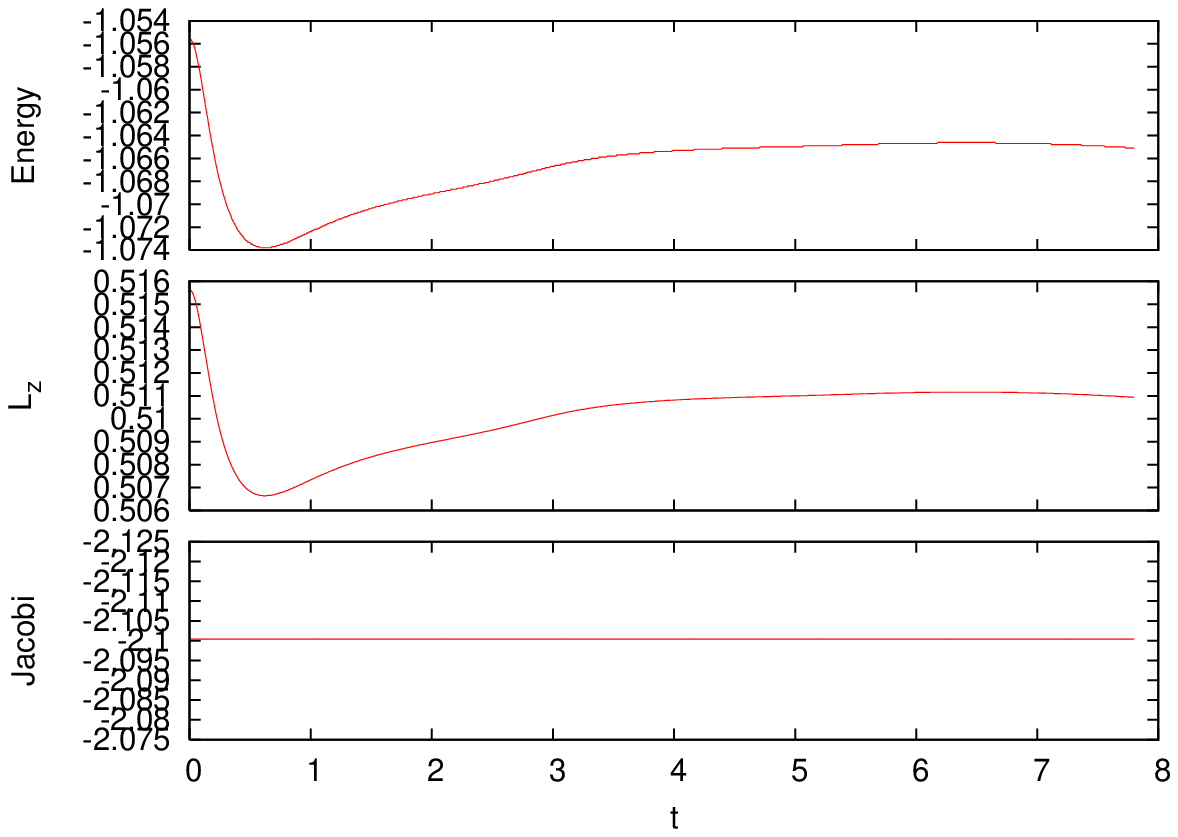} }
  \caption{ As in Figure \protect{\ref{fig:m10em4}} but for
    $M_s/M_h=10^{-3}$.}
\label{fig:m10em3}
\end{figure}

\begin{figure}
  \subfigure[inner]{
    \includegraphics[width=0.43\columnwidth]{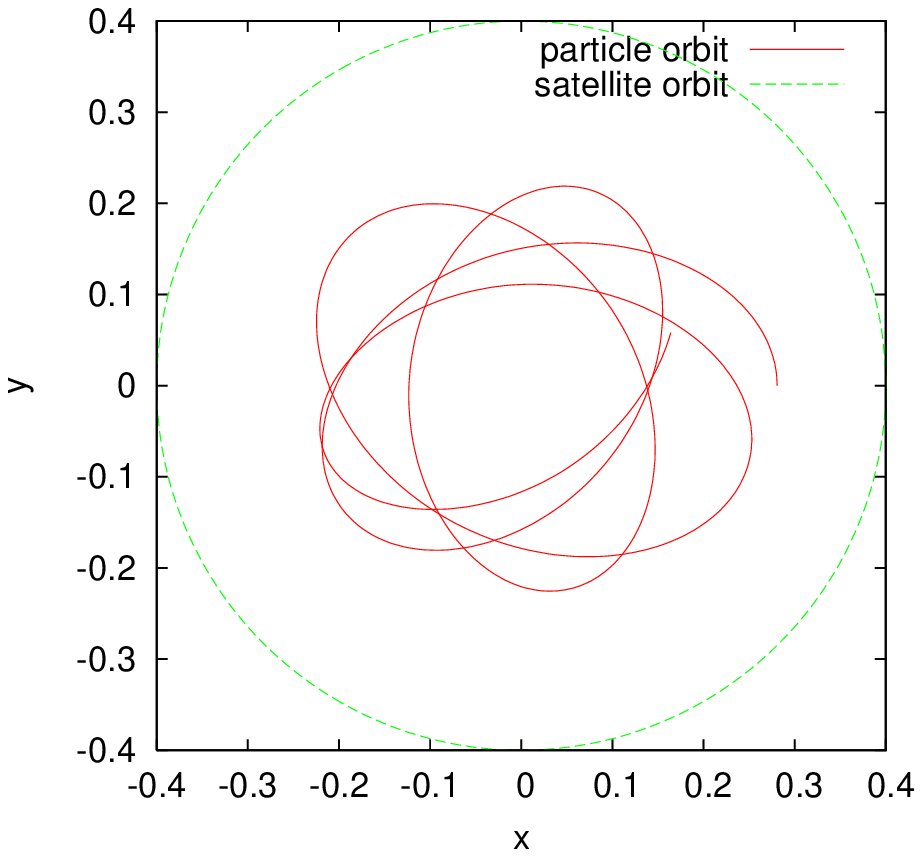}
    \includegraphics[width=0.57\columnwidth]{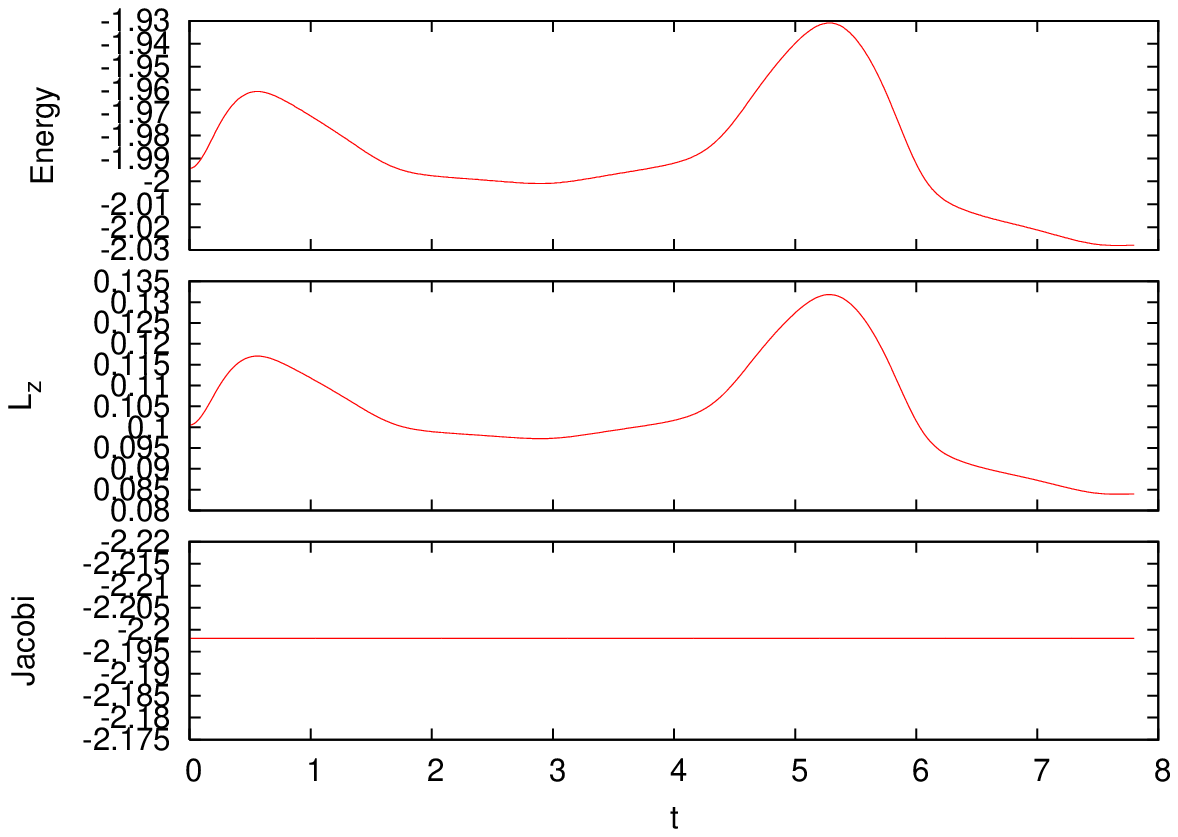} }
    \subfigure[outer]{
    \includegraphics[width=0.43\columnwidth]{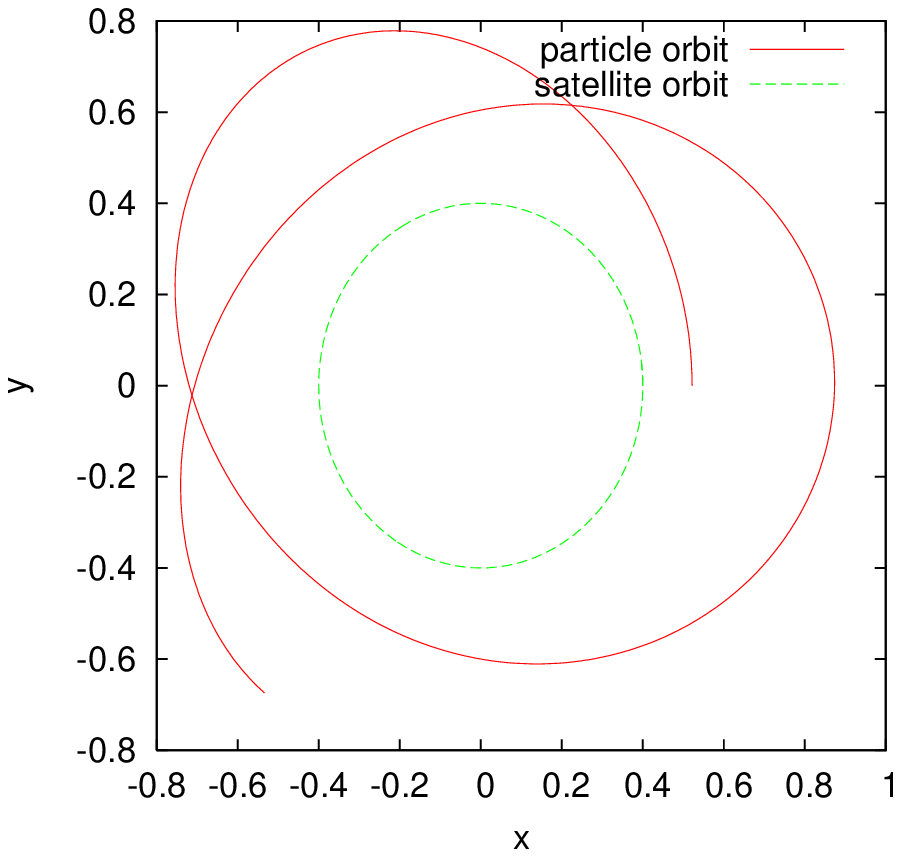}
    \includegraphics[width=0.57\columnwidth]{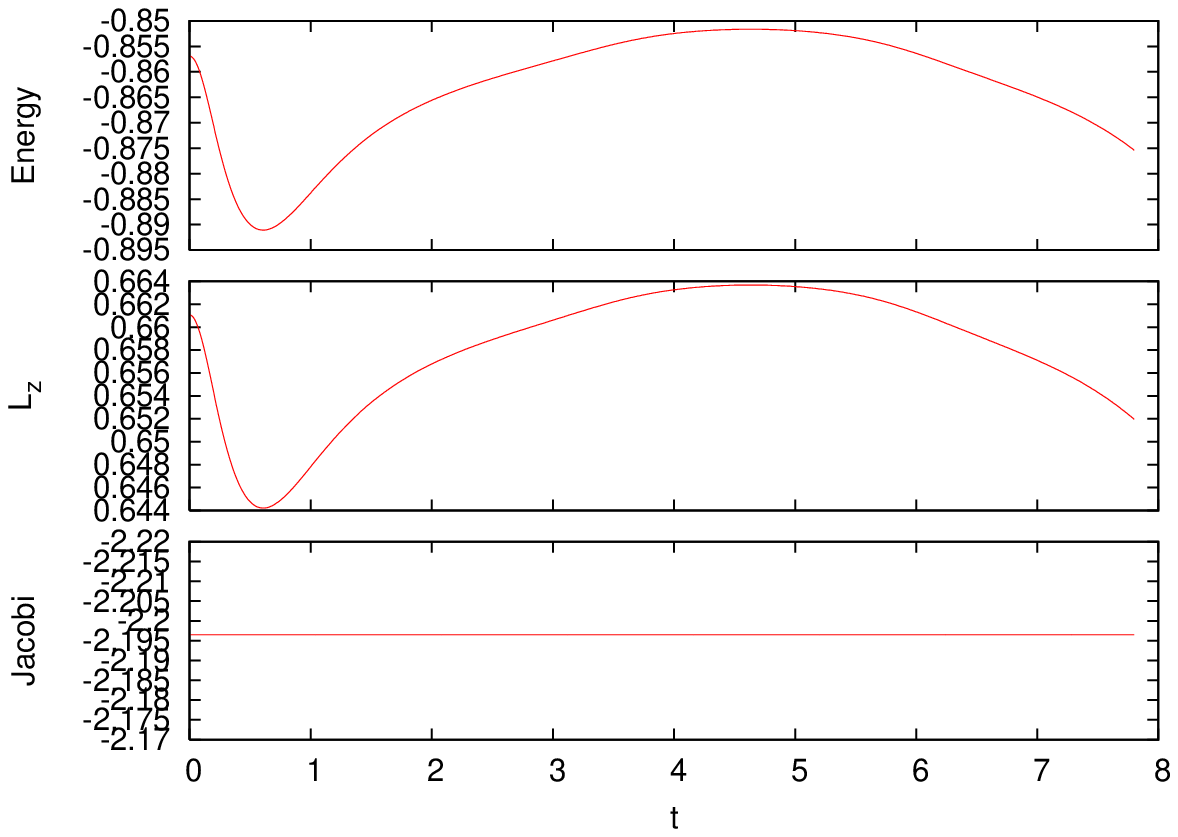} }
  \caption{ As in Figure \protect{\ref{fig:m10em4}} but for
    $M_s/M_h=10^{-2}$.  }
\label{fig:m10em2}
\end{figure}

\begin{figure}
\centerline{\includegraphics[width=0.75\columnwidth,angle=0]{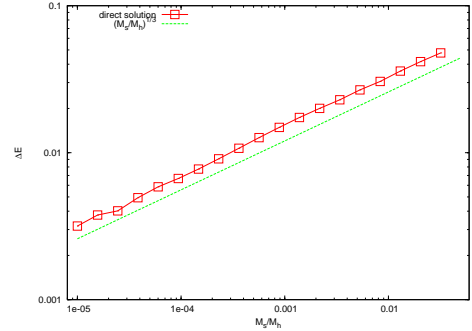}}
\caption{A numerical test of the predicted scaling of the energy
  change of escaping particles (Equation  \protect{\ref{equ:scale}}) with
  satellite mass.  The $x$-axis is the ratio of satellite mass $M_s$
  to total halo mass $M_h$ and the $y$-axis is the magnitude of the energy
  change, $\Delta E$.  The straight line is the relation $(M_s/M_h)^{1/3}$.
}
\label{fig:ScaleTest}
\end{figure}

\begin{figure}
\centerline{\includegraphics[width=0.75\columnwidth,angle=0]{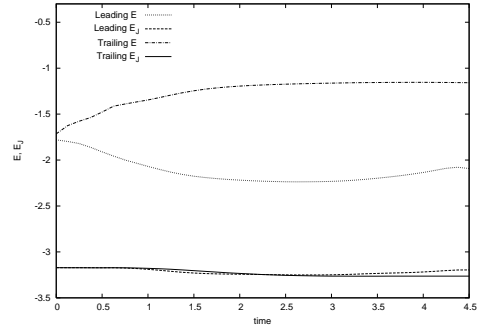}}
\caption {The time evolution of the energy, $E$, and the Jacobi
  constant, $E_J$, for two ensembles of particles ejected from the
  low-mass satellite selected from the leading and trailing tail,
  respectively.  The value $E_J$ is nearly conserved while the energy
  changes owing to work done during escape.  Changes in $E_J$ are caused
  by mass loss and the resulting evolution of the satellite's
  gravitational potential.}
\label{fig:jacobiE}
\end{figure}

Although the derivation of the scaling assumes $M_s/M_h\rightarrow0$,
we demonstrate numerically that it applies over all values of interest
by integrating the equations of motion in the rotating potential.  We
adopt $r_s=0.4$ and choose values of the Jacobi constant that are 1\%
larger than the critical value passing through $r_\times$ with zero
velocity.  The initial motion, in the rotating frame, is along (or
against) the direction of rotation for inner (outer) escapees.
Figures \ref{fig:m10em4}--\ref{fig:m10em2} show the resulting
trajectories and conserved quantities for $M_s/M_h=10^{-4}, 10^{-3},
10^{-2}$.  For inner (outer) escapees, the energy and angular momentum
decrease (increase) after the initial transient for $t<0.5$.  Figure
\ref{fig:ScaleTest} shows that the energy change for ensembles of
orbits in the leading tail chosen as follows.  The initial position is
chosen to be 2\% of $r_\times$ outside of the X-point and the
velocities are chosen to have a normal distribution in the satellite
frame with a dispersion that is 2\% of the satellite's circular velocity at
$r_\times$.  The orbits in Figures \ref{fig:m10em4}--\ref{fig:m10em2}
are representative members of these ensembles.  The magnitude of the
energy change $\Delta E$ is defined as the ensemble average of
$|E_{min}-E_{init}|$ where $E_{init}$ is the initial energy and
$E_{min}$ is the minimum energy along the orbit.  These numerical
values are consistent with the predicted scaling $(M_s/M_h)^{1/3}$.
Circumstance may eventually bring the ejected particle close to the
satellite once again, as seen in Figure \ref{fig:m10em2}.  The radial
extent of the annulus covered by the orbit increases only gradually
with increasing satellite mass, reflecting the same weak dependence on
$M_s/M_h$.  In the simulations, the mass loss and subsequent
reequilibration of the satellite potential causes small deviations in
the Jacobi constant even though the satellite orbit remains circular
as shown in Figure \ref{fig:jacobiE}.  Nonetheless, the restricted
three-body dynamics explains most of the features seen in the orbit
evolution.

\subsubsection{Non-circular orbits}
\label{sec:ecctail}

\begin{figure*}
\centerline{
\includegraphics[width=0.4\textwidth,angle=0]{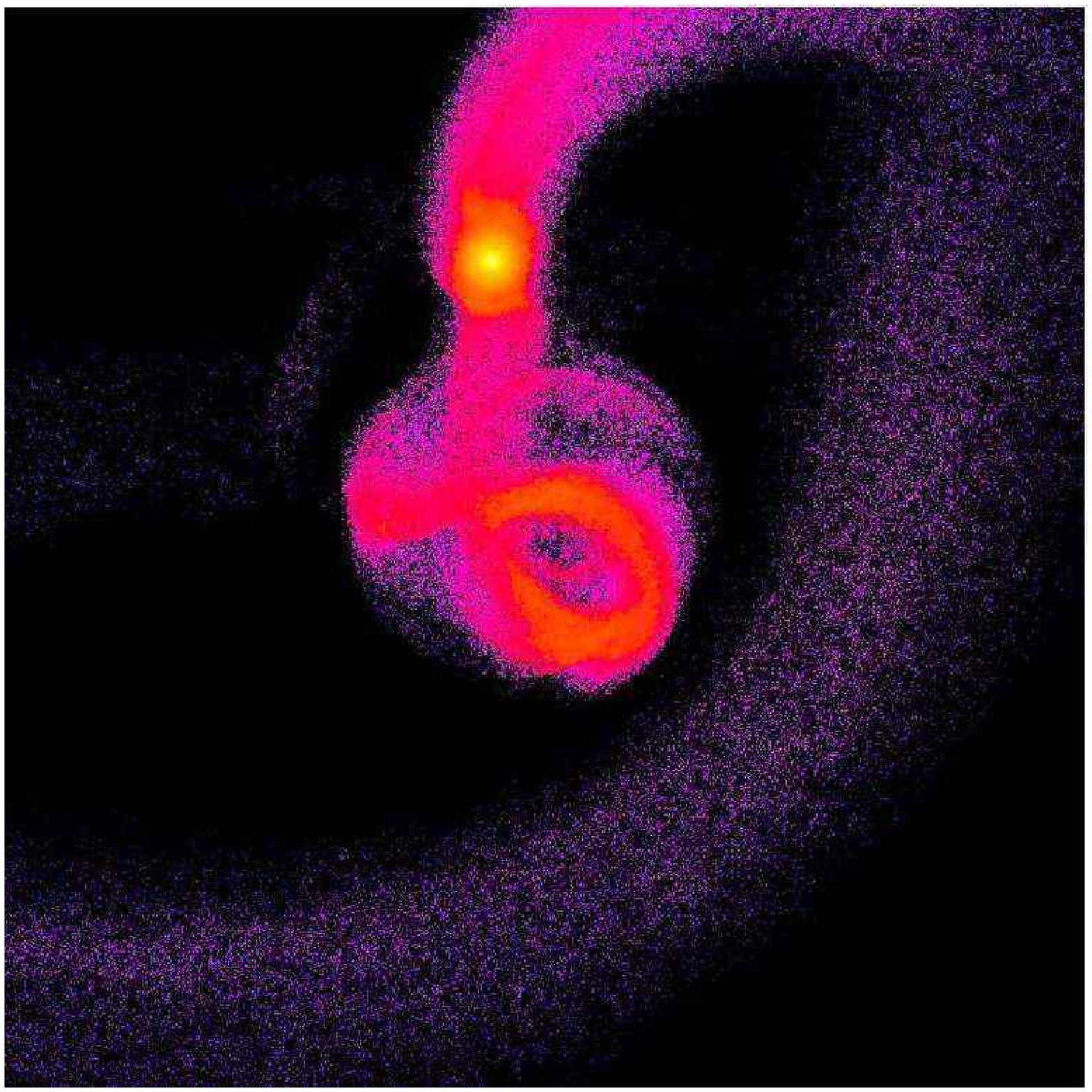}
\includegraphics[width=0.4\textwidth,angle=0]{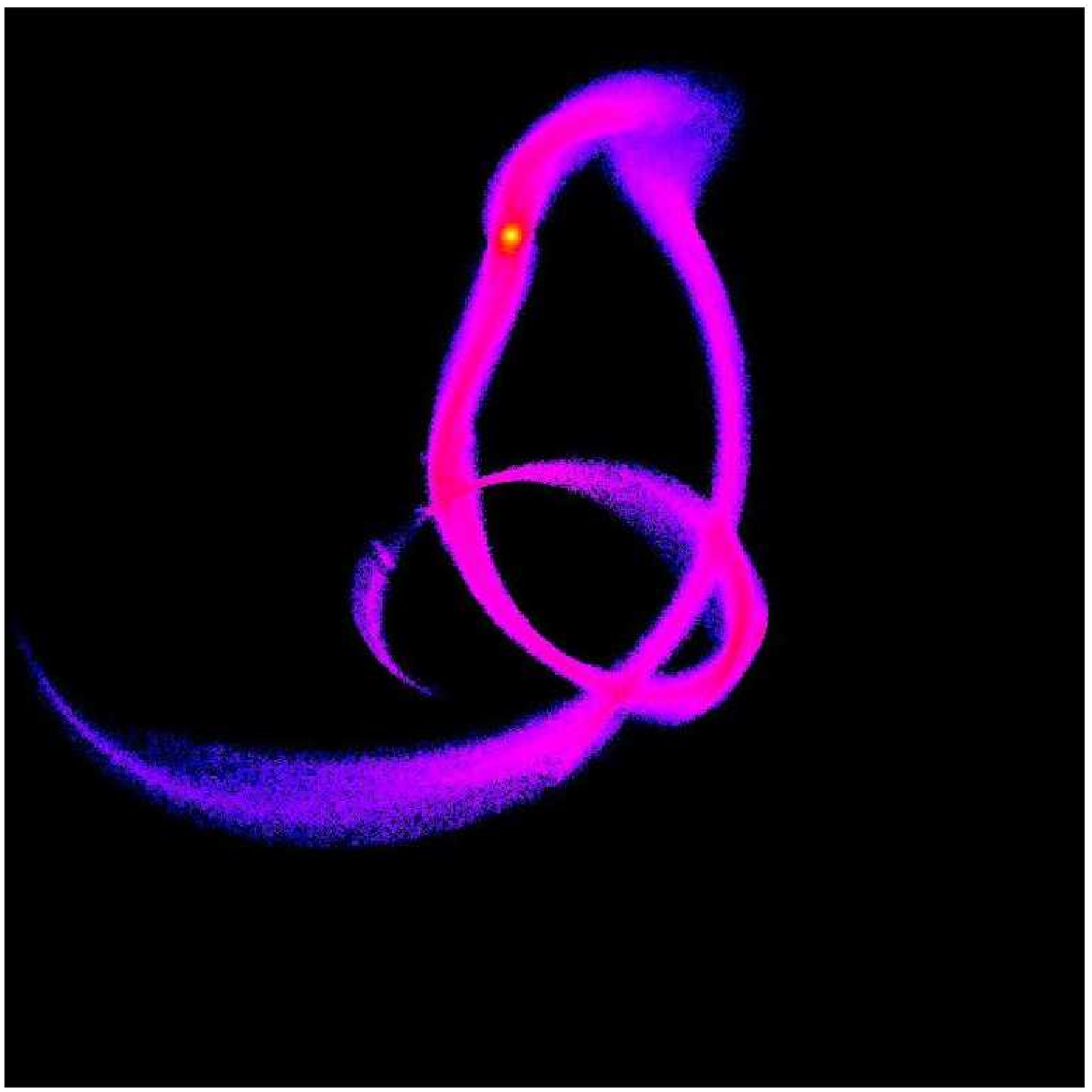}
}
\caption {Tails at time $T=5.0$ for the massive (left) and the
  low-mass (right) satellite on an eccentric, $e=0.5$, orbit.}
\label{fig:FramesEcc}
\end{figure*}

Although we still expect some of the insight gained from the restricted
three-body problem to carry over to the evolution of a satellite on an {\em
  eccentric} orbit, this more complex situation requires direct
simulation.  Figure \ref{fig:FramesEcc} shows the tidal tails of a
massive and low-mass satellite on an eccentric, $e=0.5$,
orbit.  The tidal tail morphology for eccentric satellite orbits is
significantly more complex than for circular satellite orbits and
varies more strongly with satellite mass.  In the left panel of
Figure \ref{fig:FramesEcc}, the massive satellite has dramatically
decelerated the leading tail, which now reaches the host halo centre
and forms an inner ``reservoir'' of ejecta. The deceleration by the
satellite causes the leading tail to appear close to radial.  In the
right panel of Figure \ref{fig:FramesEcc}, the multiply segmented
tail from the low-mass satellite is caused by two mechanisms.  First,
during each satellite orbit, the leading tail forms during the
approach to pericentre.  After pericentre, the tidal strain and the
mass-loss rate diminishes resulting in a gap in the tail.  Second,
deceleration by the satellite changes the orbits of the newly
disconnected leading tail, producing a distinct segment.

\begin{figure}
\centerline{
\includegraphics[width=0.475\columnwidth,angle=0]{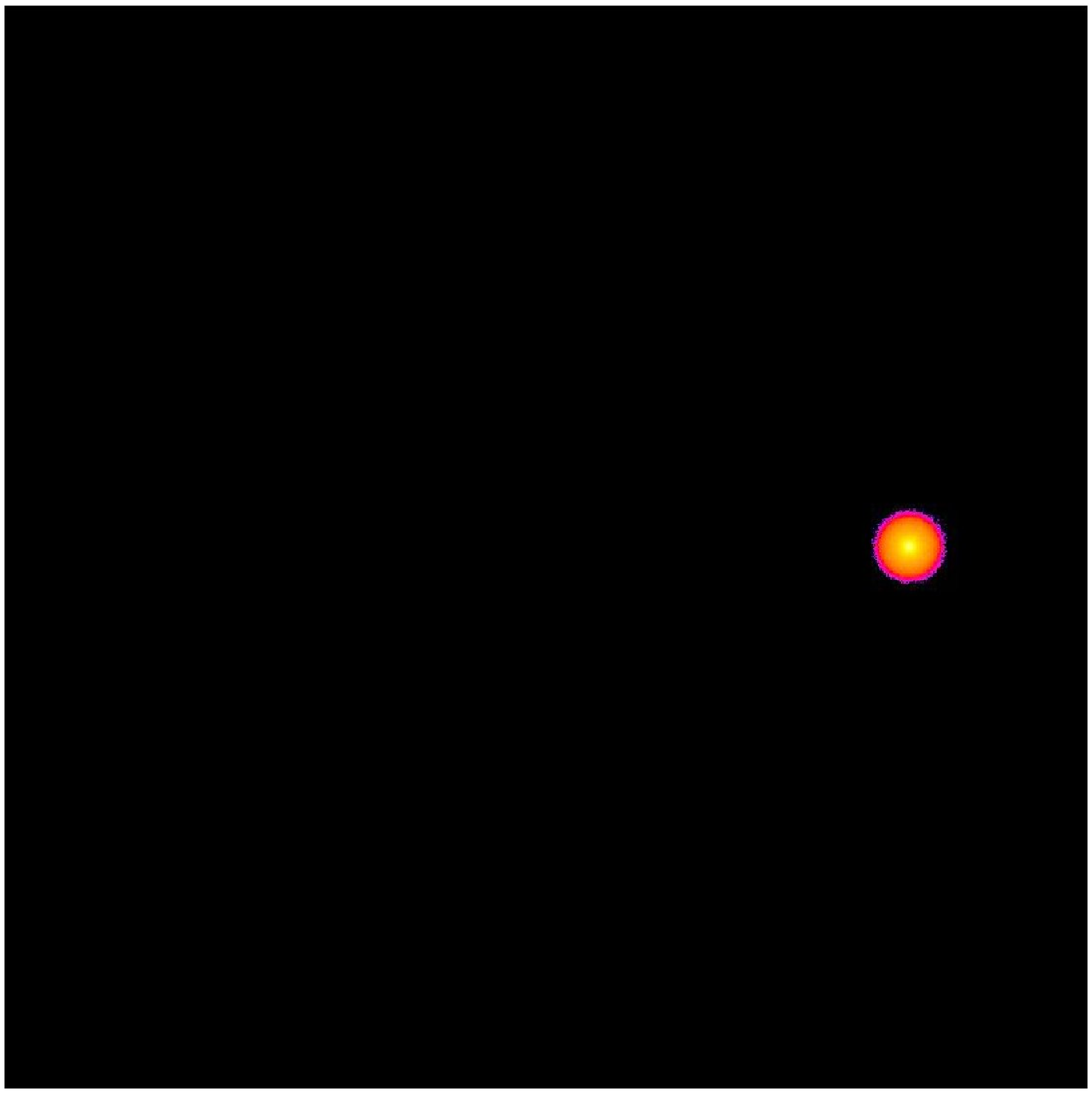}
\includegraphics[width=0.475\columnwidth,angle=0]{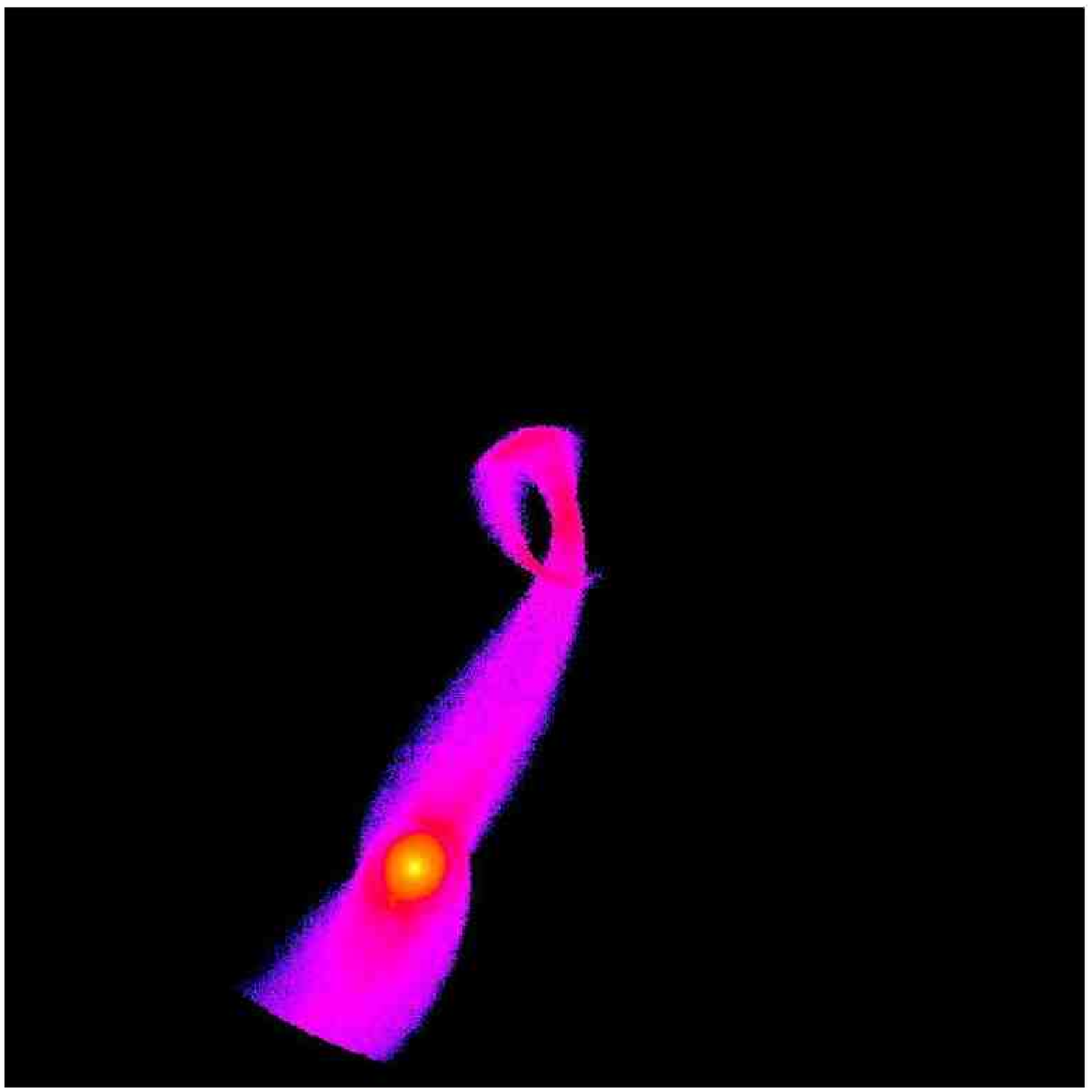} }
\centerline{
\includegraphics[width=0.475\columnwidth,angle=0]{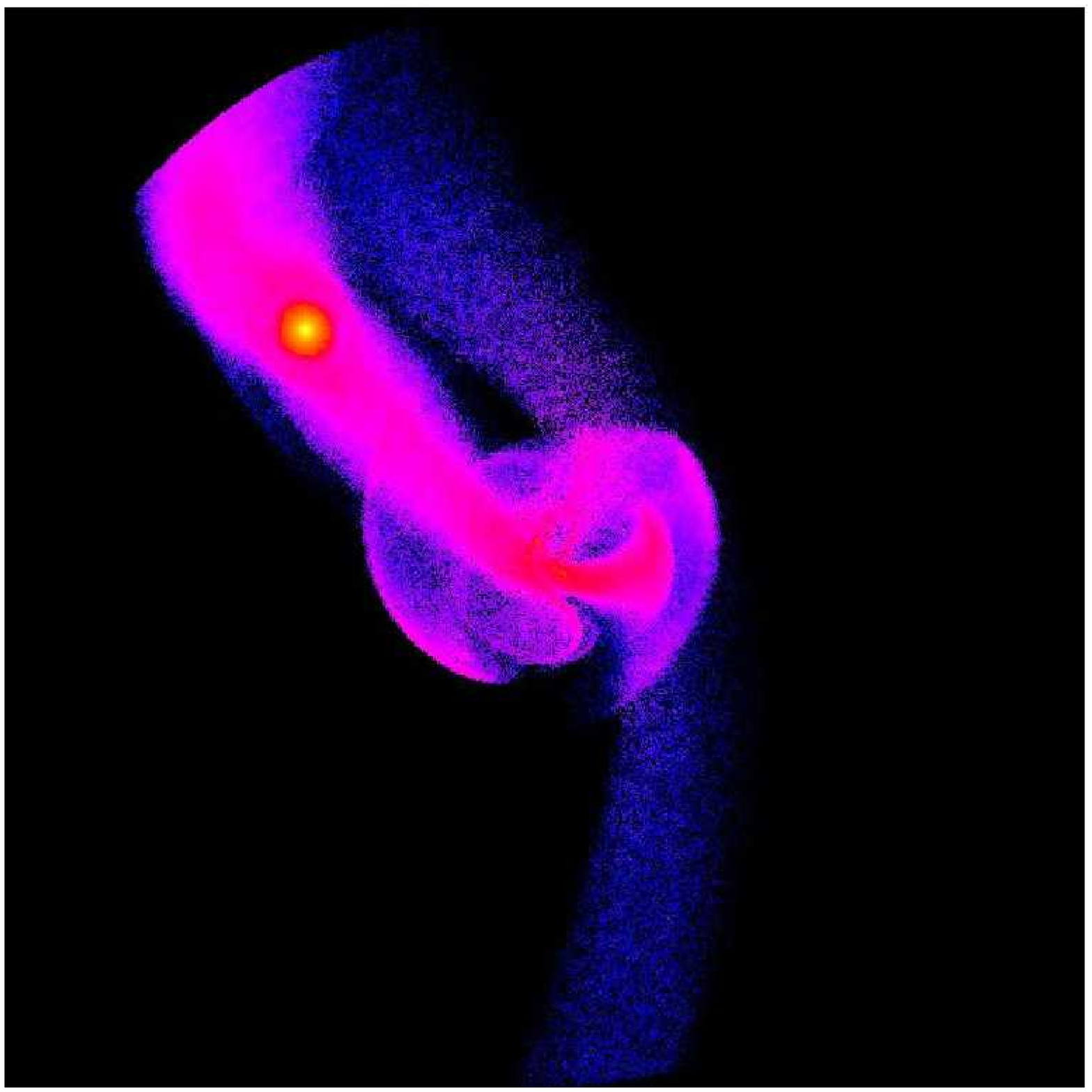}
\includegraphics[width=0.475\columnwidth,angle=0]{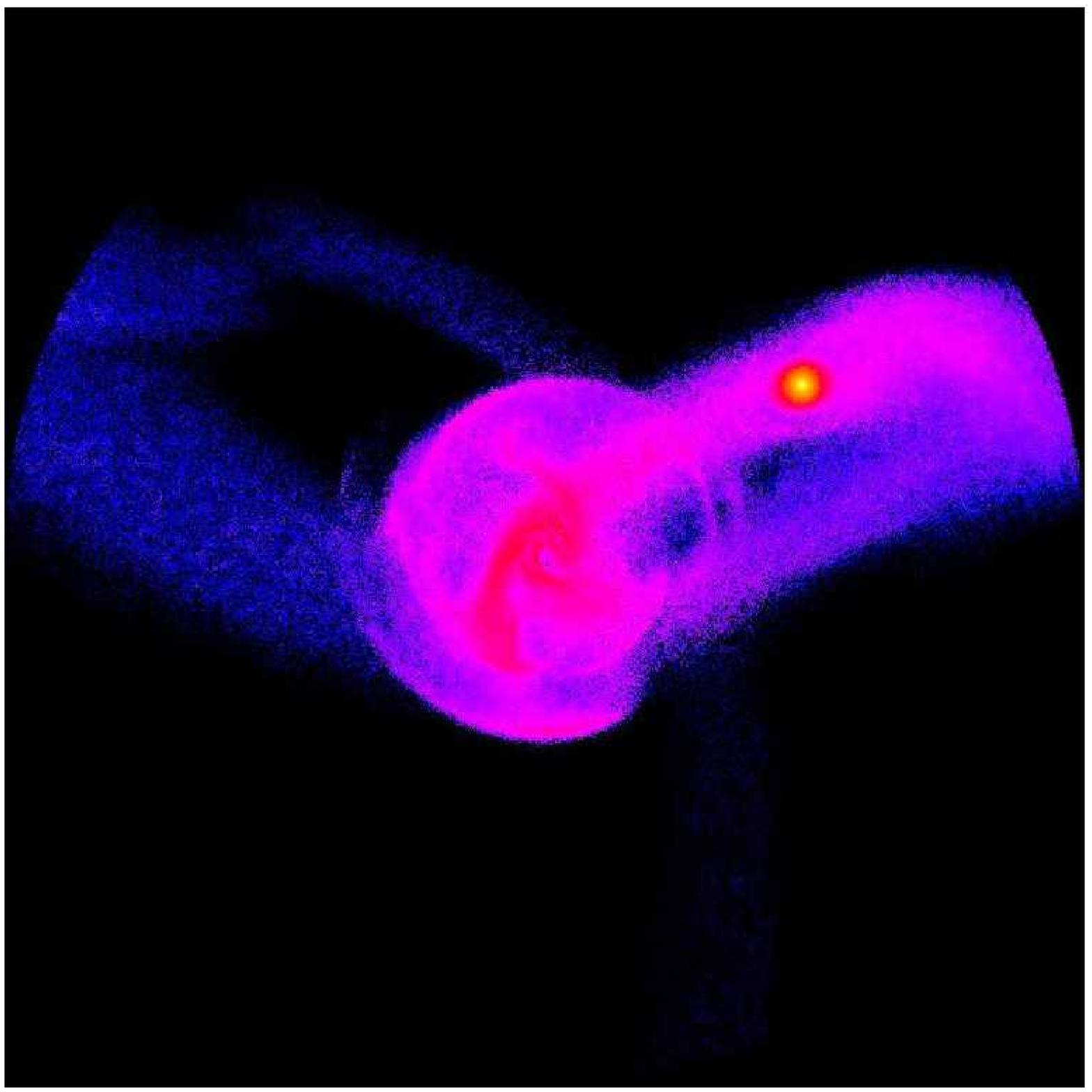} }
\caption {As in Figure \ref{fig:FramesMTC1} but for the massive satellite
  on an $e=0.74$ orbit at $T=0.0$, $1.5$, $3.0$, and $4.5$ in the
  top-left, top-right, bottom-left panels, and bottom-right panels,
  respectively.}
\label{fig:FramesMTK}
\end{figure}

\begin{figure}
\centerline{
\includegraphics[width=0.33\columnwidth,angle=0]{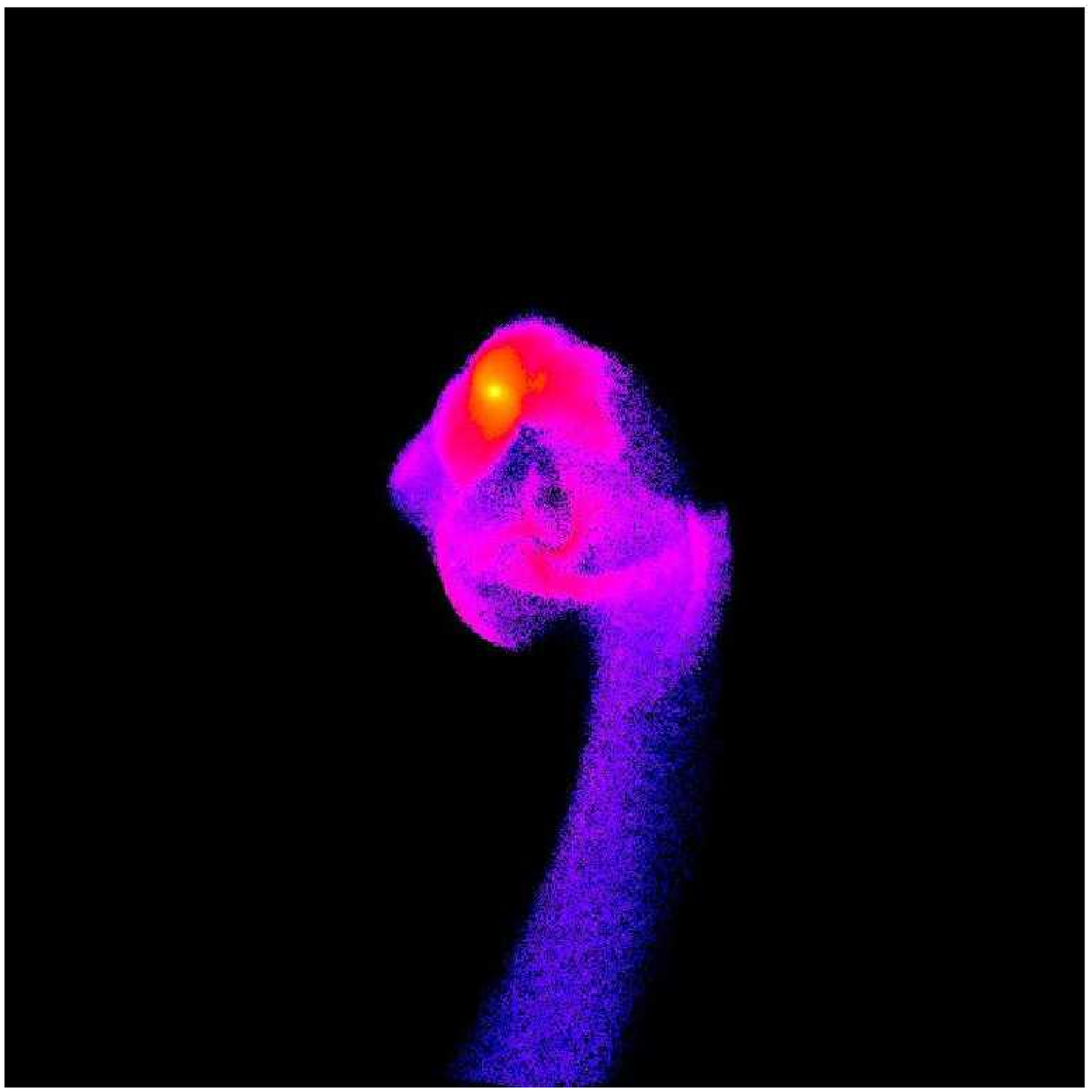}
\includegraphics[width=0.33\columnwidth,angle=0]{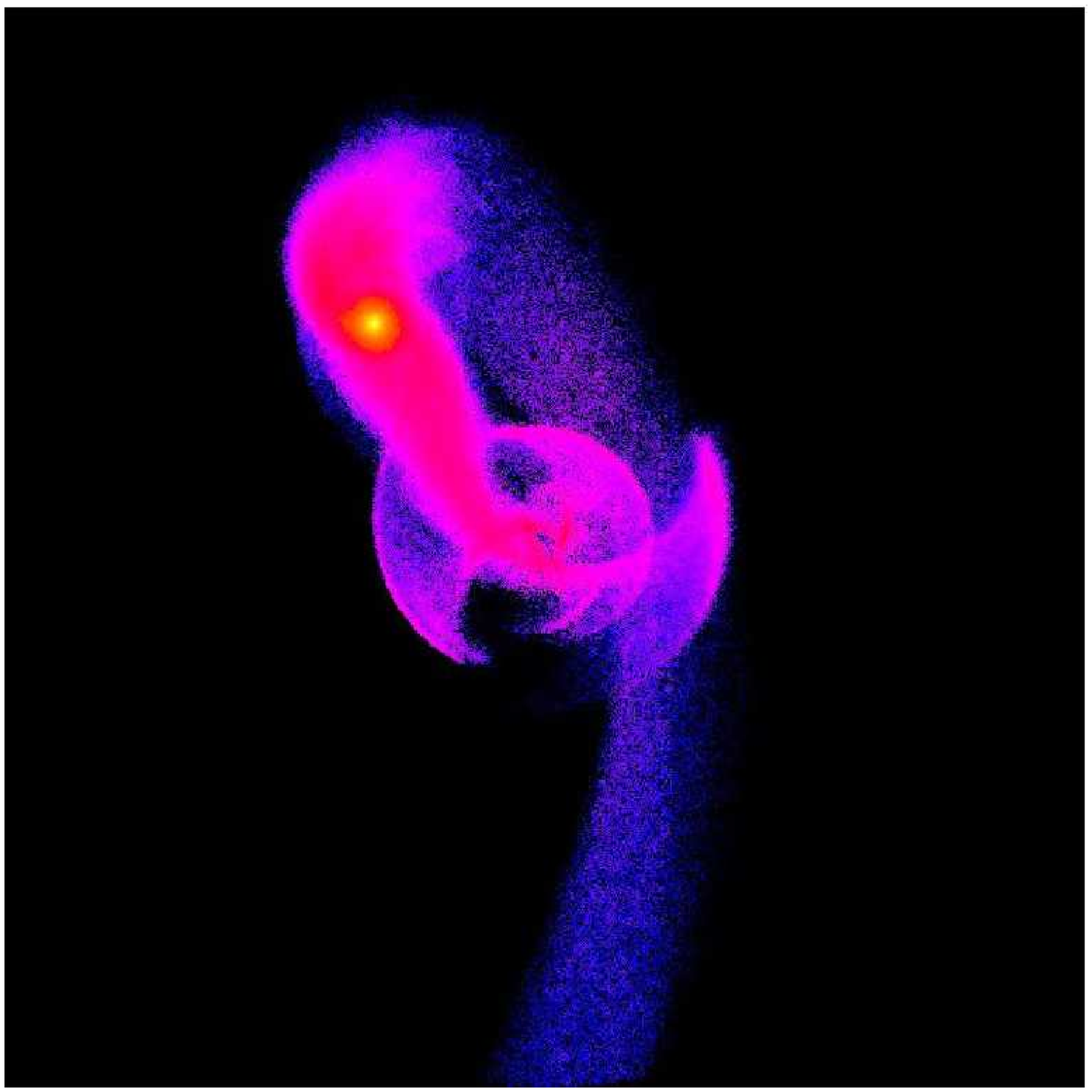}
\includegraphics[width=0.33\columnwidth,angle=0]{frame.MTK1.00024.eps}
}
\caption {The same as Figure \ref{fig:FramesMTK} but for times $T=2.5$, $2.75$,
  and T=$3.0$ from left to right, respectively, as the the satellite
  moves from pericentre to apocentre.  Owing to the deceleration by the
  satellite potential, the leading tail falls toward the centre of the
  host halo.}
\label{fig:FramesMTK2}
\end{figure}

Figure \ref{fig:FramesMTK} shows the evolution of a
massive satellite on an e=0.74 orbit.  Initially, the leading tail
points directly toward the halo centre but the strong deceleration by the 
satellite eventually fills
the inner halo with ejecta.  Figure \ref{fig:FramesMTK2} provides
a finer time sampling of the evolution between pericentre and
apocentre for the same simulation.  Instantaneously, the
morphology can be very complex and the position angle of the leading
tail can vary significantly from its nearly radial average.  There is
little correlation between the tail location and the satellite orbit.  The
location of the inner ejecta, e.g. its outer turning points, is determined
by the host halo potential, the time-varying satellite potential, and the
satellite orbit in combination.  Therefore, unlike streams from very low-mass
satellites, the tail orientation is not directly
informative.  However, through dynamical modelling, the location of the
inner ejecta may provide constraints on combinations of satellite properties
and its history, and the galaxy potential.

\section{Observational applications}
\label{sec:observation}

We have demonstrated that tail morphology depends sensitively on the
satellite mass and orbit.  For modest to high-mass satellites, the
ejected tails have orbits that differ significantly from that of
the progenitor satellite. In this section, we illustrate the
observational implications of these results.

\subsection{Projected satellite tail morphology}
\label{aitoff}

\begin{figure*}
\centerline{
\includegraphics[width=0.4\textwidth,angle=0]{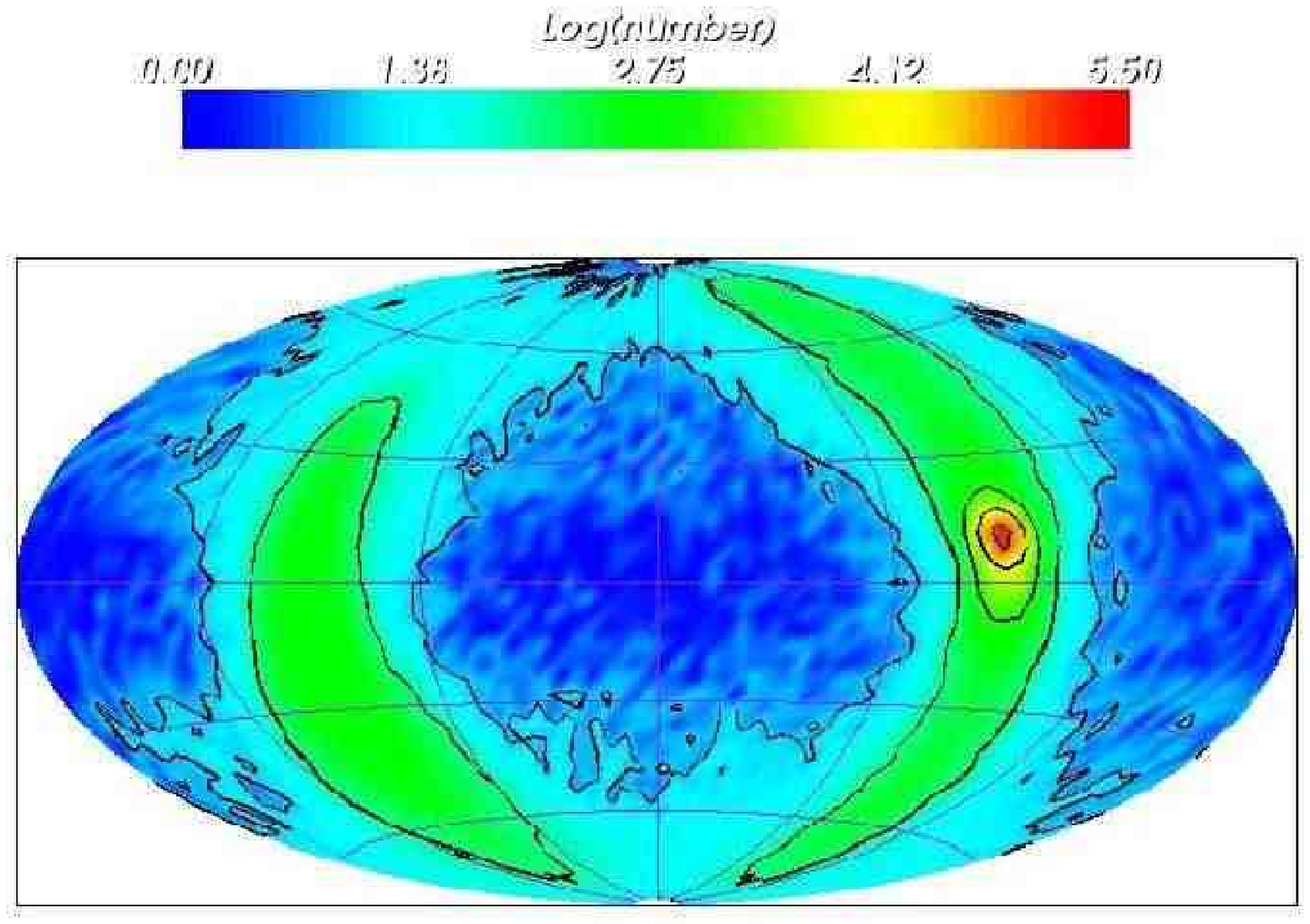}
\includegraphics[width=0.4\textwidth,angle=0]{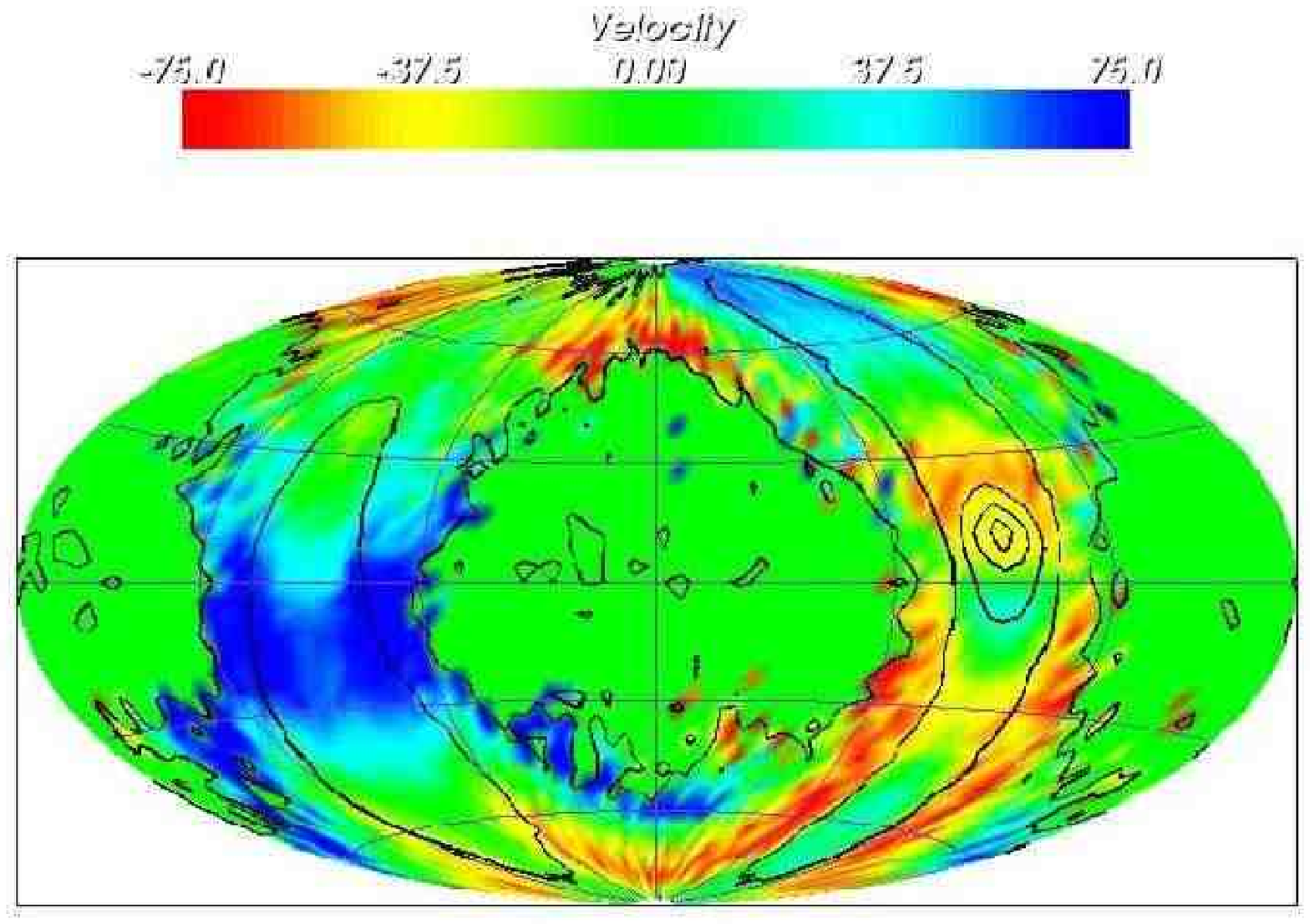} 
}
\centerline{
\includegraphics[width=0.4\textwidth,angle=0]{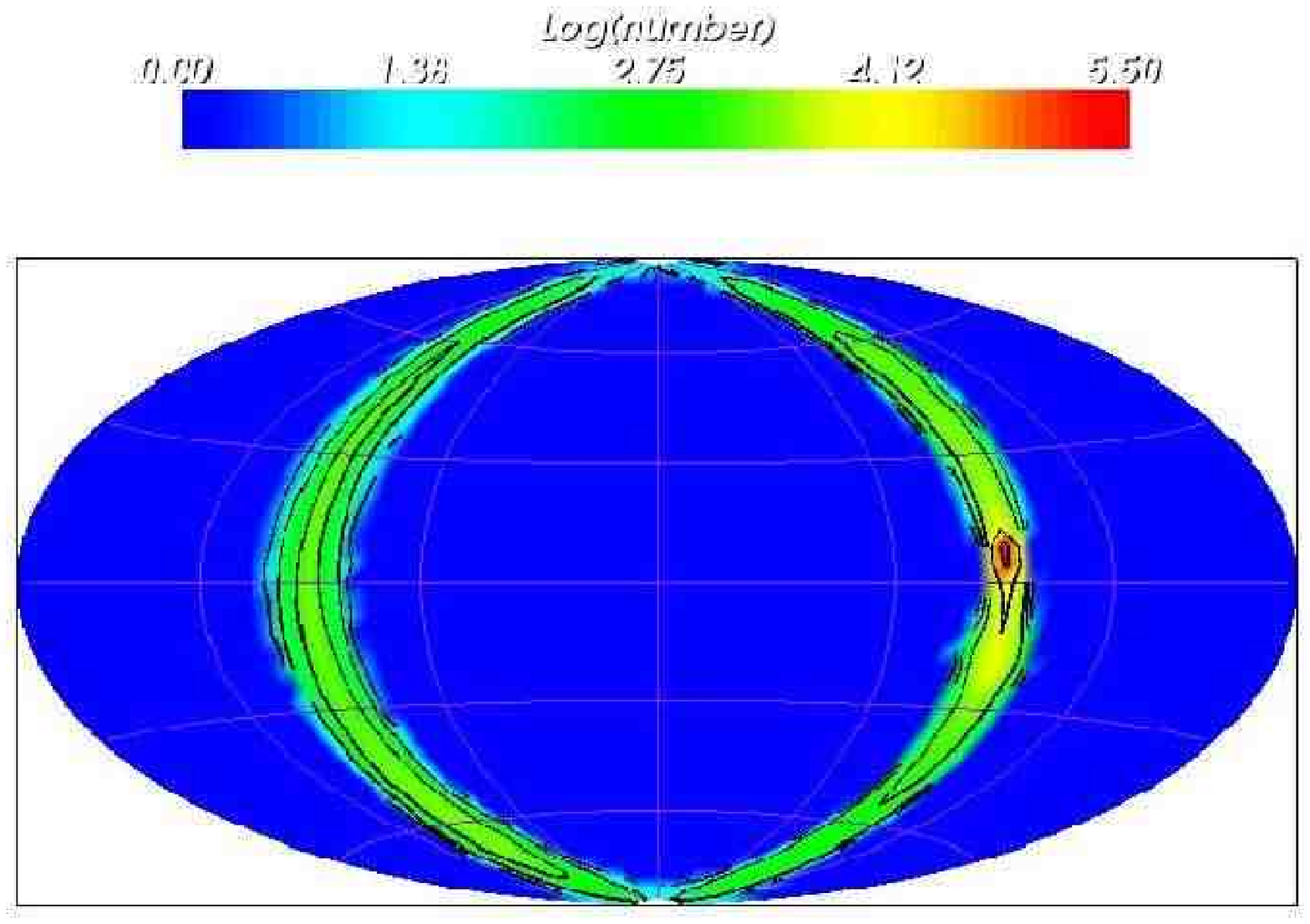}
\includegraphics[width=0.4\textwidth,angle=0]{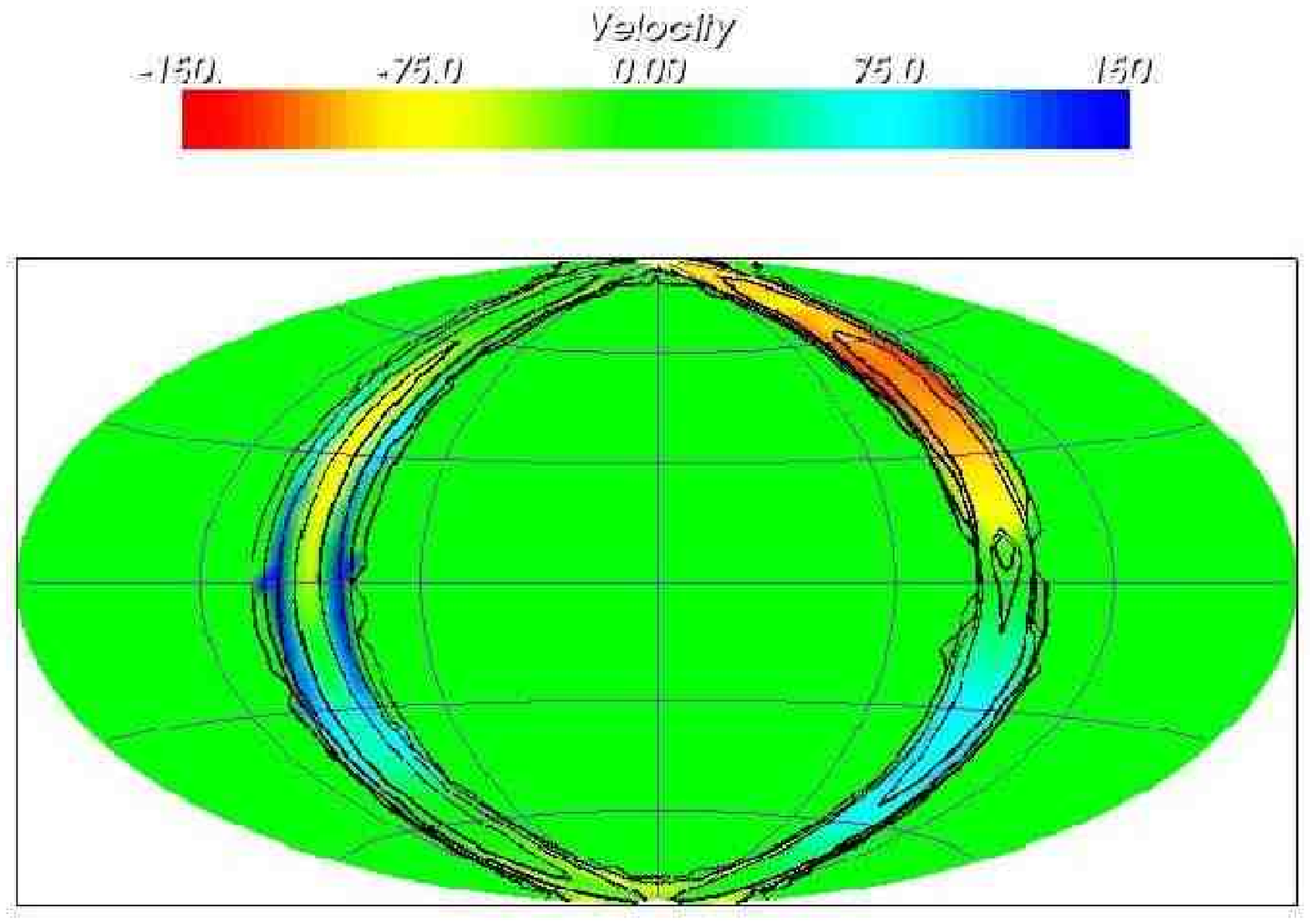} 
}
\caption {Aitoff projections of the massive satellite (top panels) and
  the low mass satellite (bottom panels) on an $e=0.5$
  orbit.  The snapshots for these projections are shown
  in Figure \ref{fig:FramesEcc}.  The observer is located at the centre
  of the host halo. The left panels show the number density of 
  particles and the right panels show the mean radial velocity.  The black
  isocontours in all four panels represent the 
  particle number density.  Colour bars show the number
  density (left) and radial velocity (right) scales. }
\label{fig:Aitoff}
\end{figure*}

The observational implications for Milky Way streams can be summarised
by projecting the tail star counts and radial velocity signatures
against the sky with the observer at the centre\footnote{A specific
  Milky Way model would take into account the solar position and an
  orbital estimate for a particular progenitor satellite.  However, in
  this study, the satellites are chosen to be only representative of
  CDM predictions and, in the same vein, the galaxy centre is an
  intuitively simple inner-galaxy view point.} of host halo.  Figure
\ref{fig:Aitoff} shows Aitoff projections of number density and mean
radial velocity for the massive satellite (top panels) and the
low-mass satellite (bottom panels) with an $e=0.5$ orbit (the same
simulations described in Figure \ref{fig:FramesEcc} 
at the same time, $T=5.0$). The Aitoff
projection covers the entire sky, $0^\circ \le l \le 360^\circ$ and
$-90^\circ \le b \le 90^\circ$, and the pixel size is $4^{o} \times
4^{o}$. The number density of the particles (left panels) and the mean
radial velocity (right panels) are coded by colour.  The contours in
all the panels represent the particle number density.  Velocity
outliers at low number density are trimmed by setting to ${\bar
  v}_r=0$ all the pixels with fewer than 10 particles.  The satellites are
located at $l\approx270^{o}$ and $b \approx 0^{o}$ and move in the
positive $b$ direction.

The radial velocity signatures of the massive and low-mass satellites
are distinctly different.  These qualitative differences are a direct
consequence of the large energy and angular momentum changes of the
ejecta orbits leading to the phase wrapping of the leading tail and the dramatic
broadening of the trailing tail (see Section \ref{sec:ecctail}).  This causes the
lower overall mean velocity values with a more rapid angular
variation around the sky.  In contrast, the mean
velocity of the leading and trailing tails for the low-mass
satellite are smooth and slowly vary around the sky.  Quite
clearly, the debris from the massive satellite will not show the
distinct kinematic and spatial signatures that have been exploited in
recent observational campaigns.

Near $b=0^\circ$ and $l=90^\circ$, one observes a region of receding
orbits surrounded in longitude by regions of approaching orbits.
Figure \ref{fig:FramesEcc} (right snapshot) shows that
line-of-sight projections will encounter strong leading and trailing
tails from same satellite at different radii.  The closer the tail to
the observer, the larger the angular extent perpendicular to the
motion of the stream.  Their velocity signature in the Aitoff projection
occurs as a single line of sight cuts through these tails at different radii.

The Aitoff projections contain most of the information that one
might obtain from combined kinematic--photometric surveys such as
RAVE \citep{Steinmetz06}.  In particular,
these results show that tail morphology depends on satellite
mass.  Therefore, a wide range of kinematic ``template'' models may be
required to best exploit the information implicit in observed halo stars.

\subsection{The effects on tidal tail radial velocity}
\label{sec:vr-l}

\begin{figure}
\centerline{\includegraphics[width=1.0\columnwidth,angle=0] {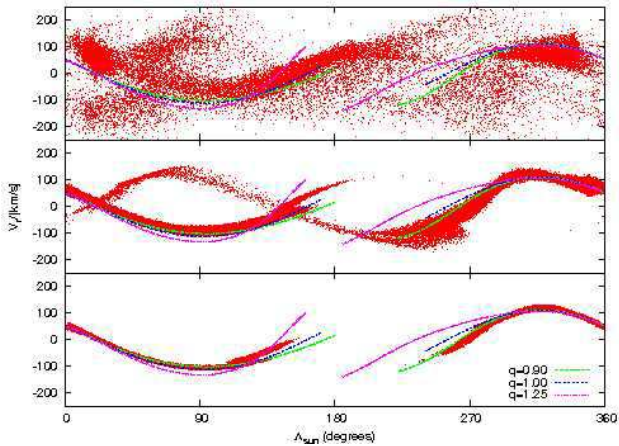}}
\caption {The radial velocity of tail particles as a function of
  orbital longitude for the massive satellite, low-mass satellite, and
  tiny mass satellite, from top to bottom, shown in Milky Way units.
  All three satellites have orbits with $e=0.5$ and are centred
  at a longitude of $0^{o}$.  The width of the distribution
  in $|v_r|$ increases with satellite mass. For comparison,
  the satellite trajectories for halos with flattenings of $q= 0.9, \:
  1.0, \: 1.25$ are also shown.  Although all three
  simulations are performed in a spherical host halo potential, the
  tail locus is better matched for a satellite trajectory with $q=
  0.9$.}
\label{fig:vr_l}
\end{figure}

Radial velocity--orbital longitude diagrams are frequently used to
characterise large-scale kinetic features in the Milky Way. Figure
\ref{fig:vr_l} shows radial velocity--orbital longitude diagrams for the
ejecta of satellites with orbits having $e=0.5$ for each of our three masses.
We convert simulation units to Milky Way units by assuming a
virial radius of 250 kpc and a total mass of $1.0 \times 10^{12}
M_{\odot}$ \citep{KZS02}.  In Figure
\ref{fig:vr_l}, the Sun has ${\bf R}=(-8.0, 0.0, 0.0)$ kpc and the
Galactic plane and the satellite's orbital plane are coincident.  
Here we adopt the Sagittarius longitudinal coordinate system described 
in \citet{MSWO03} for the orbital longitude.
All satellites have ${\bf R}=(50.0, -7.5, 0.0)$ kpc and move in the $y$
direction.
Therefore, the satellite has $l\approx0^\circ$ and longitude
increases along the trailing tail (in the $-y$ direction).  
The radial velocity is measured from the halo centre.
The spread in $|v_{r}|$ is proportional to the satellite mass, as expected
from the previous discussion and hence the mean velocity will be an
unbiased diagnostic of the satellite orbit only for very low mass satellites.
Although we have only modelled the dark matter, it is likely
that the $v_{r}-l$ space distribution for stellar and dark matter ejecta
will be similar in most cases since the internal satellite velocity
dispersion plays only a minor role in shaping the ejecta distribution.

\citet{LJM05} use M giants from the Two Micron All-Sky Survey
\citep[2MASS,][]{Skrutskie06} to map the position and velocity distributions
of tidal debris from the Sagittarius dwarf spheroidal galaxy. Assuming
that tidal tails approximately align with the satellite trajectory,
the authors note that the radial velocity distribution of tidal debris
suggests an prolate Milky Way halo with an axis ratio of $q=c/a=1.25$.  However,
our results demonstrate that the tails do \emph{not} follow the
satellite orbit.  In Figure \ref{fig:vr_l}, we also plot satellite
trajectories for three different halo flattenings to compare with the
particle distributions of our simulations evolved in a spherical host halo.
Following \citet{LJM05}, we flatten our host halo parallel to the satellite's
motion and compute point-mass satellite trajectories to compare with
our simulated $v_r-l$ diagrams.  Surprisingly, the distributions
of the low-mass satellite and the tiny-mass satellite tidal tails
most closely matches a $q=0.9$ halo.  The gravitational acceleration by the
satellite shifts the tail location in the radial velocity distribution
and this trend is degenerate with the effects of halo flattening.
For instance, the location of the leading tails decelerated by a
massive satellite is degenerate with the trajectories of tails in
an oblate halo with no satellite deceleration.  We have not attempted to
model the Milky Way in sufficient detail to estimate the halo
flattening including satellite deceleration.  However, the degeneracy
between halo flattening and the shift caused by the satellite gravitational
acceleration suggests that the
\citet{LJM05} conclusions may be biased and a more careful analysis
including the full dynamics of the halo-satellite interaction is
necessary.

\subsection{The effects on the tidal tail phase space distribution}
\label{sec:psp}

Several groups have proposed phase-space-based detection diagnostics
for moving groups associated with disrupted dwarf galaxy and star
cluster streams.  \citet{LL95} proposed using the intrinsic
correlation of moving groups' radial energy and galactocentric radius
to identify disrupted systems.  The procedure is as follows.  Assuming
a spherical gravitational potential for the outer galaxy, one
estimates the radial energy $E_{r} = v_{r}^{2}/2 + \Phi(r)$ and the
galactocentric radius $r$ of the putative ejecta stars from
observations.  Then, assuming that all of the debris from a single
satellite has the same orbital energy, $E$, and angular momentum, $L$,
conservation of energy implies a simple linear relationship in
$r^{-2}$: $E = E_r - L^2/2r^2$.  Hence, linear features in the
observed $E_r$--$r^{-2}$ diagram indicate the detection of a tidal
stream.  Recently \citet{belokurov07} used this method to support the
detection of stellar streams in the Sloan Digital Sky Survey.

\begin{figure}
\centerline{\includegraphics[width=1.0\columnwidth,angle=0]{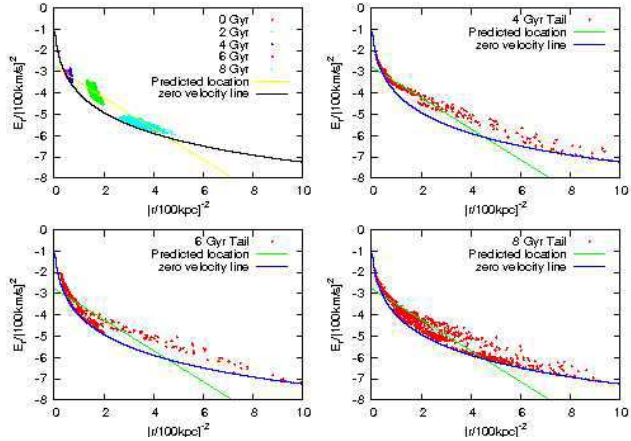}}
\caption { Radial energy, $E_{r}=v_{r}^{2}/2 + \Phi(r)$, plotted
  against inverse galactocentric radius, $r^{-2}$, for the low-mass
  satellite with an $e=0.5$ orbit. Particles with the same energy and
  angular momentum as the satellite will lie on a particular straight
  line.  The zero velocity curve describes the lowest possible energy
  at a given radius for this host halo.  The material bound to the
  satellite follows the straight locus at all times (top left).  The
  escaped particles are shown at $T=2.0$ (top right), $T=3.0$ (bottom
  left) and $T=4.0$ (bottom right), respectively.  Only one particle
  out of 250 are plotted for visibility.  Because the leading
  (trailing) tail loses (gains) energy, the ejecta deviates from the
  predicted straight line with significant scatter. Times in Gyrs are scaled
  to the Milky Way.}
\label{fig:Er_dSph2}
\end{figure}

\begin{figure}
\centerline{\includegraphics[width=1.0\columnwidth,angle=0]{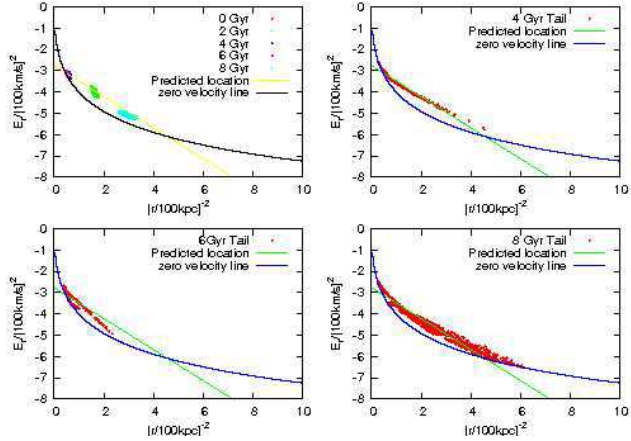}}
\caption{The same as Figure \ref{fig:Er_dSph2} but for the tiny-mass satellite.
}
\label{fig:Er_dSph3}
\end{figure}

However, as we have now seen, a massive satellite will modify the
conserved quantities of the ejecta orbits and change their location in
$E_{r}-r^{-2}$ space.  Figures \ref{fig:Er_dSph2} and
\ref{fig:Er_dSph3} show the $E_{r}-r^{-2}$ diagrams for the low-mass
and tiny-mass satellite simulations on an $e=0.5$ orbit.  For
clarity, we have reduced the point density by randomly sampling the
simulation phase space and plot the bound particles at five different times in
the upper-left panels.  We calculate the expected linear relation from
the satellite's initial position and velocity. The bound material in
low-mass and tiny-mass satellites lies along the predicted linear
relation at all times.  We plot the tail particles at three different
times in the other three panels. As one can see in Figure
\ref{fig:Er_dSph2}, the deviation of the tail particles from the
predicted locus and the scatter in $E$ at fixed $r^{-2}$ for the
low-mass satellite is large.  Especially at late times, e.g. the
bottom right panel in Figure \ref{fig:Er_dSph2}, the tail nearly fills
the region between the zero velocity curve and the predicted locus.
However, one can see from Figure \ref{fig:Er_dSph3} that tail
particles from the tiny-mass satellite do follow the predicted linear
relation. Therefore, we conclude that the \citet{LL95} diagnostic can
only detect streams from very low-mass satellites such as globular
clusters.

\begin{figure}
\centerline{\includegraphics[width=1.0\columnwidth,angle=0]{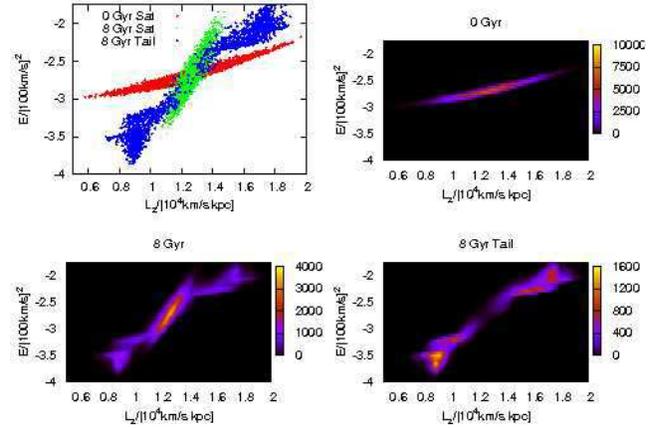}}
\caption {
   The distribution of particles bound to the low-mass satellite on
  an $e=0.5$ orbit and its ejected tail particles plotted in
  $E$--$L_z$ space. {\it Top left:} The phase-space distribution at $T=0.0$
  and $T=4.0$ for the bound satellite and its tail (subsampled as in
  Figure \ref{fig:Er_dSph2}). Density plots of the phase-space
  distribution are shown at $T=0.0$ (top right), at $T=4.0$ (bottom
  left) and at $T=4.0$ for the ejected tail particles alone (bottom
  right).
  }
\label{fig:PSP_dSph2}
\end{figure}

\begin{figure}
  \centerline{\includegraphics[width=1.0\columnwidth,angle=0]{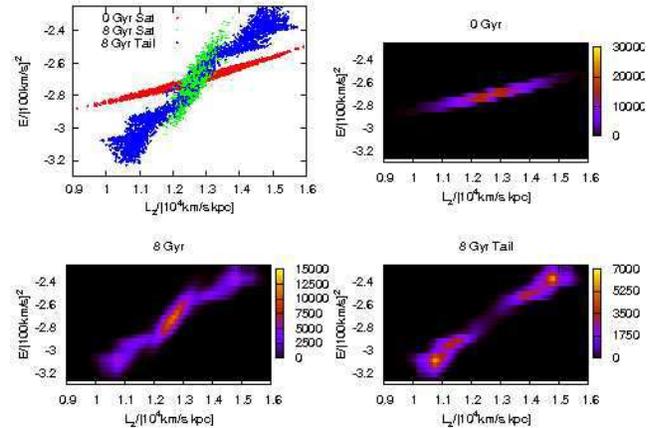}}
  \caption { The same as Figure \ref{fig:PSP_dSph2} but for the tiny-mass
  satellite}  \label{fig:PSP_dSph3} 
\end{figure} 

Motivated by the prospect of six-dimensional phase-space data from
future astrometric missions, \citet{HdZ00} proposed to identify
phase-mixed satellite debris by a cluster analysis in $(E, L, L_{z})$
space.  We explore the consequences of tail evolution on this approach
using the same two simulations in Figure \ref{fig:PSP_dSph2} for the
low-mass satellite and in Figure \ref{fig:PSP_dSph3} for the tiny-mass
satellite.  For simplicity, we assume that we know the orbital
orientation and consider only the $E$--$L_z$ projection.  The top-left
panels show the distribution at 0 Gyr ($T=0$) and at 8 Gyr ($T=4$)
when scaled to the Milky Way halo in $E$--$L_z$ space. Once again, we
randomly sampled the material to improve clarity.  The top-right and
bottom-left panels show density estimates in $E$--$L_z$ space for the
satellite at 0 Gyr and at 8 Gyr, respectively.  In the bottom-right
panels we show the density of only the tail particles at 8 Gyr.  The
overall position of the satellite and its tail changes little from 0
Gyr to 8 Gyr, although the shape of the distribution shifts.  In both
figures, there are two or more peaks at high and low energy with
respect to the satellite owing to decelerations and accelerations of
tail particles by the satellite potential. Moreover, the tidal field
is nonaxisymmetric and this leads to spatial correlations in the
energy and angular momentum of the least bound satellite particles, which in
turn leads to the production of several apparently disassociated
phase-space clumps before disruption.

\section{Discussion and summary}
\label{sec:summary}

The observational detection of ``S''- or ``Z''-shaped tidal tails in
globular clusters \citep[e.g.][]{Leon.etal:00, Odenkirchen.etal:03,
  Grillmair.Dionatos:06} promises sensitive statistical tests of the
Galaxy's gravitational potential and has renewed the quest for streams
from larger satellites.  For globular clusters, i.e. very low mass
satellites, the tidal tail morphology is easily interpreted.  
\citep{CapuzzoDolcetta.etal05,Montuori.etal07} However,
for massive satellites, the bisymmetry that leads to this simple
morphology is broken by the interaction of the host halo's
gravitational field and the self-gravity of the satellite itself.  We present
new dynamical aspects and morphologies of tidal tails produced in
satellites of significant mass, $M_s/M_h\ge 0.0001$.  There are two
dynamical principles that affect the tail production for massive
satellites.  First, the leading and trailing X-points, points where
the attractive force of the host halo and satellite are balanced at
zero velocity, do not occur at equal distances from the centre nor do
they have the same equipotential value for large-mass satellites (see Figure
\ref{fig:jacobi}).  Second, the escaped ejecta in the leading
(trailing) tail continues to be decelerated (accelerated) by the
satellite's gravity leading to large offsets of the ejecta
orbits from the satellite's original orbit (see Figure
\ref{fig:FramesEcc}).  We show that this is consistent with
Hill-Jacobi theory (generalised to dark-matter halos) for satellites
on circular orbits.  In particular, the effect of the satellite's self
gravity on the tail decreases only weakly with decreasing satellite
mass, proportional to $(M_s/M_h)^{1/3}$ (see Section \ref{sec:circtail}) and,
therefore, the acceleration by the satellite after escape is important
for dwarfs and dark halos of modest mass.

These findings have several important and useful theoretical and
observational consequences.  First, for a finite mass satellite, the
morphology of the leading and trailing tails will be different owing to
the gradient in the underlying halo potential across the satellite.
In addition, the tail ejection occurs over a range of azimuth relative
to the X-point owing to the dynamical response of the originally
prograde and retrograde orbits to the tidal and non-inertial
acceleration.  These effects should be observable in high resolution
imaging for both dwarf spheroidal and globular clusters (see Figs.
\ref{fig:FramesEcc}--\ref{fig:FramesMTK2}). 

Second, the radial velocity of tail particles will be displaced from
that of the satellite orbit.  The
magnitude of the displacement is proportional to the satellite mass.
These trends distort the ejecta from the gravitationally bound
satellite trajectory in the $v_{r}-l$ plane in much the same sense as
a satellite trajectory in a flattened halo (see Figure \ref{fig:vr_l}).
In other words, in fitting the $v_r-l$ diagram for tidal tails to
satellite orbits of different flattenings, the satellite mass is
covariant with halo flattening, i.e. the \emph{shape} parameter $q=c/a$.
Therefore, a constraint on the Milky Way halo shape using tidal
streams requires mass-dependent modelling.  Finally, the acceleration
of ejecta by a massive satellite during escape spreads the velocity
distribution and obscures the signature of a well-defined ``moving
group'' in phase space (see Figs.
\ref{fig:Er_dSph2}--\ref{fig:PSP_dSph3}).

Although we believe that the physical effects described in this study
are robust, our intentionally idealised simulations ignore several
possibly relevant processes.  First, the dynamical friction and the
self-gravitation of the tail are ignored, although in all but the most
extreme mass satellites their effects on the tail morphology will be
negligible, since the mass in the tail is very small.
Second, we assume a smooth and static spherical host halo potential.
In reality, over time, as the host halo mass grows its shape may change,
and the ejecta will be perturbed by substructure.  These time
dependent effects will not affect the applicability of the dynamics
described here but will complicate the prediction of observational
signatures.  Finally, we have not included the physics of a dissipational
baryonic component that may have slightly different kinematics than that of a
dark collisionless component.  In spite of these shortcomings, our
study elaborates the details of satellite tidal tail production and
the dynamics that bear on the interpretation of observed streams.

As an example, \citet{MD94} and \citet{Johnston96} find that
satellite tails follow the satellite orbit for dwarf galaxies whose
mass is negligible compared to the galaxy mass. 
The mass of these satellites is usually similar to or less than 
our tiny-mass satellite.
Using these simulation results, \citet{Johnston01} developed 
an efficient numerical method to
investigate the detectability and interpretation of tidal debris tails.
However, we have demonstrated here that the gravity of the satellite
for $M_s/M_h\ge0.0001$ will change the actions of the tidal ejecta to
mimic halo flattening (see Section \ref{sec:vr-l}, Figure  \ref{fig:vr_l}).
In addition to the spatial distribution, the velocity distribution of
the tail is affected by the satellite potential (see Section \ref{sec:psp}).
We would like to note that \citep{FFM06} have also noticed a systematic 
distance offset for leading (trailing) tails inside (outside) the 
orbit of a satellite owing to the satellite potential. However,
they did not focus on the tail morphology.

In summary, we have shown that the interplay between the satellite and
the host halo results in a complex tail morphology whose amplitude scales
weakly with mass.  Although these findings complicate the
interpretation of stellar streams and moving groups, the intrinsic
mass dependence provides additional leverage on both the halo and on the
progenitor satellite properties.  A statistical study of these trends
will further constrain the dark halo potential and the mass
accretion history of the Milky Way.

\section*{Acknowledgements}
We would like to thank an anonymous referee for many useful comments.
This work was supported in part by NSF AST-0205969, and NASA ATP
NAG5-12038, NAGS-13308, \& NNG04GK68G. 


\end{document}